\documentclass[
 twocolumn,
 preprintnumbers,
 amsmath,amssymb,
 aps,
 superscriptaddress,
]{revtex4}

\usepackage{color}
\usepackage{graphicx}
\usepackage{bm}
\usepackage{here}
\usepackage{url}
\usepackage{subfigure}
\usepackage{mhchem}
\usepackage{comment}
\usepackage{siunitx}

\newcommand{\argmax}{\mathop{\rm arg~max}\limits}

\begin{document}
\title{Estimating entropy production by machine learning of short-time fluctuating currents}

\author{Shun Otsubo}
\affiliation{
 Department of Applied Physics, The University of Tokyo, 7-3-1 Hongo, Bunkyo-ku, Tokyo 113-8656, Japan\\
}
\author{Sosuke Ito}
\affiliation{
Universal Biology Institute, The University of Tokyo, 7-3-1 Hongo, Bunkyo-ku, Tokyo 113-0031, Japan\\}
\affiliation{JST, PRESTO, 4-1-8 Honcho, Kawaguchi, Saitama, 332-0012, Japan\\}
\author{Andreas Dechant}
\affiliation{
WPI-Advanced Institute of Materials Research (WPI-AIMR), Tohoku University, Sendai 980-8577, Japan\\}
\author{Takahiro Sagawa}
\affiliation{
 Department of Applied Physics, The University of Tokyo, 7-3-1 Hongo, Bunkyo-ku, Tokyo 113-8656, Japan\\
}
\date{\today}

\begin{abstract}
Thermodynamic uncertainty relations (TURs) are the inequalities which give lower bounds on the entropy production rate using only the mean and the variance of fluctuating currents. Since the TURs do not refer to the full details of the stochastic dynamics, it would be promising to apply the TURs for estimating the entropy production rate from a limited set of trajectory data corresponding to the dynamics. Here we investigate a theoretical framework for estimation of the entropy production rate using the TURs along with machine learning techniques without prior knowledge of the parameters of the stochastic dynamics.  Specifically, we derive a TUR for the short-time region and prove that it can provide the exact value, not only a lower bound, of the entropy production rate for Langevin dynamics, if the observed current is optimally chosen. This formulation naturally includes a generalization of the TURs with the partial entropy production of subsystems under autonomous interaction, which reveals the hierarchical structure of the estimation.  We then construct estimators on the basis of the short-time TUR and machine learning techniques such as the gradient ascent.  By performing numerical experiments, we demonstrate that our learning protocol {\color{black}performs well even} in nonlinear Langevin dynamics. We also discuss the case of Markov jump processes, where the exact estimation is shown to be impossible in general. Our result provides a platform that can be applied to a broad class of stochastic dynamics out of equilibrium, including biological systems.
\end{abstract}


\maketitle
\section{Introduction}
In the last two decades, our understanding of thermodynamics of fluctuating small systems has grown substantially, leading to the modern formulation of stochastic thermodynamics \cite{Jarzynski1997, Sekimoto2010, Seifert2012}.
It enables us to explore the fundamental properties of non-equilibrium systems \cite{Hatano2001}, and also has been extended to information thermodynamics by incorporating information contents \cite{Allahverdyan2009, Sagawa2010, Toyabe2010, Sagawa2012, Ito2013, Horowitz2014, Hartich2014, Parrondo2015, Ito2015, Shiraishi2015, Rosinberg2016}.
One of the most fundamental discoveries is the fluctuation theorem \cite{Crooks1999, Jarzynski2000} that reveals a symmetry of the entropy production by including the full cumulants of stochastic dynamics. Stochastic thermodynamics has also been applied to biophysical situations \cite{Ritort2006, Toyabe2010a}.
\\ \indent
Recently, another fundamental relation called the thermodynamic uncertainty relation (TUR) has been proposed \cite{Barato2015, Horowitz2019}. The TUR gives a lower bound on the entropy production rate $\sigma$ with a time-averaged current observable $j_d$:
 \begin{eqnarray}
 \sigma\geq2\frac{\left<j_d\right>^2}{\tau {\rm Var}(j_d)},\label{eq: TUR}
 \end{eqnarray}
where $\left<j_d\right>$ and ${\rm Var}(j_d)$ are the mean and the variance of $j_d$, and $\tau$ is the length of the time interval over which $j_d$ is observed (see Sec.~\ref{sec: theory} for the details).
An advantage of this relation lies in the fact that it does not require information on the full cumulants of the entropy production, at the cost that it only gives a lower bound.
Since the TUR implies that the entropy production rate is nonzero, reversibility is not achieved for finite $\tau$ as long as the variance is finite \cite{Pietzonka2018, Pietzonka2016}.
Rigorous proofs are provided for continuous-time Markov jump processes in the long-time limit $\tau\rightarrow\infty$ \cite{Gingrich2016}, and later for the finite-time case \cite{Pietzonka2017, Horowitz2017} using the large deviation techniques. Since then, a variety of extensions of the TUR have been considered, for example, in discrete-time systems \cite{Proesmans2017}, periodically driven systems \cite{Barato2018}, active particles \cite{Cao2019}, overdamped \cite{Dechant2018b, Dechant2019} and underdamped Langevin equations \cite{VanVu2019, Lee2019}, processes under measurement and feedback control \cite{Vu2019, Potts2019}. Moreover, several techniques \cite{Lahiri2016, Pigolotti2017, Dechant2018a, Hasegawa2019} have been adopted for the derivation of the TUR such as the Cram\`{e}r-Rao inequality \cite{Lahiri2016, Hasegawa2019}, which leads to generalizations \cite{Dechant2018, Ito2018, Ito2018a, Liu2019, Hasegawa2019a, Terlizzi2019, Falasco2019, Wolpert2019, Timpanaro2019, Guarnieri2019} of the original TUR.\\ \indent
Since the demand for estimation of the entropy production is ubiquitous \cite{Lan2012, Martinez2016, Battle2016, Seara2018}, various estimators of the entropy production have been investigated.
While some of them are based on the fluctuation theorem \cite{Frishman2018, Roldan2010, Lander2012, Martinez2019, Kim2020},
the TUR provides a simpler strategy for estimating the entropy production rate.
For the latter, in fact, we only need to know the mean and the variance of a current by adopting the following procedure: Find a current that maximizes the right-hand side (rhs) of Eq.~(\ref{eq: TUR}), and use the rhs as an estimate \cite{Gingrich2017, Li2019, Busiello2019, Manikandan2019}.
This approach has turned out to be promising because it was numerically suggested that the estimation can become exact in Langevin processes if we use currents in the short-time limit $\tau\rightarrow0$ (i.e., the short-time TUR) \cite{Manikandan2019}.
{\color{black}The dependence on the time interval $\tau$ has also been analytically studied using a concrete Langevin model \cite{Manikandan2018}.}\\ \indent
In this paper, we propose a framework for estimation of the entropy production rate inspired by this approach.
First, we prove the short-time TUR both for Markov jump processes and Langevin dynamics, and establish their equality conditions:
we prove that the equality is always achievable in Langevin dynamics by optimally choosing a generalized current, while this is not the case in Markov jump processes.
Our formulation naturally leads to a generalized TUR for an information-thermodynamics setting, in which subsystems are autonomously interacting and their partial entropy productions are relevant \cite{Allahverdyan2009, Horowitz2014, Hartich2014, Shiraishi2015, Rosinberg2016}.\\ \indent
On the basis of the above analytical results, we construct estimators of the entropy production rate for machine learning techniques such as the gradient ascent. 
Our estimators adopt model functions that can avoid the problem of overfitting.
The performance of these estimators is evaluated in several setups: (i) two or five dimensional linear Langevin equations, (ii) a two dimensional non-linear Langevin equation, and (iii) a one-dimensional Markov jump process.
We show that our method outperforms previously  proposed estimators \cite{Li2019} for the non-linear Langevin case (ii) in terms of the convergence speed, while these estimators are comparable for the linear case (i).
This is because our estimator does not assume that the distribution is Gaussian, suggesting that our method works well for a broader class of dynamics including nonlinear and non-Gaussian cases.
We numerically confirm that the exact value of the entropy production rate can be indeed obtained by our estimation method for Langevin dynamics.\\ \indent
We also demonstrate that the exact estimation is achievable both in the equilibrium limit and in the Langevin limit of the model of Markov jump process (iii).
In addition, we show that, as another advantage of the TUR-based estimators in Markov jump processes, they are robust against the sampling interval $\Delta t$ at least in one-dimensional systems.
This property is important for applications to biological systems, where it is often hard to capture elementary processes using a detector with finite time resolution \cite{Yasuda2001, Keegstra2017}.\\ \indent
This paper is organized as follows. In Sec.~\ref{sec: theory}, we derive the short-time TUR and prove the equality condition. In Sec.~\ref{sec: estimators}, we propose learning estimators after discussing the advantage of machine learning in our setting. In Sec.~\ref{sec: numerical experiment}, we numerically evaluate the performance of the estimators in the above-mentioned setups. In Sec.~\ref{sec: conclusions}, we summarize our results and make concluding remarks. 
In Appendix~\ref{sec: adam}, we explain the details of the gradient ascent.
{\color{black}In Appendix~\ref{sec: details}, we give a complete explanation for the estimators used in this study. In Appendix~\ref{sec: move}, we show the results regarding the scalability of our approach for higher dimensional data.}
\\ \indent

\section{Thermodynamic uncertainty relation in the short-time limit}
\label{sec: theory}
In this section, we consider the equality condition of the TUR in the short-time limit.
We first formulate the short-time TUR for Markov jump processes, and consider the equality condition.
We show that although the equality condition cannot be satisfied in general Markov jump processes, it can be asymptotically satisfied in the (i) equilibrium and (ii) Langevin limits.
Indeed, we analytically prove that the equality condition can be satisfied in Langevin dynamics even in far from equilibrium.
Our formulation includes the TUR with the partial entropy production rate of subsystems which interact with each other autonomously.
We also reveal the hierarchy of the lower bound on the entropy production rate when not all the currents are used.
We note that the analytical formulation in this section does not assume steady states, while in the subsequent sections we numerically estimate the entropy production rate using trajectories sampled from steady states.\\ \indent
We first formulate the short-time TUR for Markov jump processes. We consider a system with a finite number of states, where the transitions between the states are modeled by a continuous-time Markov jump process, where the transition rate from state $y$ to state $z$ is given by $r(y, z)$. We define an integrated empirical current on a transition edge from $y$ to $z$ as
\begin{eqnarray}
J_\tau(y, z) := \int_0^\tau dt(\delta_{x(t^-), y}\delta_{x(t^+), z} - \delta_{x(t^-), z}\delta_{x(t^+), y}),
\end{eqnarray}
{\color{black}where $x(t^\pm)$ represents the state of the system before (after) the jump at time $t$.} We define the empirical current as $j_\tau(y, z) := J_\tau(y, z)/\tau$, and define a generalized current $j_d$ as a linear combination of the empirical currents:
\begin{eqnarray}
j_d = \sum_{y < z}d(y, z) j_\tau(y, z),\label{eq: general}
\end{eqnarray}
where $d(y, z)$ are some coefficients. For example, if we take the thermodynamic force
\begin{eqnarray}
F(y, z) = \ln\frac{p(y)r(y, z)}{p(z)r(z, y)},
\end{eqnarray}
as $d(y, z)$, the generalized current equals the entropy production rate \cite{Esposito2010}.
Note that, we set the Boltzmann's constant to unity $k_B = 1$ throughout this study.
\\ \indent
In this study, we only consider the case of $\tau\rightarrow0$, which enables us to discuss the equality condition analytically.
In the short-time limit, using the probability distribution $p(x)$, the mean and the variance of the integrated current can be written as
{\color{black}
\begin{eqnarray}
\left<J_\tau(y, z) \right> &=&  \left\{p(y)r(y, z) - p(z)r(z, y)\right\}\tau + O(\tau^2),\\
{\rm Var}(J_\tau(y, z)) &=& \left\{p(y)r(y, z) + p(z)r(z, y)\right\}\tau \nonumber\\
&-& \left\{p(y)r(y, z) - p(z)r(z, y)\right\}^2\tau^2 + O(\tau^2),~~~~
\end{eqnarray}}
which is derived by considering the fact that $J_\tau(y, z)$ counts 1 (resp. $-1$) when a jump from $y$ to $z$ (resp. $z$ to $y$) occurs, and its probability is $p(y)r(y, z)$ (resp. $p(z)r(z, y)$).
Therefore, the mean and the variance of $j_\tau(y, z)$ becomes
{\color{black}
\begin{eqnarray}
\left<j_\tau(y, z)\right> &=& p(y)r(y, z) - p(z)r(z, y),\\
\tau{\rm Var}(j_\tau(y, z)) &=& \frac{{\rm Var}(J_\tau(y, z))}{\tau}\\
&=&  p(y)r(y, z) + p(z)r(z, y)
\end{eqnarray}
}
to the leading order in $\tau$. The partial entropy production rate associated with a transition from $y$ to $z$ is defined as \cite{Horowitz2014}
\begin{eqnarray}
\sigma_{(y, z)} := \left\{p(y)r(y, z) - p(z)r(z, y)\right\}\log\frac{p(y)r(y, z)}{p(z)r(z, y)}.
\end{eqnarray}
We now claim the following relation as the short-time TUR for $\sigma_{(y, z)}$:
\begin{eqnarray}
\sigma_{(y, z)}\frac{\tau{\rm Var}(j_\tau(y, z))}{\left<j_\tau(y, z)\right>^2}\geq2.\label{eq: t_TUR}
\end{eqnarray}
This relation can be proved as follows:
\begin{widetext}
\begin{subequations}
\begin{eqnarray}
\sigma_{(y, z)}\frac{\tau{\rm Var}(j_\tau(y, z))}{\left<j_\tau(y, z)\right>^2} &=& \left\{p(y)r(y, z) - p(z)r(z, y)\right\}\log\frac{p(y)r(y, z)}{p(z)r(z, y)}\frac{p(y)r(y, z) + p(z)r(z, y)}{\left\{p(y)r(y, z) - p(z)r(z, y)\right\}^2}\\
&\geq&2\frac{\left\{p(y)r(y, z) - p(z)r(z, y)\right\}^2}{p(y)r(y, z) + p(z)r(z, y)}\frac{p(y)r(y, z) + p(z)r(z, y)}{\left\{p(y)r(y, z) - p(z)r(z, y)\right\}^2}\\
&=& 2,
\end{eqnarray}
\end{subequations}
\end{widetext}
where we used the inequality $(a-b)\ln a/b\geq 2(a-b)^2/(a+b)$ \cite{Shiraishi2016}.\\ \indent
We next show the short-time TUR for a subsystem by summing up the above inequality using the Cauchy-Schwartz inequality.
In the limit $\tau\rightarrow0$, the variance of the generalized current becomes
\begin{eqnarray}
\tau{\rm Var}(j_d) = \sum_{y<z}d(y, z)^2 \left\{p(y)r(y, z) + p(z)r(z, y)\right\},
\end{eqnarray}
which is based on the fact that all of $j_\tau(y, z)$ are mutually independent to the leading order in $\tau$. The partial entropy production rate \cite{Allahverdyan2009, Horowitz2014, Hartich2014, Shiraishi2015, Rosinberg2016} of a subsystem $X$ can be written as
\small
\begin{eqnarray}
\hspace{-0.2cm}
\sigma_X =\sum_{y < z, ~(y, z)\in\mathcal{X}}\left\{p(y)r(y, z) - p(z)r(z, y)\right\}\log\frac{p(y)r(y, z)}{p(z)r(z, y)},~~
\end{eqnarray}
\normalsize
where the transition edges within the subsystem $X$ are denoted as $\mathcal{X}$. Here, we assume that the transitions within the subsystem $X$ and those within the remaining system are bipartite \cite{Horowitz2014}, i.e., occur independently of each other.
{\color{black}For example, if we consider a system described by the direct product of subsystems $X$ and $Y$, the bipartite condition means the following:
\begin{eqnarray}
r\left(\left\{x, y\right\}, \left\{x', y'\right\}\right) = 0 ~~ {\rm if}~ x\neq x' ~{\rm and}~ y\neq y',
\end{eqnarray}
where $r\left(\left\{x, y\right\}, \left\{x', y'\right\}\right)$ is the transition rate from state $\left\{x, y\right\}$ to $\left\{x', y'\right\}$.
With this condition, $\mathcal{X}$ denotes the set of transitions $(\{x, y\}, \{x', y'\})$ such that $x' \neq x$ and $y' = y$.}\\ \indent
We define $\mathcal{N}$ as the set of transitions $(y, z)$ such that $d(y, z)\neq0$.
If $\mathcal{N}\subset\mathcal{X}$ is satisfied, the following relation holds:
\begin{eqnarray}
\sigma_X\frac{\tau{\rm Var}(j_d)}{\left<j_d\right>^2}\geq2\label{eq: ITUR},
\end{eqnarray}
which we call the short-time TUR for the subsystem $X$.
This inequality can be proved as follows:
\begin{widetext}
\begin{subequations}
\begin{eqnarray}
\sigma_X\frac{\tau{\rm Var}(j_d)}{\left<j_d\right>^2} &=& \sum_{y < z,~ (y, z)\in\mathcal{X}}\left\{p(y)r(y, z) - p(z)r(z, y)\right\}\log\frac{p(y)r(y, z)}{p(z)r(z, y)}\frac{\sum_{y<z}d(y, z)^2 \left\{p(y)r(y, z) + p(z)r(z, y)\right\}}{\left[\sum_{y<z}d(y, z)\left\{p(y)r(y, z) - p(z)r(z, y)\right\}\right]^2}~~~~~~\\
&\geq& \sum_{y < z,~ (y, z)\in\mathcal{N}}\frac{2d(y, z)^2\left\{p(y)r(y, z) - p(z)r(z, y)\right\}^2}{d(y, z)^2\left\{p(y)r(y, z) + p(z)r(z, y)\right\}}\frac{\sum_{y<z, (y, z)\in\mathcal{N}}d(y, z)^2 \left\{p(y)r(y, z) + p(z)r(z, y)\right\}}{\left[\sum_{y<z, (y, z)\in\mathcal{N}}d(y, z)\left\{p(y)r(y, z) - p(z)r(z, y)\right\}\right]^2}~~~~~~~~\label{eq: ITUR3}\\
&\geq&2\label{eq: ITUR4},
\end{eqnarray}
\end{subequations}
\end{widetext}
where we used $\sum a_i^2 \sum b_i^2\geq \left(\sum a_ib_i \right)^2$ (the Cauchy-Schwarz inequality) in Eq.~(\ref{eq: ITUR4}). 
The condition $\mathcal{N}\subset\mathcal{X}$ means that the generalized current is only driven by the transitions within $X$, which is a natural condition to derive the uncertainty relation.
This is an extension of the TUR in the presence of measurement and feedback control \cite{Vu2019, Potts2019} to more general settings, in which subsystems interact with each other autonomously.
If we take $X$ as the total system, the TUR (\ref{eq: ITUR}) reduces to the well-known form for finite time $\tau$ \cite{Horowitz2017}. In the following, we omit the subscript $X$ when the total entropy production rate is considered.\\ \indent
We introduce $d^*$ as the optimal $d$ that saturates the Cauchy-Schwartz inequality (\ref{eq: ITUR4}), which can be explicitly written as
{\color{black} 
\begin{eqnarray}
d^*(y, z) = c~\frac{p(y)r(y, z) - p(z)r(z, y)}{p(y)r(y, z) + p(z)r(z, y)},\label{eq: optimal_markov}
\end{eqnarray}
where $c$ is a constant which reflects a degree of freedom in $d^*$.}
On the other hand, the equality of (\ref{eq: ITUR3}) does not hold in general.
We therefore consider the two limits that asymptotically satisfy the equality when $\mathcal{N} = \mathcal{X}$ is satisfied: (i) the equilibrium limit and (ii) the Langevin limit.
The equilibrium limit is a well-known equality condition of the finite-time TUR \cite{Manikandan2018, Busiello2019}, which states that
$p(y)r(y, z) - p(z)r(z, y)$ goes to zero for all pairs of $y$ and $z$.
In this work, we newly find the Langevin limit, which states that $\Delta := 2\left\{p(y)r(y, z)-p(z)r(z, y)\right\}/\left\{p(y)r(y, z)+p(z)r(z, y)\right\}$ goes to zero while keeping 
$p(y)r(y, z) - p(z)r(z, y)$ finite for all pairs of $y$ and $z$. This can be proved by the following scaling analysis:
\begin{subequations}
\begin{eqnarray}
\ln\frac{p(y)r(y, z)}{p(z)r(z, y)} &=& \ln\left(1 + \frac{\Delta}{1-\Delta/2}\right)\\
&=& \Delta + O(\Delta^3),
\end{eqnarray}
\end{subequations}
which means that the equality of (\ref{eq: ITUR3}) can be achieved as the second order convergence as $\Delta$ goes to zero.
This result suggests a striking fact that the equality condition is always achievable in Langevin dynamics even if the state is far from equilibrium {\color{black}by taking $d = d^*$ for $(y, z)$ in $\mathcal{X}$, and $d=0$ otherwise.}
\\ \indent
Indeed, we can reproduce the above result directly in the Langevin setup as follows.
We first formulate the short-time TUR with the following overdamped Langevin equations with $M$ variables ${\bm x} = (x_1, x_2, ..., x_M)$:
\begin{eqnarray}
\dot{{\bm x}} = {\bm A}({\bm x}(t), t) + \sqrt{2}{\bm G}({\bm x}(t), t)\cdot{\bm \xi}(t),
\end{eqnarray}
where ${\bm A}({\bm x}, t)$ is a drift vector, ${\bm G}({\bm x}, t)$ is an $M\times M$ matrix, ${\bm \xi}(t)$ 
is the uncorrelated white noise satisfying $\left<\xi_i(t)\xi_j(s)\right> = \delta_{ij}\delta(t-s)$ and we use $\cdot$ to denote the Ito-convention. We set the length of the time interval as infinitesimal time $\tau=dt$.
In this setup, the empirical current in the short-time limit ${\bm j}({\bm x}) \tau := \delta({\bm x}(t) - {\bm x})\circ d{\bm x}(t)$, which is in turn defined using the Stratonovich product $\circ$, can be transformed as
\small
\begin{subequations}
\begin{eqnarray}
j_i({\bm x})\tau &=&\! \frac{\delta({\bm x}(t+\tau)-{\bm x})-\delta({\bm x}(t)-{\bm x})}{2}dx_i(t)\nonumber\\
&&+ \delta({\bm x}(t)-{\bm x})dx_i(t)\\
&=& \!\frac{1}{2}\sum_j [ \nabla_j \delta({\bm x}(t)-{\bm x}) ] dx_j(t)dx_i(t)\nonumber\\
&&+ \delta({\bm x}(t)-{\bm x})dx_i(t) + O(\tau^\frac{3}{2})\\
&=&\!\sum_{j, l} [ \nabla_j \delta({\bm x}(t)-{\bm x})] G_{il}G_{jl}\tau \nonumber\\
&&\! + \! \delta({\bm x}(t) \! - \! {\bm x})(A_i \tau \!+ \!\sum_l \! \sqrt{2}G_{il}dw_l) \! +\! O(\tau^\frac{3}{2}),~~~~
\end{eqnarray}
\end{subequations}
\normalsize
where $d{\bm w}(t) := {\bm \xi}(t) \tau$. We note that ${\bm j}({\bm x})$ is a stochastic variable that depends on the realization of ${\bm x}(t)$. Its ensemble average satisfies
\begin{subequations}
\begin{eqnarray}
\left<j_i ({\bm x})\right>&=& \! \int d{\bm x} (t) P({\bm x}(t),t) j_i({\bm x})  \\
&=&\! -\sum_j\nabla_j \left[ B_{ij} P({\bm x},t)\right] + A_i P({\bm x},t)\\
&=:& \tilde{\jmath}_i ({\bm x}, t),
\end{eqnarray}
\end{subequations}
where we defined ${\bm B}:= {\bm G}{\bm G}^\mathsf{T}$ whose (i, j) element is written as $B_{ij}$.\\
\indent
Next, we calculate the ensemble average of the generalized current and its variance. For the Langevin case, the generalized current for the vector ${\bm d}$ is defined as
\begin{subequations}
\begin{eqnarray}
j_{\bm d} \tau &:=& \! \sum_i d_i({\bm x} (t), t)\circ dx_i(t)\\
&=&  \!  \sum_{i,j,l} \nabla_j\left(d_i\right)G_{il}G_{jl} \tau \nonumber\\
&& +\sum_i d_i(A_i \tau + \sum_l \sqrt{2}G_{il}dw_l).~~~~~
\end{eqnarray}
\end{subequations}
The calculation of its ensemble average can be conducted in a similar manner to that of $\left<{\bm j}({\bm x})\right>$:
\small
\begin{subequations}
\begin{eqnarray}
\left<j_{\bm d}\right>= \int d{\bm x} {\bm d}^\mathsf{T} \tilde{\bm \jmath} .
\end{eqnarray}
\end{subequations}
\normalsize
The variance of the generalized current is calculated as
\small
\begin{subequations}
\begin{eqnarray}
 \tau{\rm Var}(j_{\bm d}) &:=& \left(\left<j_{\bm d}^2\right> - \left<j_{\bm d}\right>^2\right)\tau\\
&=&\!  \int d{\bm x}(t)\frac{P({\bm x}(t),t)}{\tau} \left[\sum_{i,j,l} \nabla_j\left(d_i\right)G_{il}G_{jl} \tau \right. \nonumber\\
&&\! \left. +\sum_i d_i(A_i \tau + \sum_l \sqrt{2}G_{il}dw_l) \right]^2 \! - \! \left<j_{\bm d}\right>^2\tau  \nonumber\\
&=& 2\int d{\bm x}P{\bm d}^\mathsf{T}{\bm B}{\bm d}.
\end{eqnarray}
\end{subequations}
Then, the short-time TUR can be derived using the expression of the entropy production rate \cite{Spinney2012}
\begin{eqnarray}
\sigma = \int d{\bm x}\frac{\tilde{\bm \jmath}^\mathsf{T}{\bm B}^{-1}\tilde{\bm \jmath}}{P}
\end{eqnarray}
as
\small
\begin{subequations}
\begin{eqnarray}
\sigma\frac{\tau{\rm Var}(j_{\bm d})}{\left<j_{\bm d}\right>^2} &=&
\frac{2 \left[ \int d{\bm x} \frac{\tilde{\bm \jmath}^\mathsf{T}{\bm B}^{-1}{\bm B} {\bm B}^{-1}\tilde{\bm \jmath} }{P} \right] \left[\sum_{k,l} \int d{\bm x} P {\bm d}^\mathsf{T}{\bm B}{\bm d}  \right]}{\left(\int d{\bm x}  {\bm d}^\mathsf{T} \tilde{\bm \jmath} \right)^2}~~~~~~~\label{eq: proof_start}\\
&\geq&\frac{2\left(\int d{\bm x}\tilde{\bm \jmath}^\mathsf{T}{\bm B}^{-1} {\bm B}{\bm d}  \right)^2}{\left(\int d{\bm x}  {\bm d}^\mathsf{T} \tilde{\bm \jmath} \right)^2}\\
&=& 2,\label{eq: proof_end}
\end{eqnarray}
\end{subequations}
\normalsize
where in the third line we use the Cauchy-Schwartz inequality by considering the inner product: 
\begin{eqnarray}
\left<{\bm f} | {\bm g}\right> := \int d{\bm x} {\bm f}^\mathsf{T} {\bm B} {\bm g}.
\end{eqnarray}
The equality of the TUR can always be achieved by taking
\begin{subequations}
\begin{eqnarray}
d^*_i({\bm x}, t) &=& c~\frac{\sum_k\tilde{\jmath}_k({\bm x}, t)B_{ki}({\bm x}, t)^{-1}}{P({\bm x}, t)}\\
&=& c~\sum_{k}\nu_k({\bm x}, t)B_{ki}({\bm x}, t)^{-1}\label{eq: optimal},
\end{eqnarray}
\end{subequations}
where we defined the mean local velocity ${\bm \nu}({\bm x}, t) := \tilde{\bm \jmath}({\bm x}, t)/P({\bm x}, t)$, {\color{black}and $c$ is a constant.}
Thus, we have reproduced the result predicted by the scaling analysis in Markov jump processes.
{\color{black}Here, if we choose $c$ as $1$,} the optimal coefficient ${\bm d}^*({\bm x}, t)$ equals the thermodynamic force, and thus the generalized current becomes the entropy production rate itself, which is in accordance with the discussion in Ref.~\cite{Manikandan2019}.\\ \indent
The short-time TUR also holds for the partial entropy production rate \cite{Rosinberg2016} with this setup. In the Langevin dynamics, we regard the $i$th element of the coordinate as a subsystem.
Concretely, with the condition $d_j(x) = 0$ for $j\neq i$, we can prove the short-time TUR for a subsystem $i$ in a similar manner to Eq.~(\ref{eq: proof_start}) - (\ref{eq: proof_end}):
\small
\begin{subequations}
\begin{eqnarray}
\hspace{-0.8cm}
\sigma_i\frac{\tau{\rm Var}(j_{d_i})}{\left<j_{d_i} \right>^2} &=&
\frac{2\int d{\bm x} \frac{\tilde{\jmath}_iB_{ii}^{-1}\tilde{\jmath}_i}{P}\times\int d{\bm x} d_i B_{ii} d_i P}{\left(\int d{\bm x} d_i\tilde{\jmath}_i\right)^2}~~~~~\\
&\geq& 2.
\end{eqnarray}
\end{subequations}
\normalsize
\\ \indent
The short-time TUR with the partial entropy production rate reveals the hierarchical structure of the lower bound on the entropy production rate in the following sense.
When only generalized currents driven by a subsystem $i$ are used to calculate the lower bound, the following magnitude relation holds:
\begin{eqnarray}
\sigma \geq \sigma_i \geq\frac{2\left<j_{d_i}\right>^2}{\tau{\rm Var}\left(j_{d_i}\right)}.\label{eq: hierarchy}
\end{eqnarray}
Therefore, if accessible currents are limited, and they do not include sufficient information about the total system, 
the maximization of the lower bound can yield only $\sigma_i$ rather than $\sigma$.\\ \indent

\section{Estimators of entropy production rate}
\label{sec: estimators}
In this section, we present our framework for estimating the entropy production rate with limited amount of trajectory data using the short-time TUR and machine learning.
We first explain the overall idea of employing machine learning for this study, and introduce the method called gradient ascent. {\color{black}Then, we briefly introduce two learning estimators for Langevin dynamics.
Here, we aim to clarify their characteristics compared to previously proposed methods \cite{Frishman2018, Li2019}, and the details of their implementation are provided in Appendix \ref{sec: details}.}
We also formulate the estimation for Markov jump processes, and introduce a learning estimator and an estimator with a direct method.
\subsection{General idea and the gradient ascent}
We first discuss the motivation to use machine learning with the short-time TUR and introduce the gradient ascent.
We can construct an estimator of the entropy production rate by finding the optimal coefficient $d^*$ that maximizes the lower
bound of the TUR, i.e.,
\begin{eqnarray}
d^* &:=& \argmax_{d} \tilde{\sigma}[d]\label{eq: optimization}\\
\tilde{\sigma}[d]&:=&\frac{2\left<j_d\right>^2}{\tau{\rm Var}(j_d)}.\label{eq: rhs}
\end{eqnarray}
Then, $\tilde{\sigma}[d^*]$ is an estimator for both Markov jump and Langevin dynamics. 
In particular, $\tilde{\sigma}[d^*]$ gives the exact value in Langevin dynamics in the limit of $\tau\rightarrow 0$ as shown in Sec.~\ref{sec: theory}.
\\ \indent
If availability of trajectory data is limited in practical situations, it is not possible to calculate $\left<j_d\right>$ and ${\rm Var}(j_d)$, and thus it is not possible to numerically obtain the exact value of $d^*$. 
We remark that, while some estimators $\widehat{\left<j_d\right>}$ and $\widehat{{\rm Var}(j_d)}$ can be calculated from a finite-length trajectory, 
they generally differ from $\left<j_d\right>$ and ${\rm Var}(j_d)$.
Hereafter, we use the hat symbol to denote that the quantities are estimators calculated from the finite-length trajectory.
If we determine $d$ that maximizes a naively constructed quantity from Eq.~(\ref{eq: rhs}), 
\begin{eqnarray}
\widehat{\sigma}[d]:=\frac{2\widehat{\left<j_d\right>}^2}{\tau\widehat{{\rm Var}(j_d)}},\label{eq: rhs_hat}
\end{eqnarray}
then $\widehat{\sigma}[d]$ tends to be much bigger than the true entropy production rate because $d$ is overfitted to each realization of trajectories.
Therefore, our task is to construct a more sophisticated way of estimation that makes $d$ close to the optimal coefficient $d^*$, while avoiding overfitting as much as possible.\\ \indent
For that purpose, we employ ideas from machine learning. We first divide the whole trajectory data into two parts: training and test data. 
We only use the training data for calculating $\widehat{\left<j_d\right>}$ and $\widehat{{\rm Var}(j_d)}$ and then consider the maximization of $\widehat{\sigma}[d]|_{\rm train}$ constructed from the training data by using (\ref{eq: rhs_hat}).
We update $d$, starting from a random vector field, to increase the value of $\widehat{\sigma}[d]|_{\rm train}$. This process is called learning, and we check the progress of learning by monitoring the value of $\widehat{\sigma}[d]|_{\rm test}$ that is constructed from the test data by using (\ref{eq: rhs_hat}). The learning curve of $\widehat{\sigma}[d]|_{\rm test}$ often has a peak structure (see Appendix~\ref{sec: details}), which suggests that $d$ becomes overfitted to the training data after the peak. Therefore, we can expect that $d$ that gives the maximum of $\widehat{\sigma}[d]|_{\rm test}$ has high generalization performance, and thus we can adopt the $d$ for the estimation of the entropy production rate.\\ \indent
We next explain how to update $d$. 
For Langevin dynamics, ${\bm d}({\bm x})$ is a field over ${\bm x}$ and thus has an infinite number of parameters.
Thus we approximate ${\bm d}({\bm x})$ by a certain function $\tilde{\bm d}({\bm x}; {\bm a})$ with analytic expression and with a finite number of parameters ${\bm a}$ in Langevin dynamics. 
On the other hand, it is not necessary to consider such an approximation for Markov jump processes, because $d(y, z)$ already consists of a finite number of parameters.
We update the parameters using the method called the gradient ascent by regarding $\widehat{\sigma}[\tilde{d}]$ as the objective function $f({\bm a}) = \widehat{\sigma}[\tilde{d}]$, where the rhs depends on ${\bm a}$ through $\tilde{d}$.
The basic update rule of the gradient ascent is as follows:
\begin{eqnarray}
{\bm a} \leftarrow {\bm a} + \alpha\partial_{{\bm a}}f({\bm a}),
\end{eqnarray}
where $\alpha$ is the step size. 
Since the parameters are updated towards the direction in which $f({\bm a})$ increases the most,
the gradient ascent is an efficient algorithm for finding the maximum of the objective function $f({\bm a})$. Although the original functional of $\tilde{\sigma}[d]$ has only a single maximum {\color{black}up to the constant overall factor $c$} as shown in Eq. (\ref{eq: optimal_markov}) and (\ref{eq: optimal}), an approximated function $\widehat{\sigma}[\tilde{d}]$ can have a lot of local maxima, and thus the gradient ascent does not necessarily find the global maximum. Nevertheless, we observe that the gradient ascent works quite well in all the examples in our numerical experiment, which suggests that $\widehat{\sigma}[\tilde{d}]$ also has a simple landscape if we appropriately choose the analytic expression of $\tilde{\bm d}({\bm x}; {\bm a})$.\\ \indent
We note that a parameter that should be predetermined before the learning is called a hyperparameter; For example, the step size $\alpha$ of the gradient ascent is a hyperparameter. We specifically implement an algorithm called Adam \cite{Kingma2014} for the gradient ascent in this study, and we give a more detailed explanation on the Adam and hyperparameter tuning in Appendix~\ref{sec: adam}.

\subsection{Estimators for Langevin dynamics}
{\color{black} In this subsection, we formulate the estimation problem for Langevin dynamics. Then, we briefly introduce two learning estimators, and compare them with previously proposed methods: KDE \cite{Li2019} (kernel density estimation) and SFI \cite{Frishman2018} (stochastic force inference).
We give an overview of these methods here, while the details are provided in Appendix~\ref{sec: details}.\\ \indent
We first formulate the estimation problem. We consider the situation that we only have access to a finite-length trajectory $\{{\bm x}_0, {\bm x}_{\Delta t}, ..., {\bm x}_{n\Delta t}\}$, which is sampled from a stationary dynamics. In the case of Langevin dynamics, we regard each ${\bm d}\left(({\bm x}_{(i+1)\Delta t} + {\bm x}_{i\Delta t})/2\right)\cdot({\bm x}_{(i+1)\Delta t} - {\bm x}_{i\Delta t})/\Delta t$ as a realization of the short-time generalized current, and calculate its mean and variance to get $\widehat{\sigma}[{\bm d}]$.\\ \indent
As explained in the previous subsection, we construct learning estimators for Langevin dynamics by assuming concrete functions for the coefficient ${\bm d}({\bm x})$.
Two types of model functions are considered in this study. One is a histogram-like function which takes values on the space discretized into bins, and the other is a linear combination of Gaussian functions.
We call the learning estimators with these model functions the binned learning estimator $\widehat{\sigma}[{\bm d}_{\rm bin}]$ and the Gaussian learning estimator $\widehat{\sigma}[{\bm d}_{\rm Gauss}]$ respectively.
Here, we emphasize that the learning estimators do not assume any distributions for data points, which guarantees their high performance for nonlinear dynamics with non-Gaussian distributions.\\ \indent
In order to improve its data efficiency, a regularization term is added to the objective function $f({\bm a})$ of the binned learning estimator, and the estimator with regularization is denoted as $\widehat{\sigma}^\lambda[{\bm d}_{\rm bin}]$, where $\lambda$ is a parameter governing the magnitude of the regularization term.
In this study, we adopt the Gaussian learning estimator $\widehat{\sigma}[{\bm d}_{\rm Gauss}]$ for two dimensional data and the binned learning estimator $\widehat{\sigma}^\lambda[{\bm d}_{\rm bin}]$ for higher dimensional data. This is because the Gaussian learning estimator is found to be better than the binned learning estimator in terms of the data efficiency, while the Gaussian learning estimator is computationally costly for higher dimensional data (see Appendix~\ref{sec: details} for the comparison).\\ \indent
Our approach can be viewed as a method to fit the thermodynamic force field ${\bm F}({\bm x}):=\sum_{k}\nu_k({\bm x})B_{ki}({\bm x})^{-1}$ with these model functions, since the optimal coefficient field ${\bm d^*({\bm x})}$ is proportional to ${\bm F}({\bm x})$ as shown in Eq.~(\ref{eq: optimal}). In this sense, our approach is related to previously proposed methods, KDE \cite{Li2019} and SFI \cite{Frishman2018}, both of which estimate the thermodynamic force field in different ways. In KDE, the thermodynamic force at position ${\bm x}$ is estimated directly by using all the data points. KDE avoids overfitting by smoothing the estimate of the thermodynamic force field with a kernel function whose bandwidth is determined on the basis of the assumption that data points follow a Gaussian distribution.
In SFI, on the other hand, the thermodynamic force field is obtained by fitting the mean local velocity ${\bm \nu}({\bm x})$ and the diffusion matrix ${\bm B}({\bm x})$ with parameterized model functions respectively. SFI deals with the problem of overfitting by deriving a practical criterion to determine the number of parameters.\\ \indent
Since the quantitative comparison between our approach and SFI, both of which depend on the model functions, is not easy, we just clarify their qualitative difference here. Although the learning estimators cannot estimate ${\bm \nu}({\bm x})$ and ${\bm B}({\bm x})$ separately, the learning estimators have an advantage in that they can take any functions as the model function of ${\bm F}({\bm x})$. On the other hand, in SFI, the model functions of ${\bm \nu}({\bm x})$ and ${\bm B}({\bm x})$ are restricted to those which can be described by a linear combination of fixed basis functions.
Related to this point, we show that the representation ability of the model function indeed makes a difference in the data efficiency in Appendix \ref{sec: move}.
There is also a difference in the way to avoid overfitting. Our approach deals with the problem of overfitting simply by the data splitting scheme as described in Appendix~\ref{sec: adam}. This is enabled by the fact that we have the objective function $f({\bm a})$ to maximize, which is not the case for SFI.\\ \indent
We compare the learning estimators and KDE quantitatively in the next section. There are mainly two estimators for KDE, \footnotesize$\widehat{\dot{S}}_{\rm ss}^{\rm temp}$\normalsize and $\widehat{\sigma}[\widehat{\bm F}_{\rm sm}]$. These estimators estimate the thermodynamic force field by the kernel density estimation, and we describe the obtained field as $\widehat{\bm F}_{\rm sm}({\bm x})$. Concretely, \footnotesize$\widehat{\dot{S}}_{\rm ss}^{\rm temp}$\normalsize is defined by a temporal average:
\small
\begin{eqnarray}
&&\widehat{\dot{S}}_{\rm ss}^{\rm temp}:= \frac{1}{\tau_{\rm obs}}\int_0^{\tau_{\rm obs}}\widehat{\bm F}_{\rm sm}({\bm x}(t))\circ d{\bm x}(t)\\
&&=\frac{1}{N\Delta t}\sum_{i=1}^{N}\widehat{\bm F}_{\rm sm}\left(\frac{{\bm x}_{i\Delta t} + {\bm x}_{(i-1)\Delta t}}{2}\right)\left[{\bm x}_{i\Delta t} - {\bm x}_{(i-1)\Delta t}\right].~~~~~
\end{eqnarray}
\normalsize
On the other hand, $\widehat{\sigma}[\widehat{\bm F}_{\rm sm}]$ is based on the TUR, and simply defined by substituting $\widehat{\bm F}_{\rm sm}$ into $\widehat{\sigma}[{\bm d}]$. We adopt the short-time TUR for $\widehat{\sigma}[\widehat{\bm F}_{\rm sm}]$ in this study, while the finite-time TUR is used in the original paper \cite{Li2019} (and thus just a lower bound on the entropy production rate is obtained).\\ \indent}

\subsection{Estimators for Markov jump processes}
In Markov jump processes, we also imagine the situation that we only have access to trajectory data sampled from a stochastic jump dynamics at every discrete-time step $\Delta t$, i.e., a sequence of states $\{x_0, ..., x_{n\Delta t}\}$, which is often the case in real experiments. Thus, there is a loss of information regarding transitions which occur between the sampling time. Unlike the case of Langevin dynamics, we need to reconstruct the underlying jump dynamics to calculate the generalized current. Therefore, we first heuristically interpolate states between $x_{i\Delta t}$ and $x_{(i+1)\Delta t}$: $\{x_{i\Delta t}, x_{(i+1)\Delta t}\} \rightarrow \{x^i_0 \left(= x_{i\Delta t}\right), x^i_1, ..., x^i_{m_i} \left(= x_{(i+1)\Delta t}\right)\}$. Although this is a nontrivial task in general, such a reconstruction is always possible in one dimensional systems, for example, by connecting $x_{i\Delta t}$ and $x_{(i+1)\Delta t}$ with the shortest path. We note that such a reconstruction is not necessary if we take $\Delta t$ sufficiently small. Then, we regard each $\sum_{j=0}^{m_i-1}d(x^i_j, x^i_{j+1})/\Delta t$ as a realization of the short-time generalized current, and calculate $\widehat{\sigma}[d]$.\\ \indent
Since the coefficient $d(y, z)$ in Markov jump processes already consists of a finite number of parameters, it is not always necessary to assume an analytic function for $d$ if the number of transition edges is numerically tractable. In this study, we construct a learning estimator directly from the definition of $\widehat{\sigma}[d]$. We denote the estimator as $\widehat{\sigma}^M[d]$, and compare it with a simple estimator $\widehat{\sigma}^M_{\rm simple}$ that is based on the estimation of transition rates. Concretely, we define $\widehat{\sigma}^M_{\rm simple}$ by using a whole reconstructed jump sequence $\{x_0, x_1, ..., x_m\}$:
\begin{eqnarray}
\widehat{\sigma}_{\rm simple}^{\rm M} &:=& \sum_{y < z}\left\{\widehat{j}(y, z) - \widehat{j}(z, y)\right\}\ln\frac{\widehat{j}(y, z)}{\widehat{j}(z, y)},\\
\widehat{j}(y, z) &:=& \frac{1}{n\Delta t} \sum_{i=0}^{m-1}\chi_{y, z}(x_i, x_{i+1}),
\end{eqnarray}
where $\chi_{y, z}(x_{i}, x_{i+1}) := \delta_{y, x_i}\delta_{z, x_{i+1}}$.
\\ \indent

\section{Numerical experiments}
\label{sec: numerical experiment}
\begin{figure}[t]
\includegraphics[width = 0.9\linewidth]{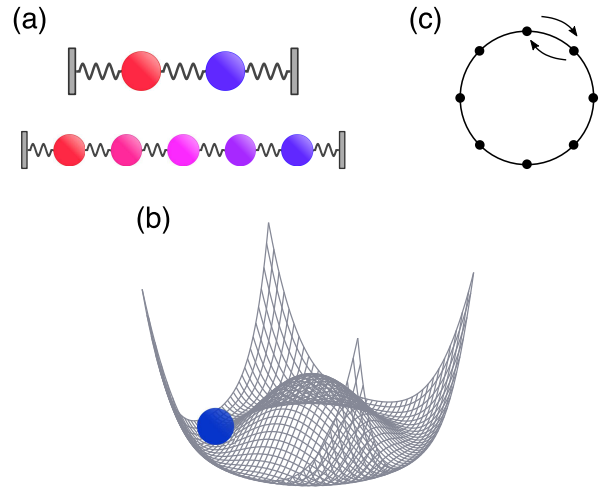}
\caption{Schematics of the models: (a) the $N$-beads model with $N = 2$ and $N = 5$, (b) the Mexican-hat potential model, (c) the one-dimensional hopping model.}
\label{fig: models}
\end{figure}
We now perform numerical experiments. Specifically, we compare the learning estimators $\widehat{\sigma}[{\bm d}_{\rm Gauss}]$ or $\widehat{\sigma}^\lambda[{\bm d}_{\rm bin}]$ with the KDE estimators \footnotesize$\widehat{\dot{S}}_{\rm ss}^{\rm temp}$\normalsize and $\widehat{\sigma}[\widehat{{\bm F}}_{\rm sm}]$ in Langevin processes, and compare $\widehat{\sigma}^{\rm M}[d]$ with $\widehat{\sigma}_{\rm simple}^{\rm M}$ in a Markov jump process.
Their performance is evaluated using finite-length trajectory data sampled from the steady states which are simulated  by the following four models: (i) a two-beads Langevin model, (ii) a five-beads Langevin model, (iii) a two-dimensional Langevin model with a Mexican-hat potential and (iv) a one-dimensional hopping model.
For (i) and (ii), the learning estimators $\widehat{\sigma}^\lambda[{\bm d}_{\rm bin}]$ or $\widehat{\sigma}[{\bm d}_{\rm Gauss}]$
show a performance comparable to those of \footnotesize$\widehat{\dot{S}}_{\rm ss}^{\rm temp}$\normalsize and $\widehat{\sigma}[\widehat{{\bm F}}_{\rm sm}]$. For (iii), on the other hand, the Gaussian learning estimator $\widehat{\sigma}[{\bm d}_{\rm Gauss}]$ outperformes \footnotesize$\widehat{\dot{S}}_{\rm ss}^{\rm temp}$\normalsize and $\widehat{\sigma}[\widehat{{\bm F}}_{\rm sm}]$, because the model is nonlinear and the stationary distribution deviates from a Gaussian distribution. For (iv), we first show that the optimal estimation with the short-time TUR converges to the true entropy production rate in both the Langevin limit and the equilibrium limit. Then, the learning estimator $\widehat{\sigma}^{\rm M}[d]$ is shown to converge fast compared to the direct method $\widehat{\sigma}_{\rm simple}^{\rm M}$. We also show that the learning estimator is robust against the choice of the sampling interval of trajectory data.\\ \indent

\subsection{$N$-beads model}
\label{sec: Nbeads}
\begin{figure*}
\begin{center}
\begin{tabular}{cc}
 	\multicolumn{2}{c}{
	\subfigure[]{
		\includegraphics[width = 0.435\linewidth]{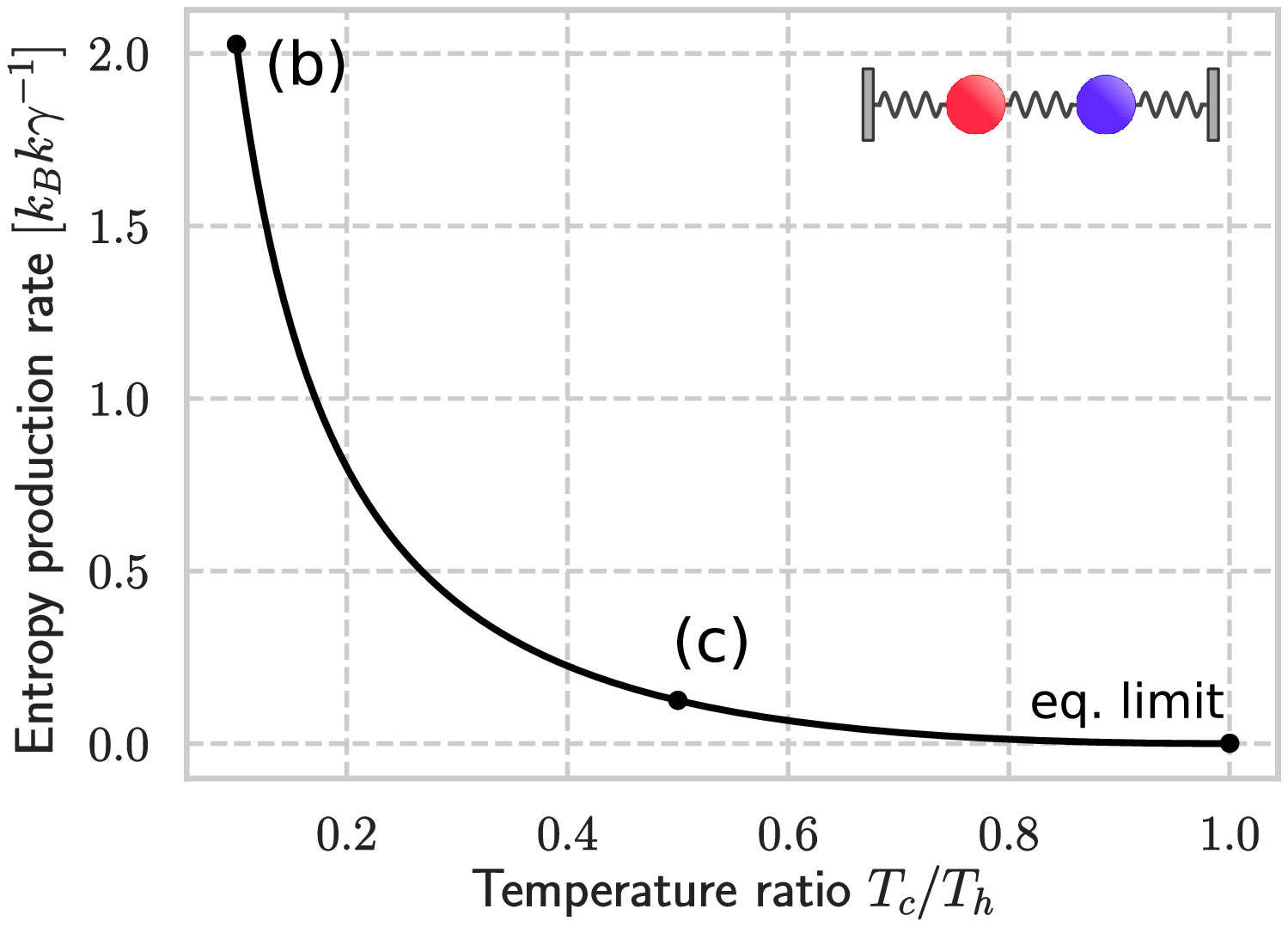}\label{fig: 2beads_true}}
	}\\
	\subfigure[$T_c/T_h = 0.1$]{
		\includegraphics[width = 0.435\linewidth]{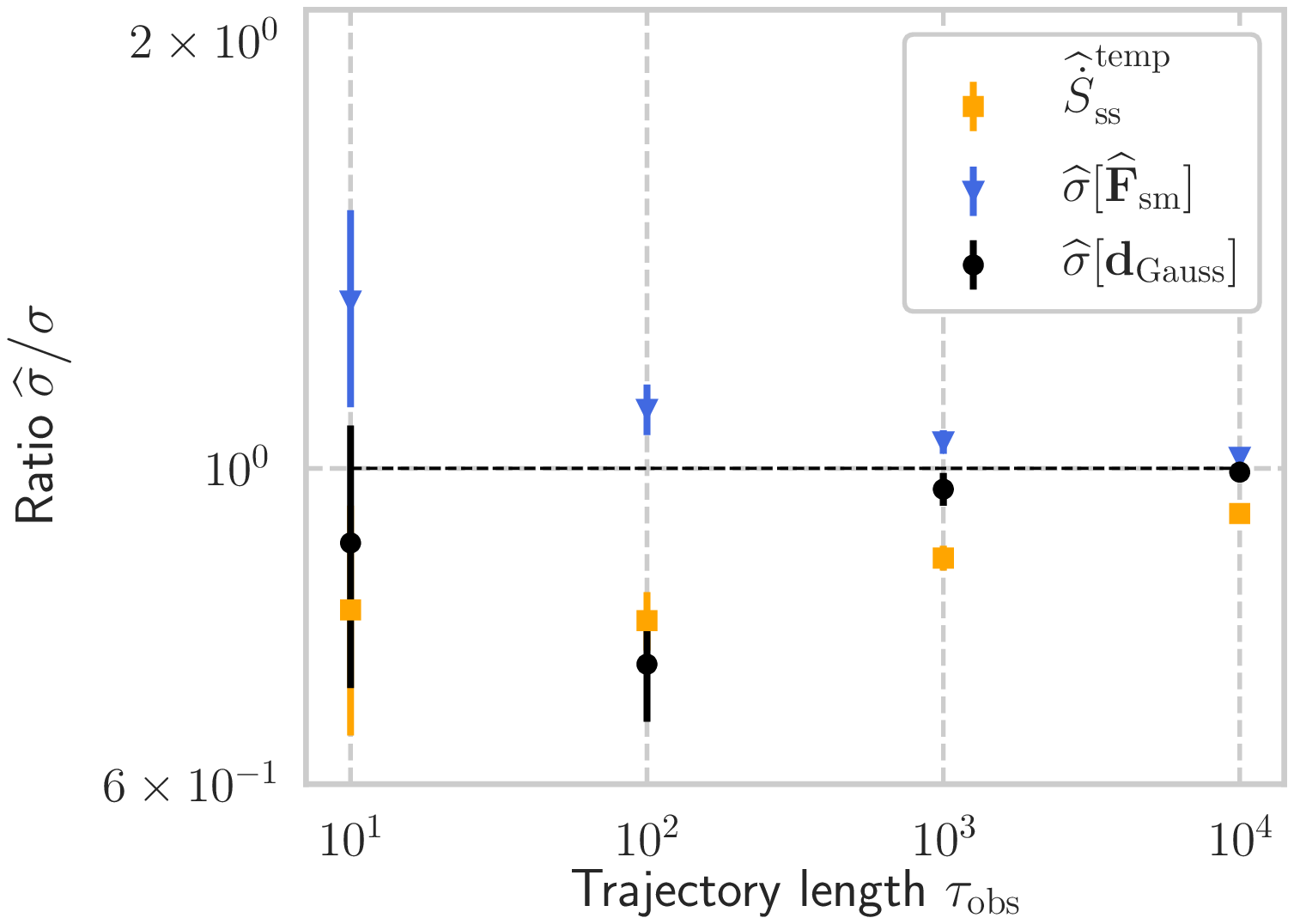}\label{fig: 2beads_main_r01}}&
	\subfigure[$T_c/T_h = 0.5$]{
		\includegraphics[width = 0.435\linewidth]{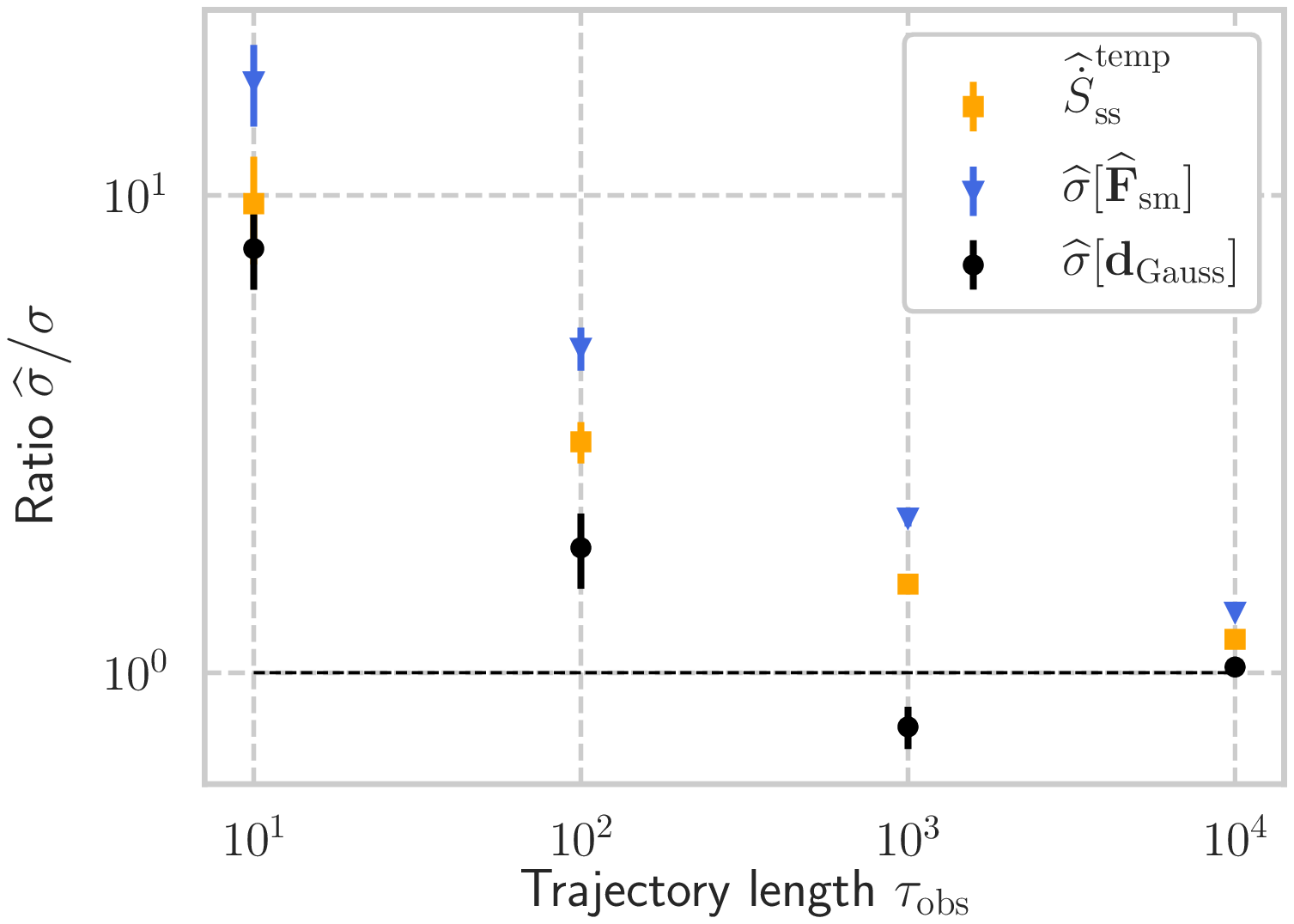}\label{fig: 2beads_main_r05}}
\end{tabular}
\caption{Numerical experiment with the two-beads model: (a) The dependence of the entropy production rate on the temperature ratio $T_c/T_h$. (b)(c) Performance of the estimators at (b) $T_c/T_h = 0.1$ and (c) $T_c/T_h = 0.5$ with \footnotesize$\widehat{\dot{S}}_{\rm ss}^{\rm temp}$\normalsize (yellow squares), $\widehat{\sigma}[\widehat{{\bm F}}_{\rm sm}]$ (blue triangles) and $\widehat{\sigma}[{\bm d}_{\rm Gauss}]$ (black circles). The mean and its standard deviation of ten independent trials are plotted. The Gaussian learning estimator uses the hyperparameters listed in TABLE~\ref{table: hyperparameters}, and the other system parameters are set as $k = \gamma = 1$ and $T_h = 250$. {\color{black}The sampling interval of the trajectories is set as $\Delta t = 10^{-3}$, and thus the number of data points is $10^3\tau_{\rm obs}$, half of which is used for the training, and the other half for the estimation in the case of $\widehat{\sigma}[{\bm d}_{\rm Gauss}]$.}}
\label{fig: 2beads}
\end{center}
\end{figure*}
\begin{figure*}
\begin{center}
\begin{tabular}{cc}
	\multicolumn{2}{c}{
	\subfigure[]{
		\includegraphics[width = 0.44\linewidth]{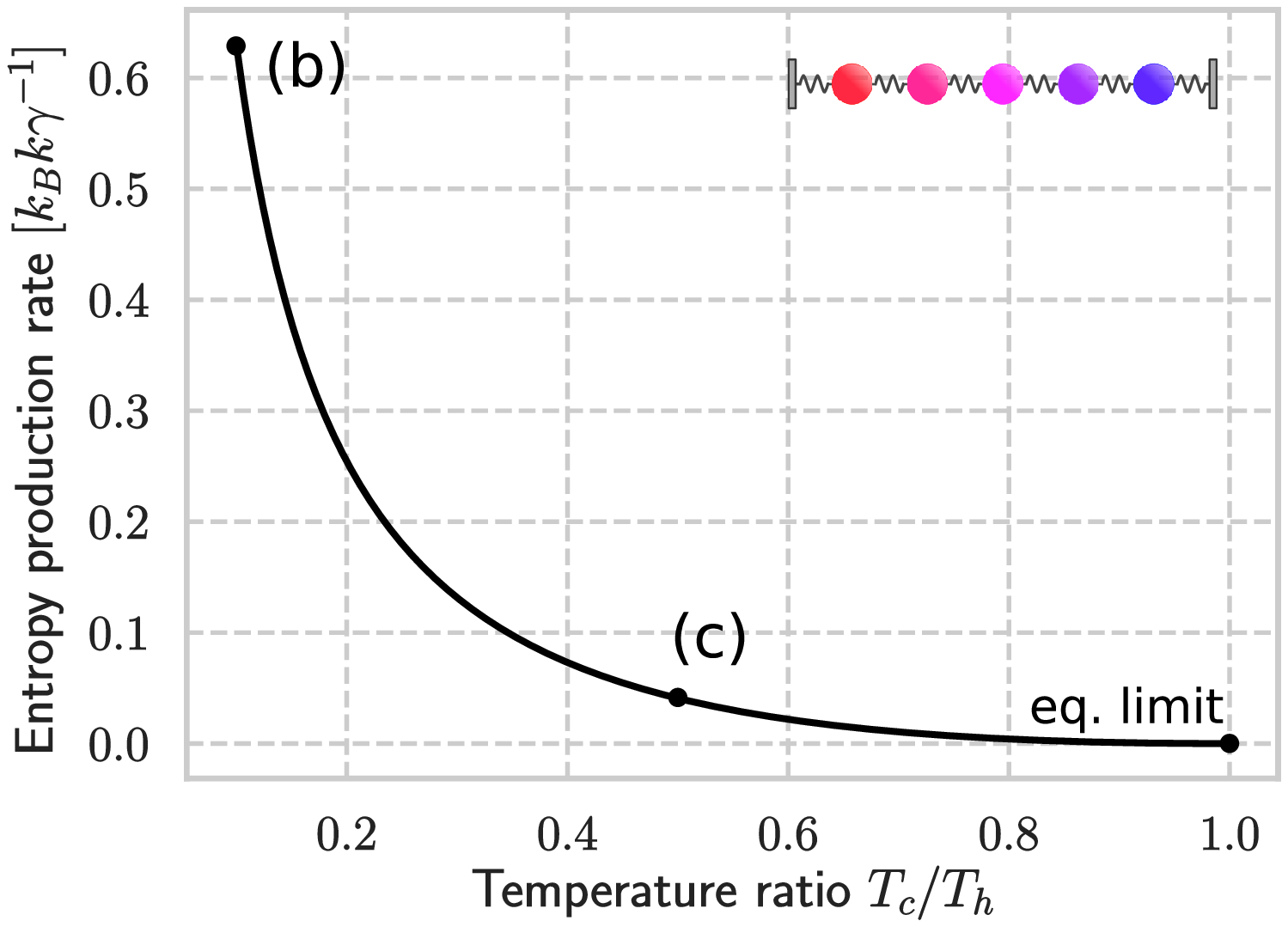}\label{fig: 5beads_true}}
	}\\
	\subfigure[$T_c/T_h = 0.1$]{
		\includegraphics[width = 0.44\linewidth]{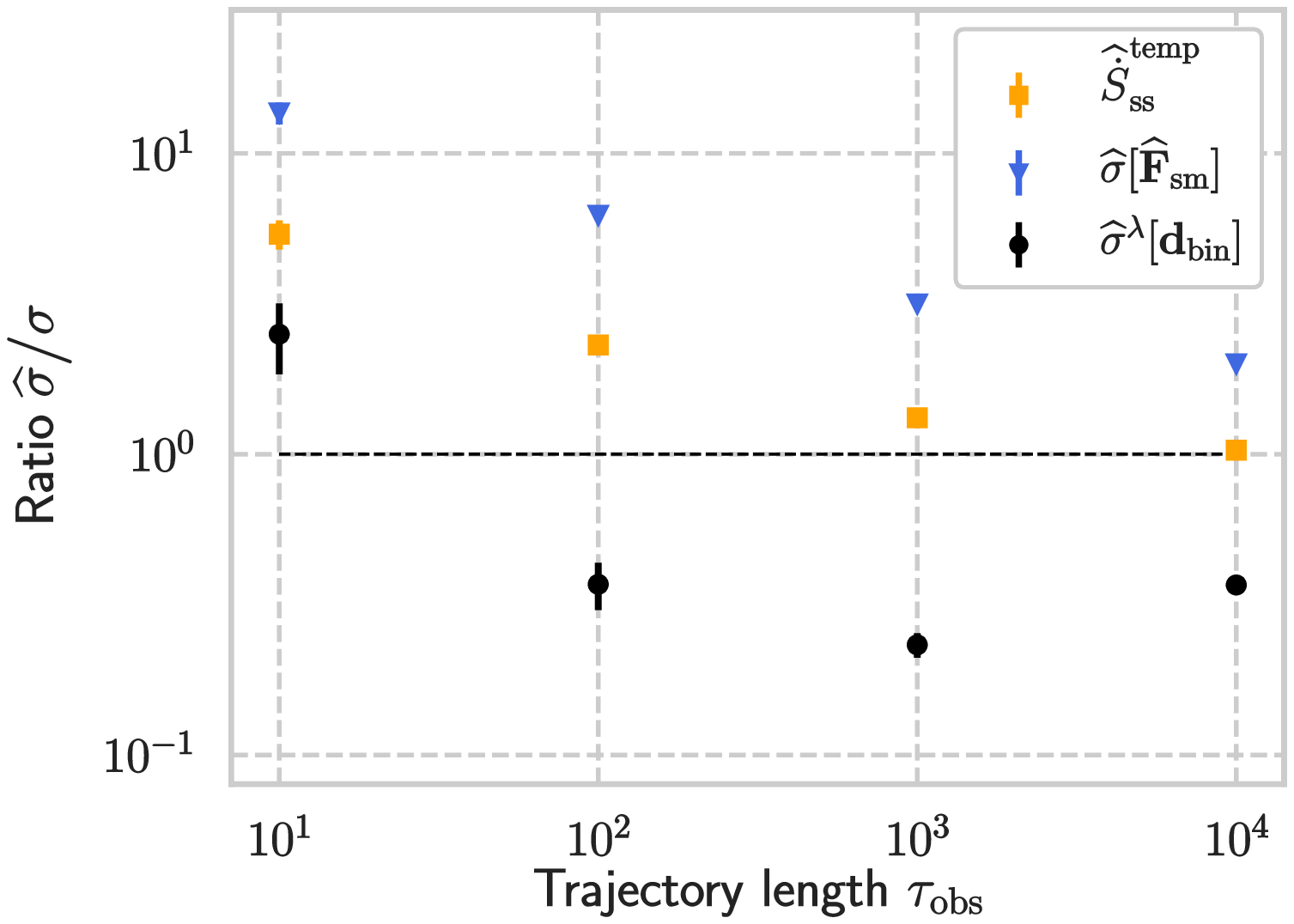}\label{fig: 5beads_main_r01}}&
	\subfigure[$T_c/T_h = 0.5$]{
		\includegraphics[width = 0.44\linewidth]{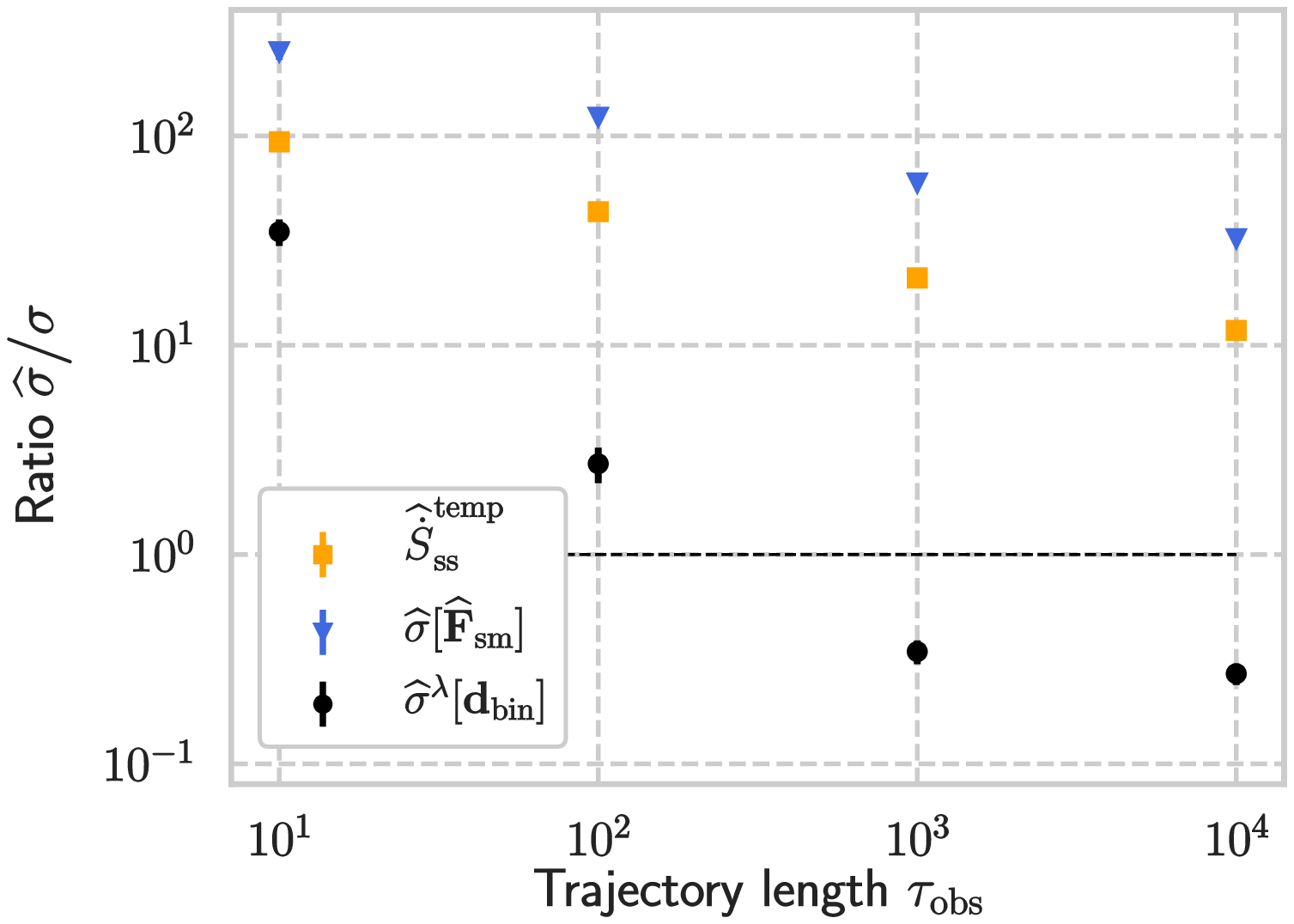}\label{fig: 5beads_main_r05}}
\end{tabular}
\caption{Numerical experiment with the five-beads model: (a) The dependence of the entropy production rate on the temperature ratio $T_c/T_h$. (b)(c) Performance of the estimators at (b) $T_c/T_h = 0.1$ and (c) $T_c/T_h = 0.5$ with \footnotesize$\widehat{\dot{S}}_{\rm ss}^{\rm temp}$\normalsize (yellow squares), $\widehat{\sigma}[\widehat{{\bm F}}_{\rm sm}]$ (blue triangles) and $\widehat{\sigma}^\lambda[{\bm d}_{\rm bin}]$ (black circles). The mean and its standard deviation of ten independent trials are plotted. The binned learning estimator uses the hyperparameters listed in TABLE~\ref{table: hyperparameters}, and the other system parameters are set as $k = \gamma = 1$ and $T_h = 250$. {\color{black}The sampling interval of the trajectories is set as $\Delta t = 10^{-3}$, and thus the number of data points is $10^3\tau_{\rm obs}$, half of which is used for the training, and the other half for the estimation in the case of $\widehat{\sigma}^\lambda[{\bm d}_{\rm bin}]$.}
}
\label{fig: 5beads}
\end{center}
\end{figure*}
We first consider the $N$-beads model illustrated in Fig.~\ref{fig: models}(a), which was introduced in a previous study \cite{Li2019}.
Specifically, we use the two-beads and five-beads models to evaluate the performance of the estimators for Langevin equations. 
We show that the learning estimators $\widehat{\sigma}^\lambda[{\bm d}_{\rm bin}]$ or $\widehat{\sigma}[{\bm d}_{\rm Gauss}]$ shows the convergence comparable to those of the KDE estimators \footnotesize$\widehat{\dot{S}}_{\rm ss}^{\rm temp}$\normalsize and $\widehat{\sigma}[\widehat{{\bm F}}_{\rm sm}]$, while the learning estimators become better in computation time as the trajectory length increases.\\ \indent
In the model, the dynamics of $N$ beads are observed, which are connected to each other and to the boundary walls by springs with stiffness $k$.
The beads are immersed in viscous fluids at different temperatures: $T_h$ and $T_c$ in the two-beads model, and {\color{black}$T_i = T_h + (T_c - T_h)(i-1)/4, ~(i = 1, 2, 3, 4, 5)$ in the five-beads model}. The viscous fluids induce injection or absorption of energy to (from) the beads through the friction $\gamma$, leading to a steady heat flow between the fluids and the beads.\\ \indent
The displacements of the beads from their equilibrium positions are described by the Langevin equation.
In the case of the two-beads model,
\begin{eqnarray}
\dot{\bm x} &=& A{\bm x} + F{\bm \xi}_t, \\
A &=& \left(\begin{matrix}
-2k/\gamma & k/\gamma\\
k/\gamma & -2k/\gamma
\end{matrix}\right),\\
F&=& \left(\begin{matrix}
\sqrt{2T_h/\gamma} & 0\\
0 & \sqrt{2T_c/\gamma}
\end{matrix},
\right),
\end{eqnarray}
where ${\bm x} = \left(x, y\right)^\mathsf{T}$ is the vector of displacements, and ${\bm \xi}_t$ is the independent Gaussian white noise satisfying
$\left<\xi_{t, i}\xi_{t', j}\right> = \delta_{ij}\delta(t-t')$. The equation for the five-beads model can be written in a similar form, {\color{black}which is defined by $A_{ij} = \delta_{i, j}\left(-2k/\gamma\right) + (\delta_{i, j+1} + \delta_{i+1, j})k/\gamma$ and $F_{ij} = \delta_{i,j}\sqrt{2T_i/\gamma}$.} Since the Langevin equations are linear, the steady-state probability distributions become Gaussian distributions. Therefore, they are analytically tractable, and the entropy production rate can be calculated as
\begin{eqnarray}
\sigma = \frac{k\left(T_h-T_c\right)^2}{4\gamma T_hT_c}
\end{eqnarray}
for the two-beads model, and as
\begin{eqnarray}
\sigma = \frac{k(T_h-T_c)^2(111T_h^2 + 430T_hT_c + 111T_c^2)}{495T_hT_c(3T_h + T_c)(T_h + 3T_c)\gamma}
\end{eqnarray}
for the five-beads model (see Ref.~\cite{Li2019} for the details).\\ \indent
In Fig.~\ref{fig: 2beads}, we show the results of our numerical experiment with the two-beads model. We generate trajectory data of length $\tau_{\rm obs}$ which are sampled every $\Delta t = 10^{-3}$ (thus the number of data points is $10^3\tau_{\rm obs}$) with parameter setting: $k = \gamma = 1$ and $T_h = 250$. Figure~\ref{fig: 2beads_true} shows the dependence of the entropy production rate on the temperature ratio $T_c/T_h$, where $T_c/T_h = 1$ corresponds to the equilibrium limit. In Fig.~\ref{fig: 2beads_main_r01} and \ref{fig: 2beads_main_r05}, we compare the convergence of each estimator as we increase the trajectory length at the temperature ratio $T_c/T_h = 0.1$ and $T_c/T_h = 0.5$.
The hyperparameter tuning for the Gaussian learning estimator is conducted as described in the Supplemental Material, and the values listed in TABLE.~\ref{table: hyperparameters} are adopted.\\ \indent
In both temperature ratios, the Gaussian learning estimator shows the best convergence, while the difference among these estimators is not significant. The convergence at $T_c/T_h = 0.5$ is worse than that at $T_c/T_h = 0.1$ for all the estimators, because the mean local velocities become small when the system is close to equilibrium.\\ \indent 
In Fig.~\ref{fig: 5beads}, we show the results of a numerical experiment with the five-beads model in the same manner as the two-beads model with parameters: $\Delta t = 10^{-3}$, $k = \gamma = 1$ and $T_h = 250$. Since the computational cost of the Gaussian learning estimator is large for high-dimensional data, the binned learning estimator is adopted here.\\ \indent
\footnotesize$\widehat{\dot{S}}_{\rm ss}^{\rm temp}$\normalsize shows the best convergence at the temperature ratio $T_c/T_h = 0.1$, while the binned learning estimator  seems to be better at $T_c/T_h = 0.5$. The convergence of each estimator is much worse than that in the two-beads model because of the high dimensionality. Interestingly, the convergence of $\widehat{\sigma}[\widehat{{\bm F}}_{\rm sm}]$ is not as good as \footnotesize$\widehat{\dot{S}}_{\rm ss}^{\rm temp}$\normalsize in both parameter settings, which is contrary to the results reported in Ref.~\cite{Li2019}. This is because
$\widehat{\sigma}[\widehat{{\bm F}}_{\rm sm}]$ is based on the short-time TUR in this study, while the finite-time TUR is used in the previous study.
This result might suggest that the finite-time and the long-time TUR based estimator have some advantages over the short-time TUR based estimator in terms of the convergence speed for high-dimensional data, while more exhaustive research is necessary to clarify this conclusion.\\ \indent
We remark on the computation speed of these estimators, which we investigate in detail in the Supplemental Material. First, the computational complexities of the learning estimators are $O(N)$ in terms of the sample size $N := \tau_{\rm obs}/\Delta t$, while \footnotesize$\widehat{\dot{S}}_{\rm ss}^{\rm temp}$\normalsize and $\widehat{\sigma}[\widehat{{\bm F}}_{\rm sm}]$ scale as $O(N^2)$. We confirmed that, as we increase the length of trajectories generated by the $N$-beads model, the computation time of these estimators increase as predicted, and the learning estimators become better than \footnotesize$\widehat{\dot{S}}_{\rm ss}^{\rm temp}$\normalsize and $\widehat{\sigma}[\widehat{{\bm F}}_{\rm sm}]$ in computation time.  However, there is a drawback for the learning estimators that they need the hyperparameter tuning additionally. Nonetheless, in the case of large trajectory data, we argue that the learning estimators are better in computation time, because the advantage that comes from the scaling of $N$ is significant. 

\subsection{Mexican-hat potential model}
\label{sec: mx}
\begin{figure*}
\begin{center}
\begin{tabular}{cc}
	\subfigure[]{
		\includegraphics[width = 0.44\linewidth]{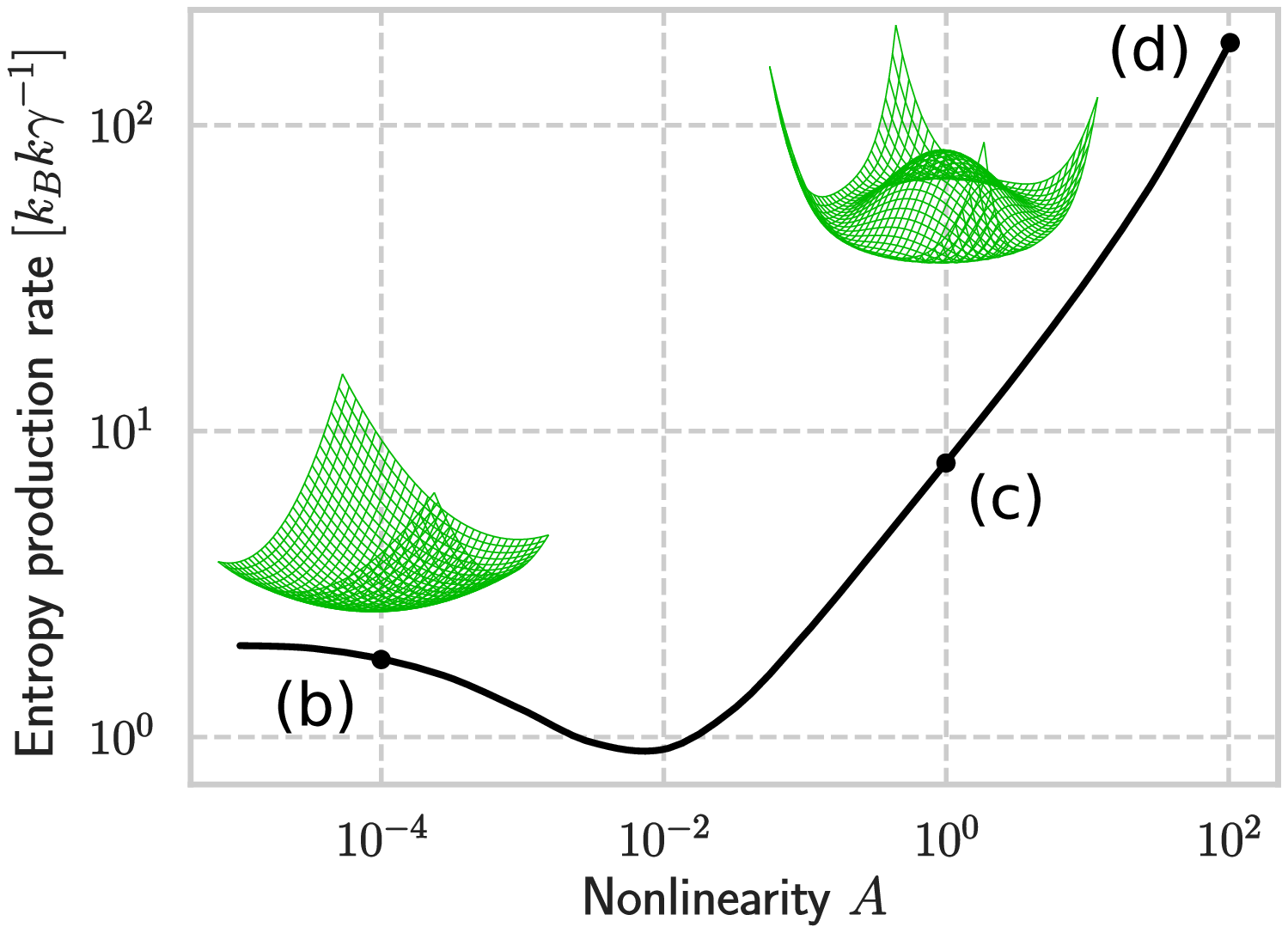}\label{fig: mx_true}}&
	\subfigure[$A = 10^{-4}$]{
		\includegraphics[width = 0.44\linewidth]{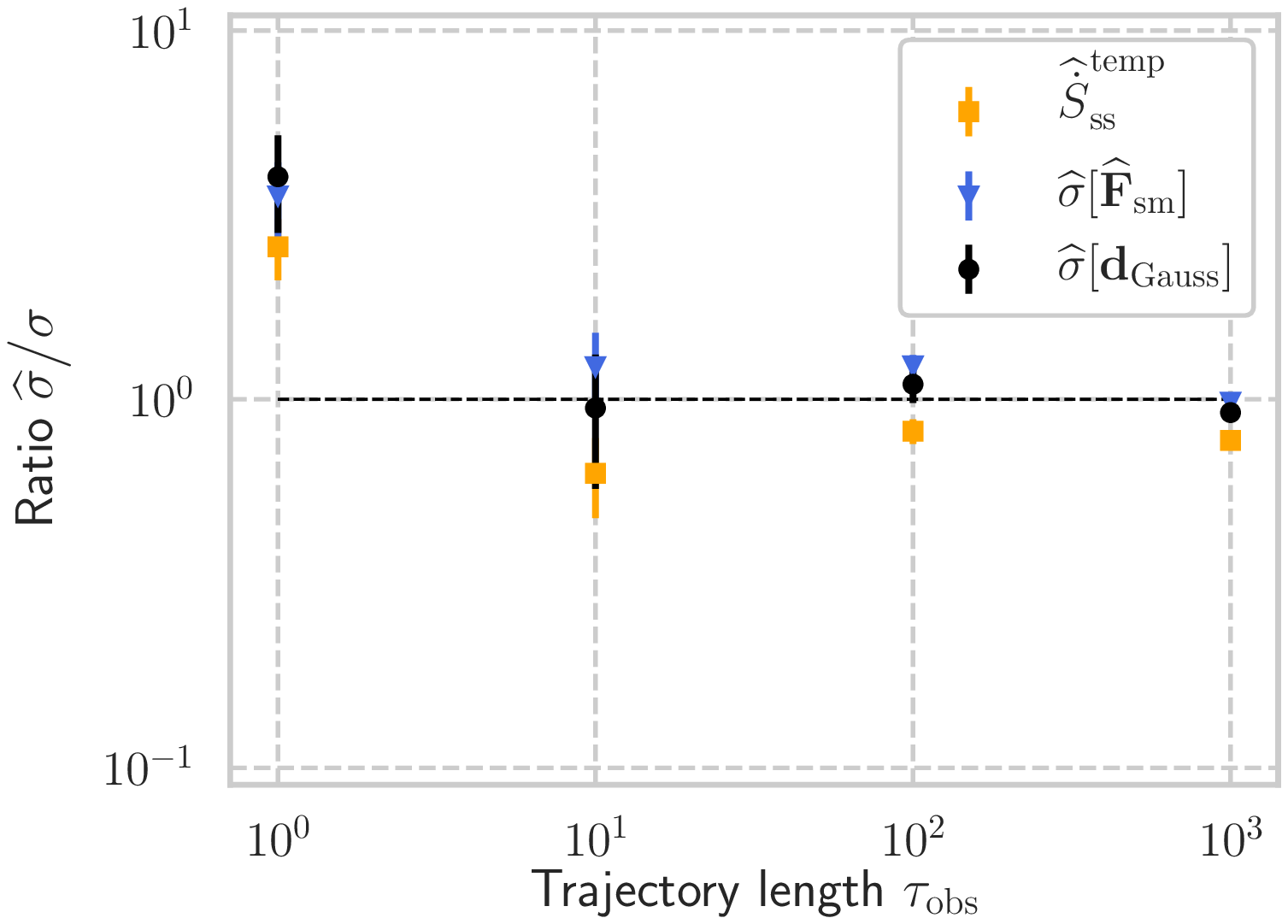}\label{fig: mx_main_A0001}}\\
	\subfigure[$A = 1$]{
		\includegraphics[width = 0.44\linewidth]{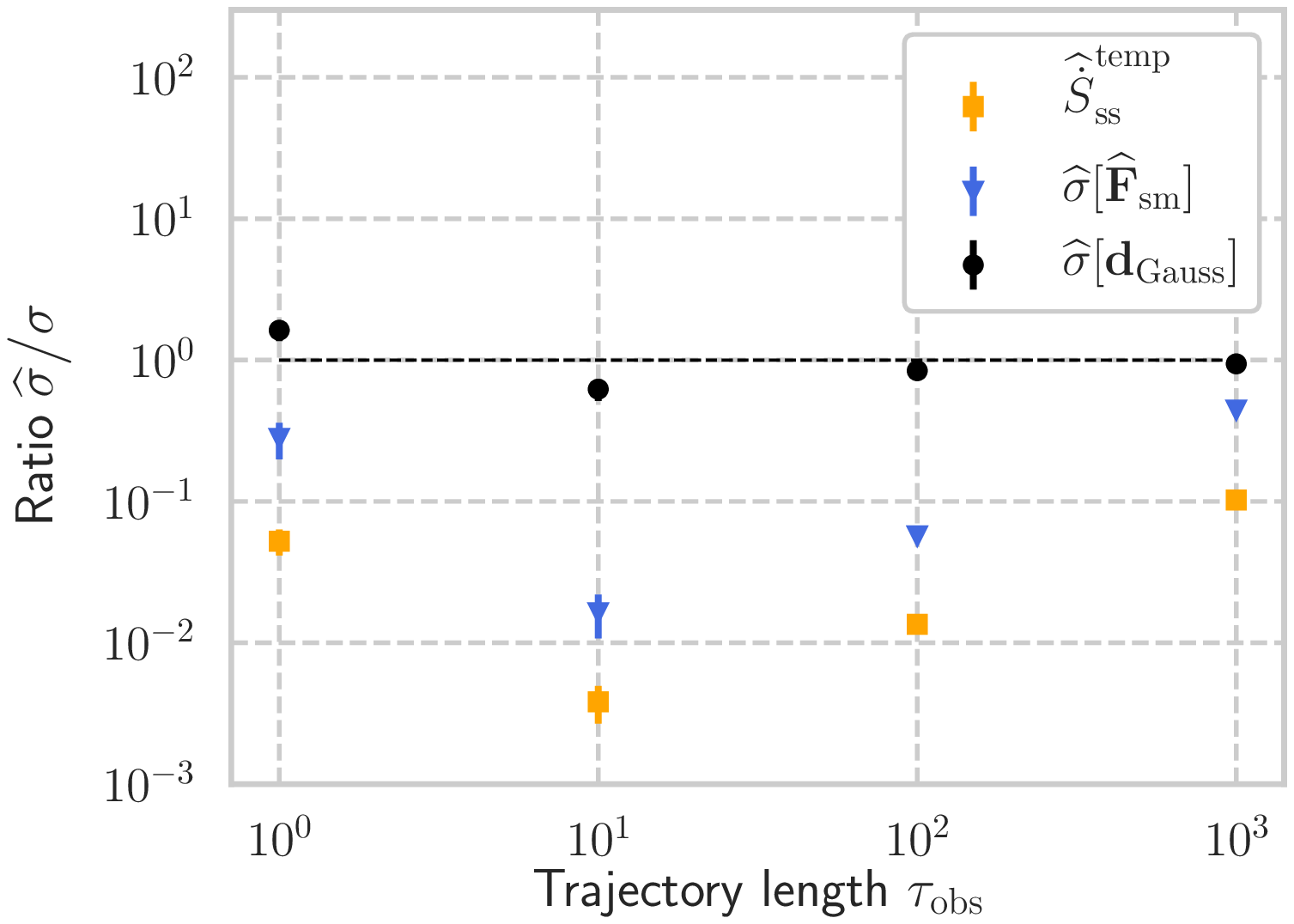}\label{fig: mx_main_A1}}&
	\subfigure[$A = 10^2$]{
		\includegraphics[width = 0.44\linewidth]{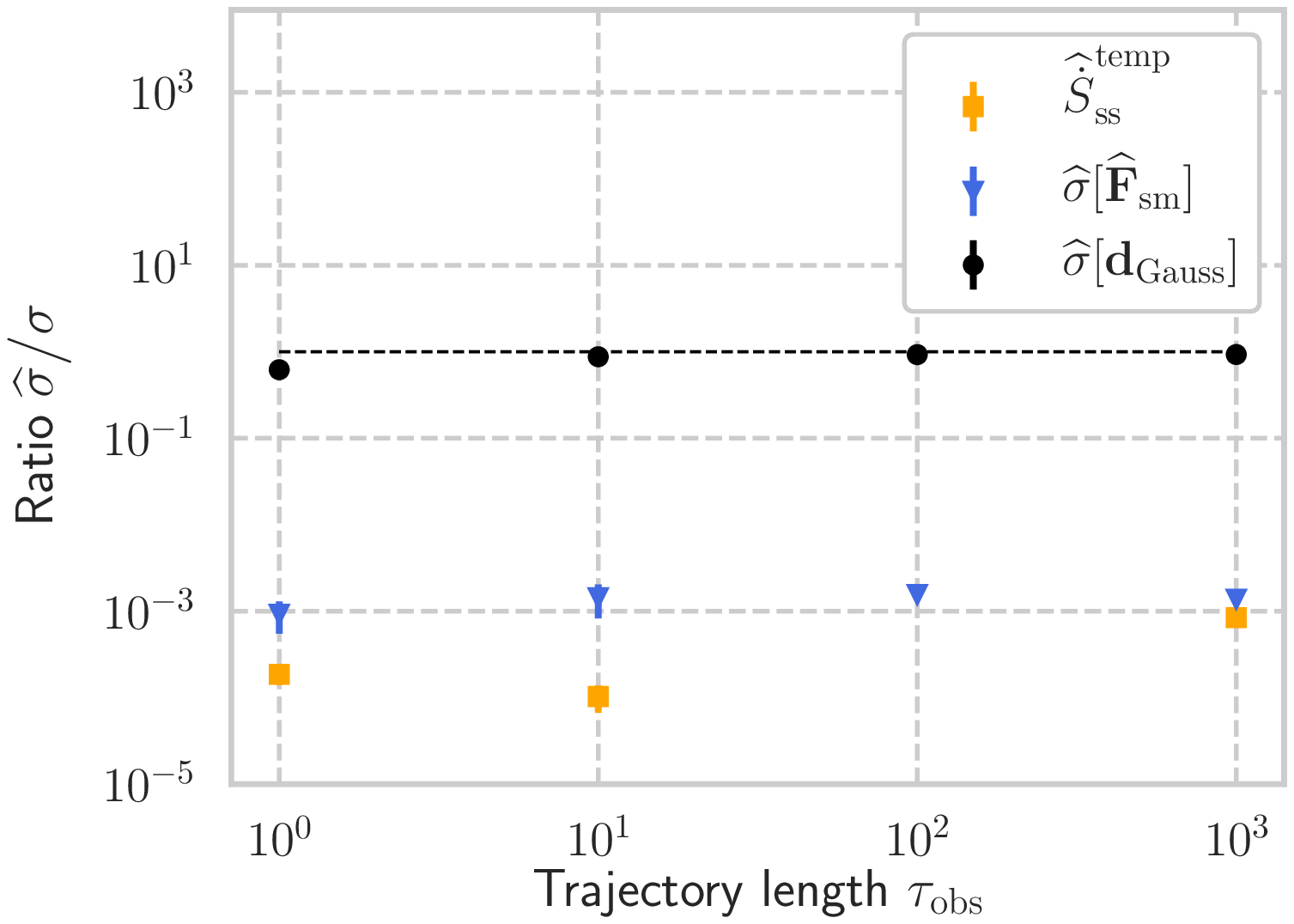}\label{fig: mx_main_A100}}
\end{tabular}
\caption{Numerical experiment with the Mexican-hat potential model: (a) The dependence of the entropy production rate on the nonlinearity $A$ . We draw the potential shapes at $A = 10^{-4}$ and $A = 10^2$. (b)(c)(d) Performance of the estimators at (b) $A = 10^{-4}$, (c) $A = 1$ and (d) $A = 100$ with \footnotesize$\widehat{\dot{S}}_{\rm ss}^{\rm temp}$\normalsize (yellow squares), $\widehat{\sigma}[\widehat{{\bm F}}_{\rm sm}]$ (blue triangles) and $\widehat{\sigma}[{\bm d}_{\rm Gauss}]$ (black circles). The mean and its standard deviation of ten independent trials are plotted. The Gaussian learning estimator uses the hyperparameters listed in TABLE~\ref{table: hyperparameters}, and the other system parameters are set as $k = \gamma = 1$ and $T_h = 250$. {\color{black}The sampling interval of the trajectories is set as $\Delta t = 10^{-4}$, and thus the number of data points is $10^4\tau_{\rm obs}$, half of which is used for the training, and the other half for the estimation in the case of $\widehat{\sigma}[{\bm d}_{\rm Gauss}]$.}
In (d), a point of $\widehat{\dot{S}}^{\rm temp}_{\rm ss}$ is missing at $\tau_{\rm obs} = 10^2$ because the value is negative.}
\label{fig: mx}
\end{center}
\end{figure*}

\begin{figure*}
\begin{center}
\begin{tabular}{cc}
	\subfigure[Optimal coefficient field]{
		\includegraphics[width = 0.44\linewidth]{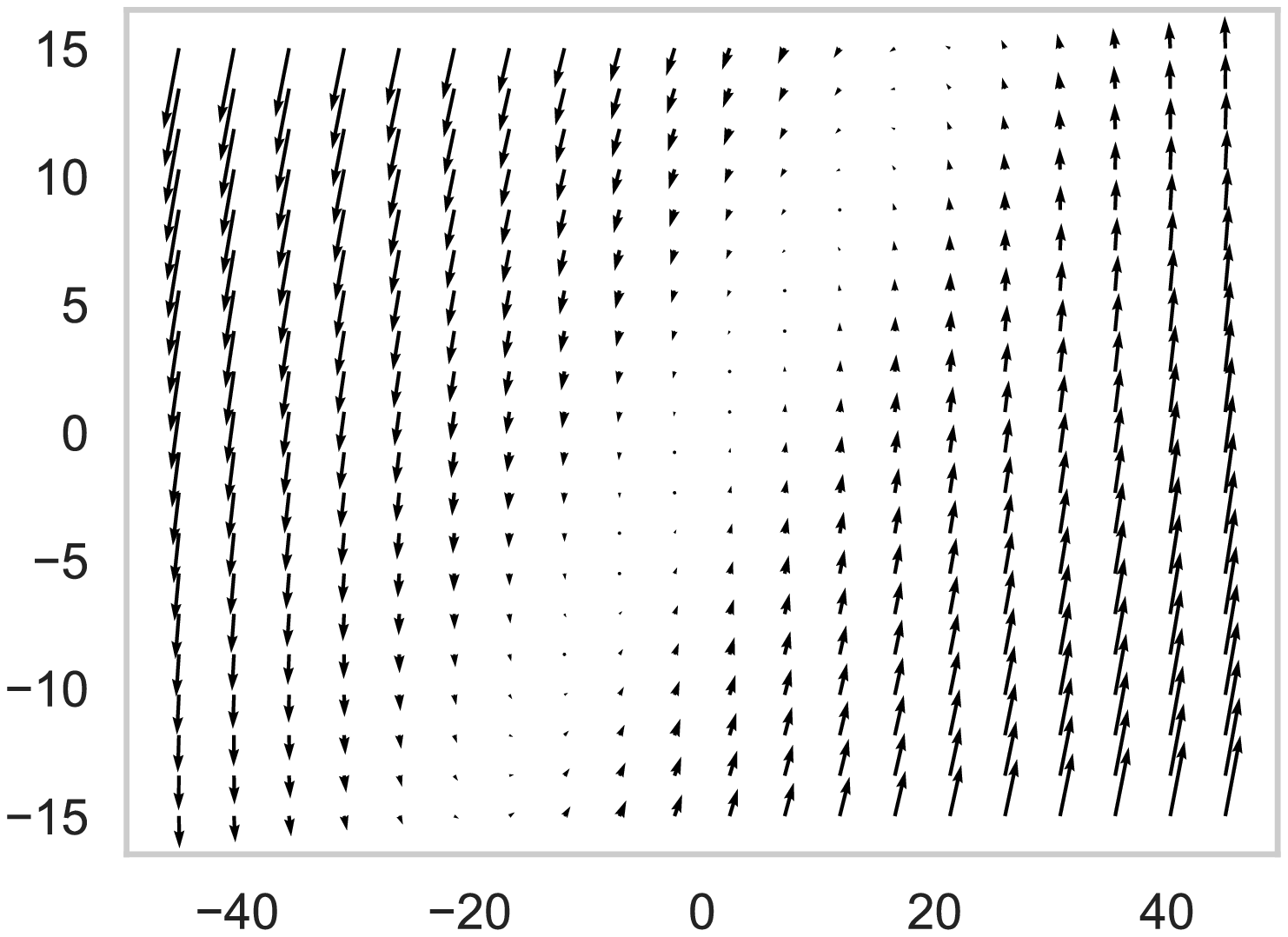}\label{fig: 2beads_optimal_coefficient}}&
	\subfigure[Numerically obtained coefficient field]{
		\includegraphics[width = 0.44\linewidth]{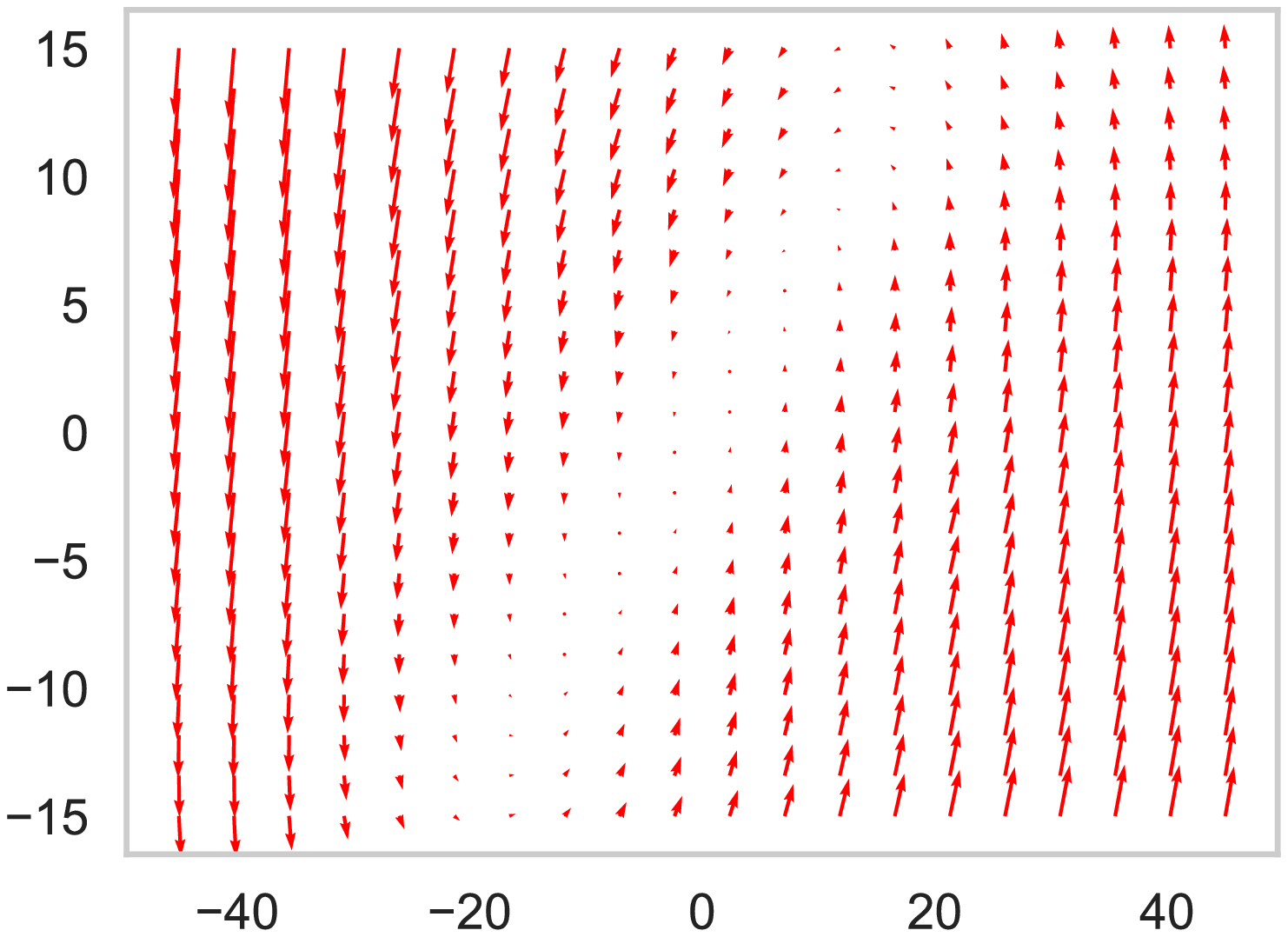}\label{fig: 2beads_gauss_coefficient}}\\
	\subfigure[Optimal coefficient field]{
		\includegraphics[width = 0.44\linewidth]{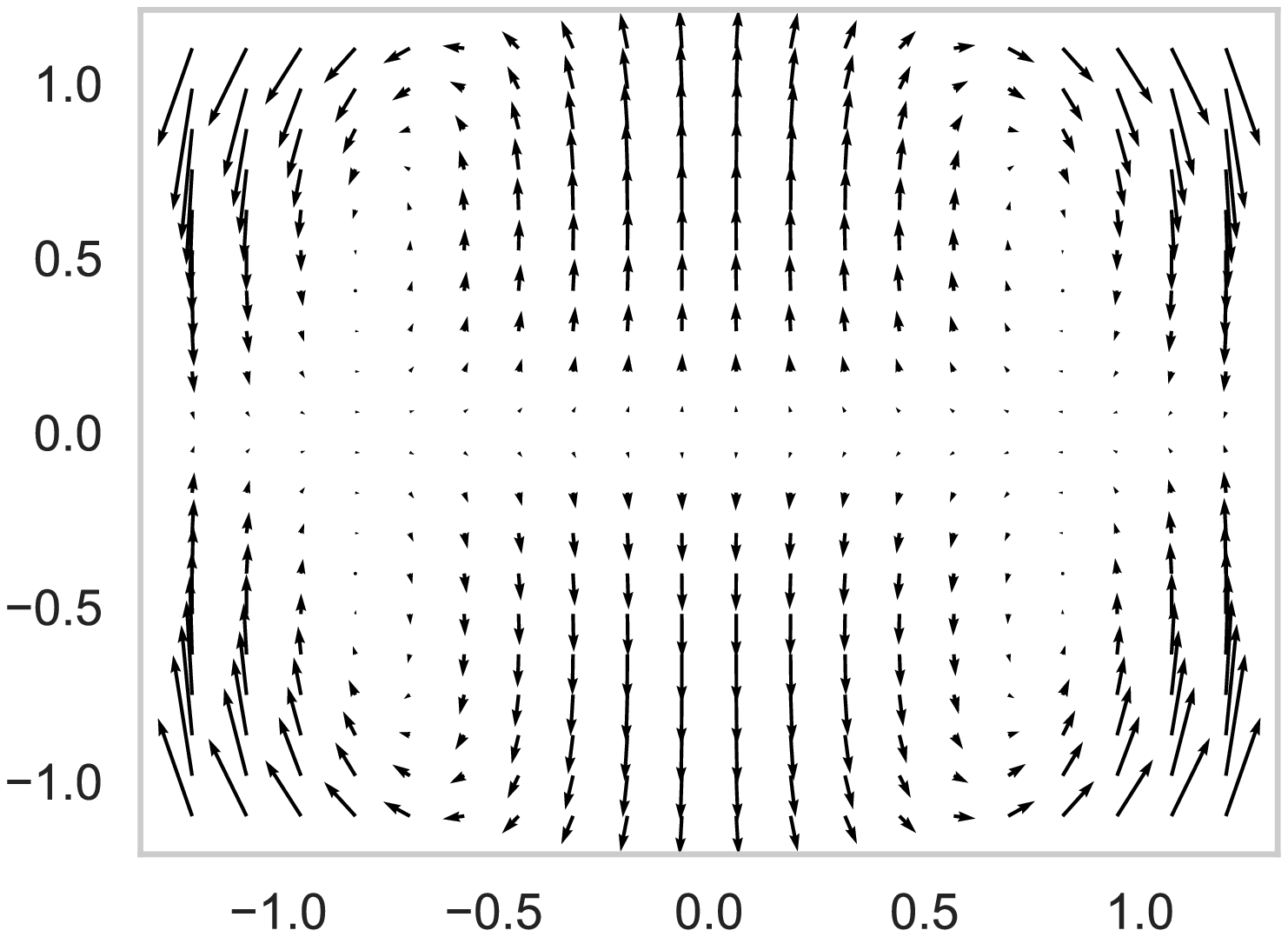}\label{fig: mx_optimal_coefficient}}&
	\subfigure[Numerically obtained coefficient field]{
		\includegraphics[width = 0.44\linewidth]{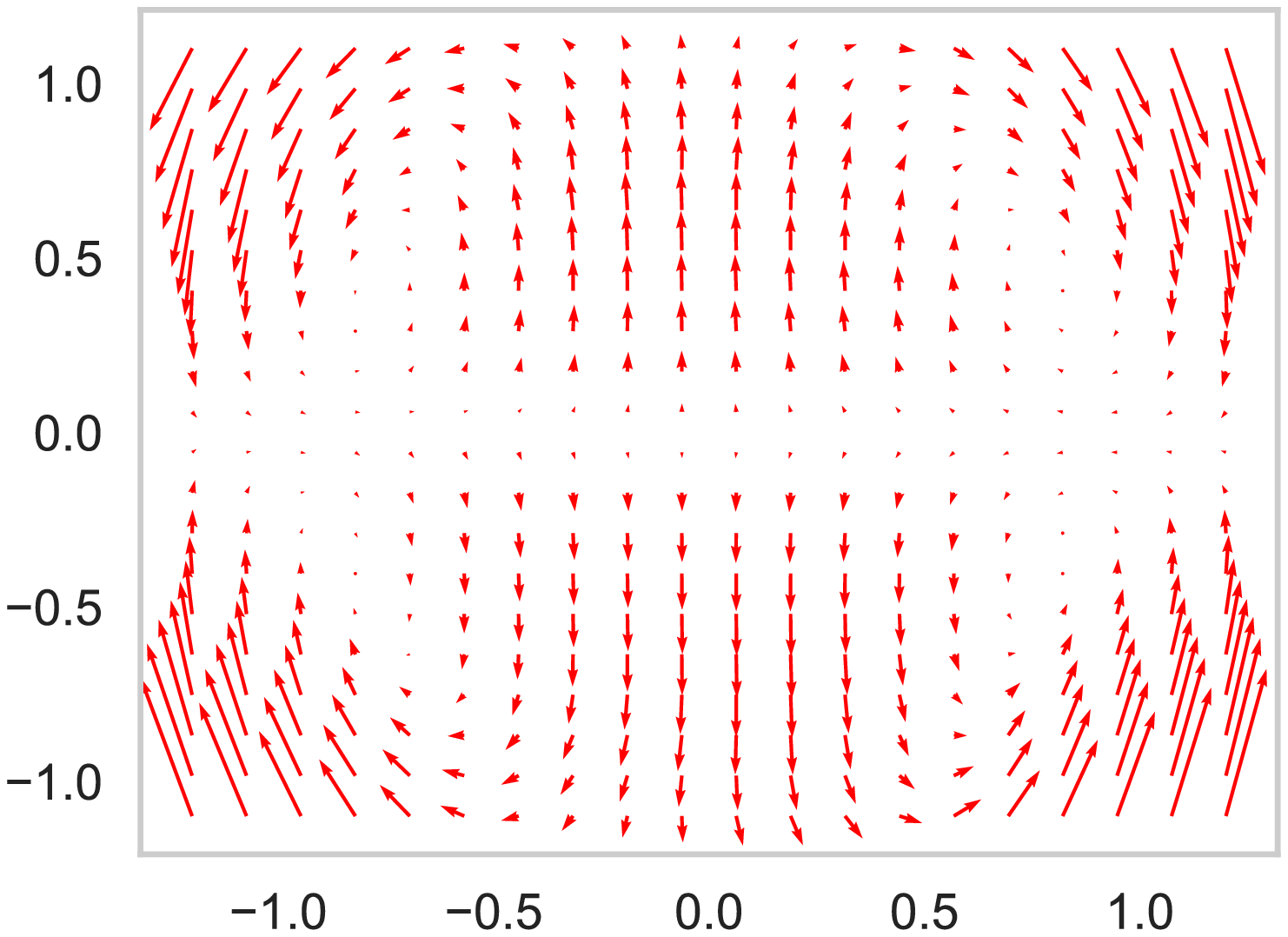}\label{fig: mx_gauss_coefficient}}
\end{tabular}
\caption{Comparison between the optimal and numerically obtained coefficient fields ${\bm d}({\bm x})$: (a) Analytically obtained coefficient field ${\bm d}^*({\bm x})$ for the two-beads model ($T_c/T_h = 0.1$). (b) Coefficient field ${\bm d}_{\rm Gauss}({\bm x})$ obtained by learning of the Gaussian learning estimator for the two-beads model ($T_c/T_h = 0.1$). (c) Optimal coefficient field ${\bm d}^*({\bm x})$ obtained by exact diagonalization of the discretized Fokker-Planck equation for the Mexican-hat potential model ($A = 10^2$). (d) Coefficient field ${\bm d}_{\rm Gauss}({\bm x})$ obtained by learning of the Gaussian learning estimator for the Mexican-hat potential model ($A = 10^2$). The horizontal axis is $x$ and the vertical axis is $y$. The Gaussian learning estimator is trained with trajectory data of length $\tau_{\rm obs} = 10^3$ for the two-beads model and $\tau_{\rm obs} = 10^2$ for the Mexican-hat potential model. The hyperparameters listed in TABLE~\ref{table: hyperparameters} are adopted for the learning, and the other system parameters are set in the same way as those in Figs.~\ref{fig: 2beads} and \ref{fig: mx}.
}
\label{fig: coefficient}
\end{center}
\end{figure*}
We next compare the performance of the estimators using trajectory data of non-linear Langevin dynamics, where the stationary distribution deviates from a Gaussian distribution. We show that the Gaussian learning estimator converges the fastest, while the KDE estimators do not work well especially at the parameter settings with large nonlinearity.\\ \indent
We here consider the following Langevin equation:
\begin{eqnarray}
\dot{\bm x} &=& -\frac{1}{\gamma}\nabla U + F{\bm \xi}_t,\label{eq: nonlinear}\\
U &=& Ak(r^4 - r^2) + k(x^2 + y^2 - xy),\\
F&=& \left(\begin{matrix}
\sqrt{2T_h/\gamma} & 0\\
0 & \sqrt{2T_c/\gamma}
\end{matrix}
\right),
\end{eqnarray}
where $r$ is the distance from the origin $r = \sqrt{x^2 + y^2}$, and ${\bm \xi}_t$ is the Gaussian white noise satisfying
$\left<\xi_{t, i}\xi_{t', j}\right> = \delta_{ij}\delta(t-t')$. We can imagine a Brownian particle whose motion in $x$ and $y$ directions are coupled with two thermal reservoirs at different temperatures $T_h$ and $T_c$, respectively (a similar model is used in \cite{Manikandan2019}). In addition, the particle is confined in a Mexican-hat type potential as illustrated in Fig.~\ref{fig: models} (b). The parameter $A$ represents the nonlinearity of the model, and the model converges to the two-beads model at $A = 0$. At finite $A > 0$, the stationary distribution deviates from a Gaussian distribution due to the small hill at the center of the potential.\\ \indent
In Fig.~\ref{fig: mx}, we show the results of the numerical experiment with the Mexican-hat potential model. We generate trajectory data of length $\tau_{\rm obs}$, which are sampled every $\Delta t = 10^{-4}$ with parameters $k  = \gamma = 1$, $T_h = 250$ and $T_c = 25$. In Fig.~\ref{fig: mx_true}, we show the dependence of the entropy production rate on the nonlinearity $A$.
In order to evaluate the performance of the estimators, we calculate the true value of the entropy production rate by using the stationary distribution obtained by exact diagonalization of the transition matrix, where the transition matrix is obtained by discretizing the corresponding Fokker-Planck equation \cite{Gingrich2017}. In Fig.~\ref{fig: mx_main_A0001}, (c) and (d), we compare the convergence of the estimators at different nonlinearity $A$.\\ \indent
The Gaussian learning estimator shows the best convergence in all the parameter settings. Especially, for the larger nonlinear cases, the KDE estimators do not work well due to the assumption that the stationary distribution is Gaussian.
We can also see that the result of $A = 10^{-4}$ (Fig.~\ref{fig: mx_main_A0001}) is close to that in the two-beads model (Fig.~\ref{fig: 2beads_main_r01}) as expected.
In short, all the results so far show the effectiveness of the learning estimators.\\ \indent
Finally, we show that our learning method indeed obtains the coefficient field ${\bm d}({\bm x})$ close to the optimal one ${\bm d}^*({\bm x}) \propto{\bm F}({\bm x})$. In Fig.~\ref{fig: coefficient}, the optimal and numerically obtained coefficient fields are shown for the two-beads model ($T_c/T_h = 0.1$) and the Mexican-hat potential model ($A = 10^{2}$).
{\color{black}Here, in order to compare ${\bm d}({\bm x})$ with the thermodynamic force field ${\bm F}({\bm x})$, we rescale the obtained field ${\bm d}({\bm x})$ by $2\widehat{\left<j_{\bm d}\right>}/\tau\widehat{{\rm Var}\left(j_{\bm d}\right)}$.
This is because when ${\bm d}({\bm x}) = c{\bm F}({\bm x})$, the generalized current satisfies $\left<j_{\bm d}\right> = c\sigma$, $\tau{\rm Var}\left(j_{\bm d}\right) = 2c^2\sigma$, and thus $2\left<j_{\bm d}\right>/\tau{\rm Var}\left(j_{\bm d}\right)$ equals $1/c$.}
The numerically obtained coefficient fields resemble the optimal ones especially around the center for which there are sufficient data. We note that only the results of the Gaussian learning estimator are shown here, while the binned learning estimator is also confirmed to obtain the coefficient field accurately when applied to the two dimensional model.\\ \indent
{\color{black}We can further investigate the higher-order statistics of the time-integrated entropy production by calculating the integrated generalized current using the obtained thermodynamic force field.} This is a slightly different approach from the one presented in Ref.~\cite{Manikandan2019}, while our method would be useful due to the simplicity of the protocol. We leave such application as an interesting future issue.

\subsection{One-dimensional hopping model}
\begin{figure*}
\begin{center}
\begin{tabular}{cc}
	\subfigure[$N_{\rm state} = 10$]{
		\includegraphics[width = 0.44\linewidth]{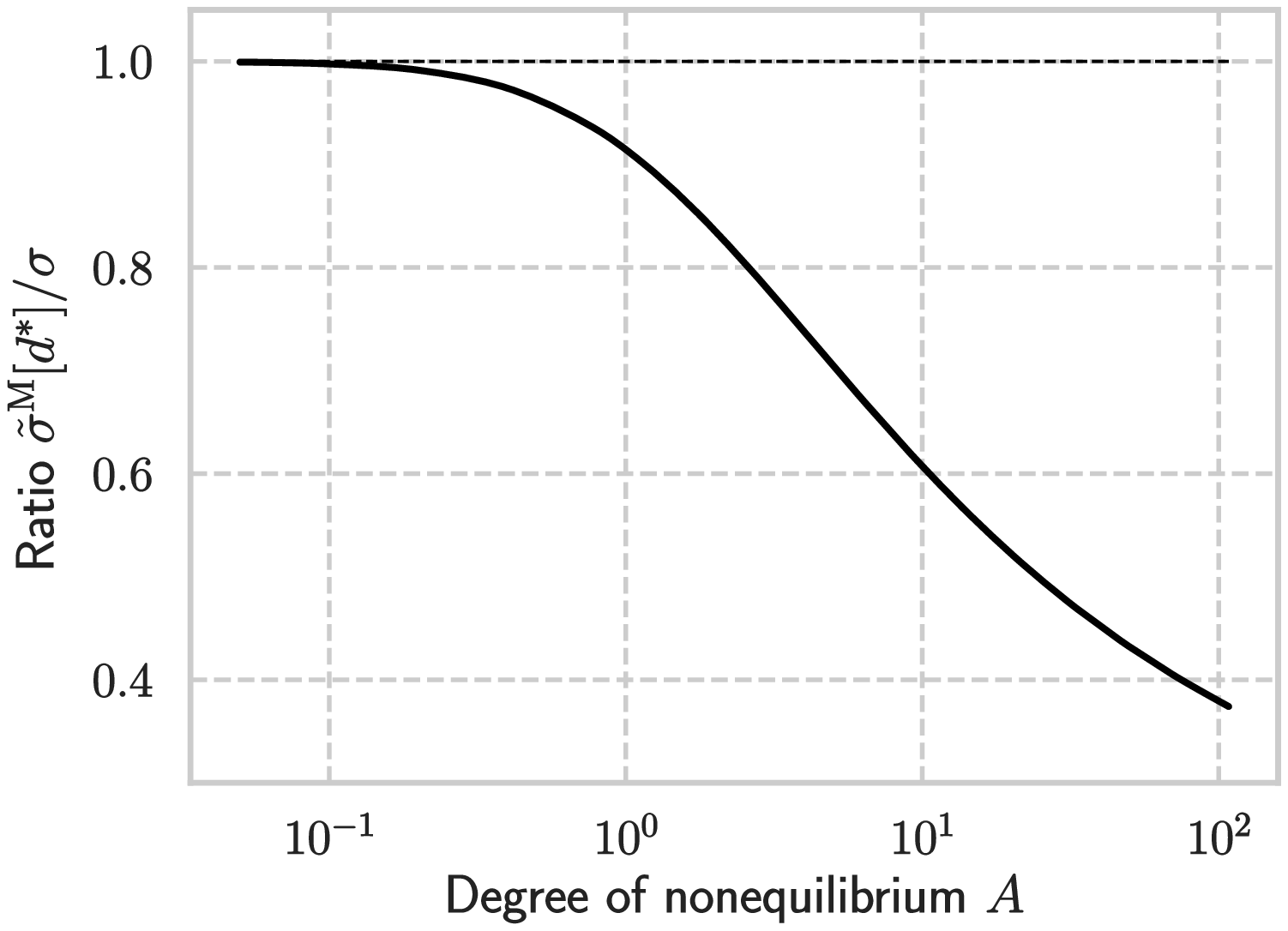}\label{fig: mj_eq}}&
	\subfigure[$A = 10$]{
		\includegraphics[width = 0.44\linewidth]{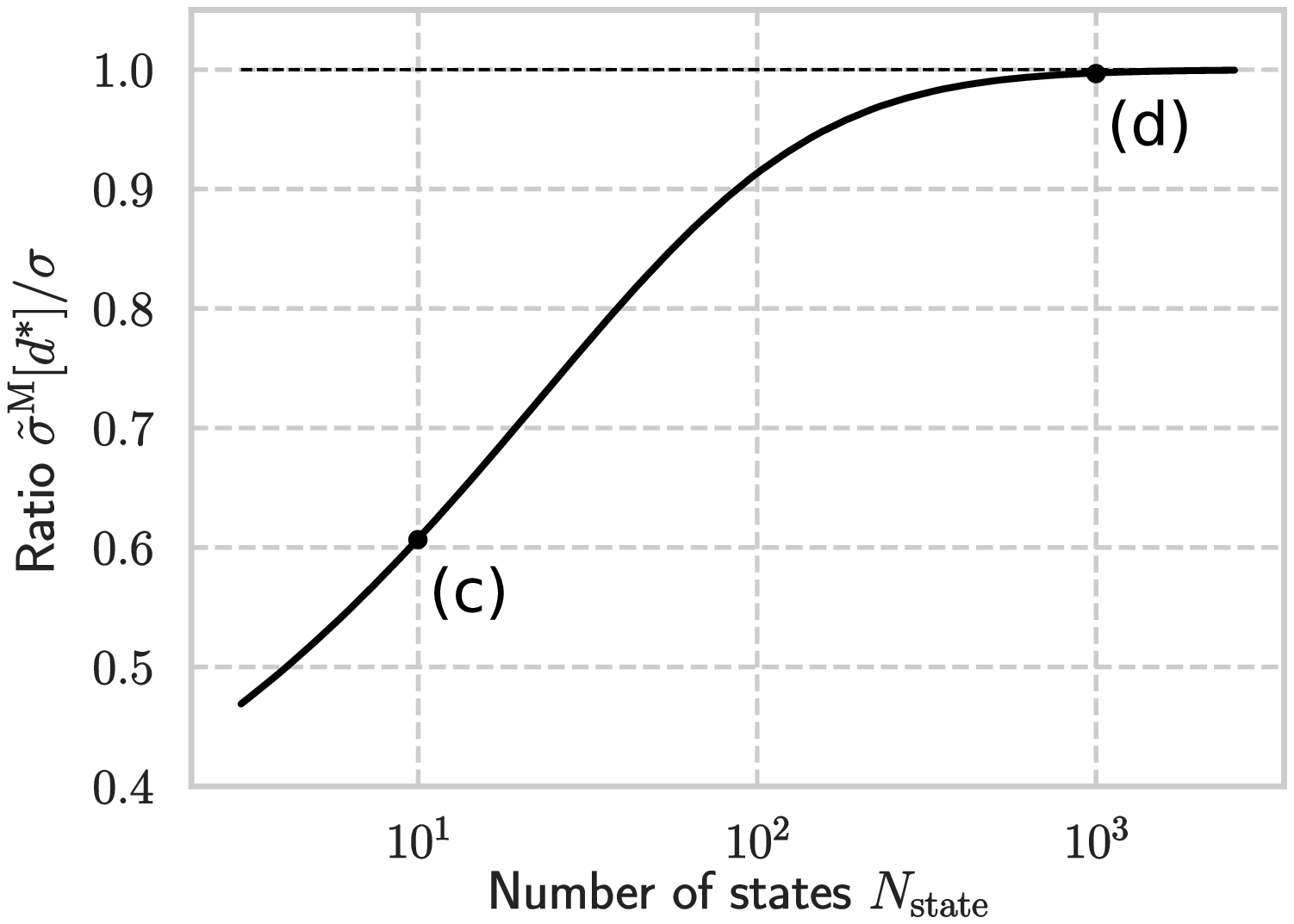}\label{fig: mj_Nstate}}\\
	\subfigure[$A = 10, N_{\rm state} = 10$]{
		\includegraphics[width = 0.44\linewidth]{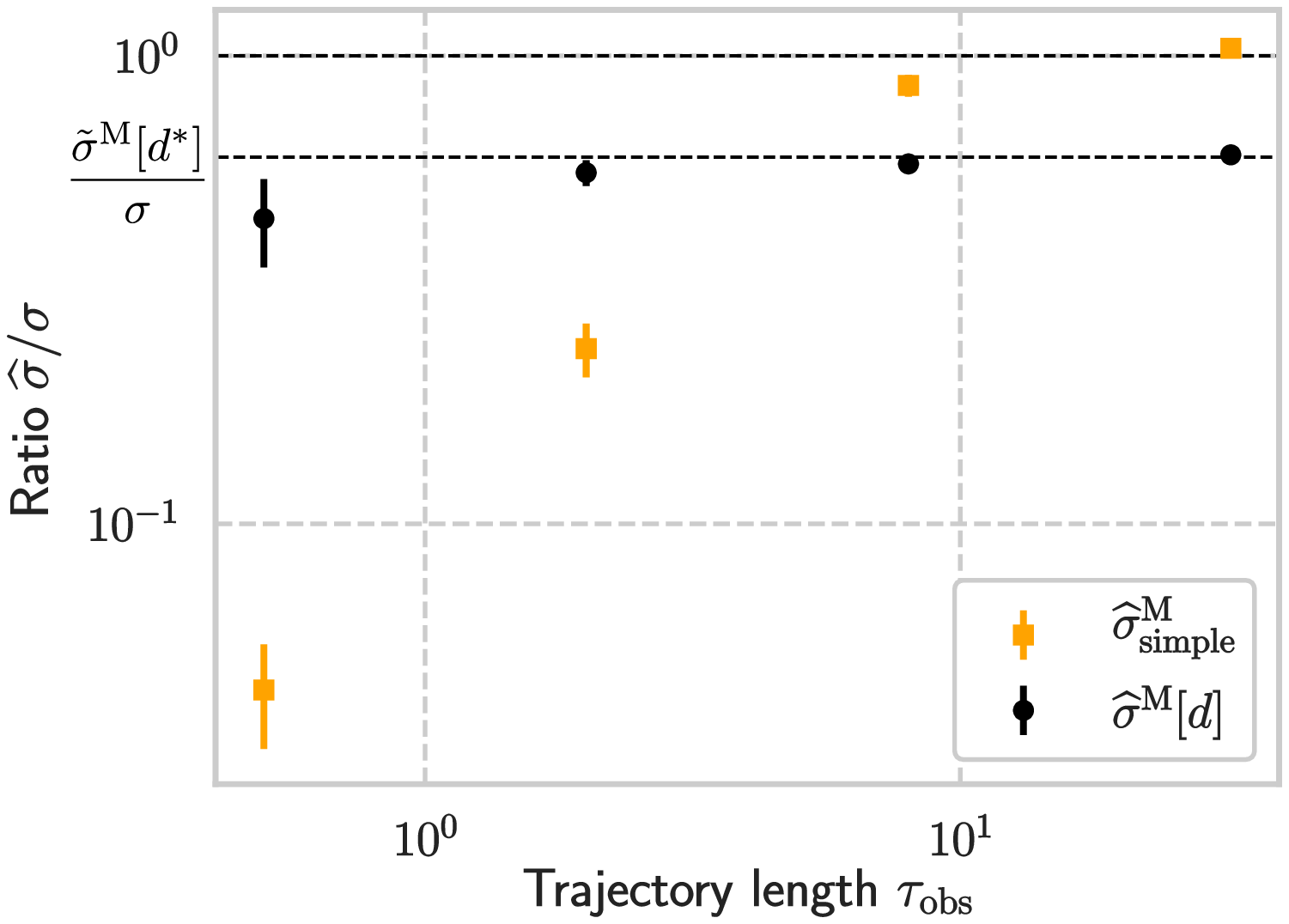}\label{fig: mj_tau}}&
	\subfigure[$A = 10, N_{\rm state} = 10^3$]{
		\includegraphics[width = 0.44\linewidth]{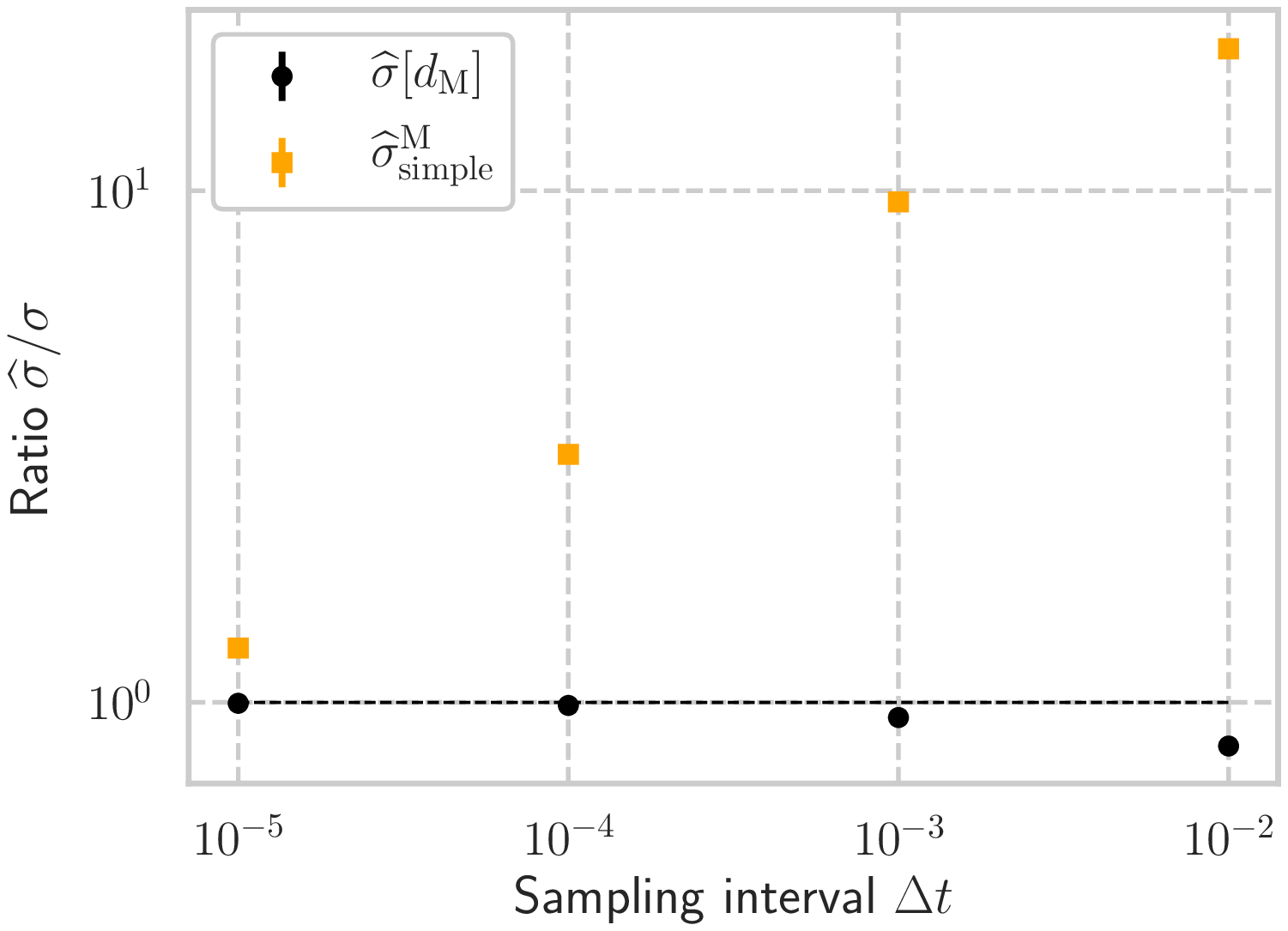}\label{fig: mj_dt}}\\
\end{tabular}
\caption{Numerical experiment with the one-dimensional hopping model: (a) The degree of nonequilibrium $A$ versus the ratio of the optimal estimation to the true entropy production rate, i.e., $\tilde{\sigma}^{\rm M}[d^*]/\sigma$ at $N_{\rm state} = 10$. (b) The number of states $N_{\rm state}$ versus the above-mentioned ratio at $A = 10$. In (a) and (b), the curves are drawn by interpolating several points which are calculated on the basis of the exact diagonalization of the transition matrix. (c) Performance of the estimators at $N_{\rm state} = 10, A = 10$. 
{\color{black}The sampling interval of the trajectories is set as $\Delta t = 10^{-3}$ here, and thus the number of data points is $10^3\tau_{\rm obs}$, half of which is used for the training, and the other half for the estimation in the case of $\widehat{\sigma}^{\rm M}[d]$.}
(d) The sampling interval dependence of the estimators with a fixed trajectory length $\tau_{\rm obs} = 100$ at $N_{\rm state} = 1000, A = 10$. The mean and its standard deviation of ten independent trials are plotted in (c) and (d). The hyperparameters listed in TABLE~\ref{table: hyperparameters} are adopted for $\widehat{\sigma}^{\rm M}[d]$, and the other system parameters are set as $D = 1$ and $f = 3$.}
\label{fig: mj}
\end{center}
\end{figure*}
Lastly, we consider a Markov jump process, in which we can take (i) the equilibrium limit and (ii) the Langevin limit.
We first show that the optimal estimation $\tilde{\sigma}^{\rm M}[d^*]$ converges to the true value $\sigma$ in both the limits as predicted in Sec.~\ref{sec: theory}. Then, we compare the performance of the learning estimator $\widehat{\sigma}^{\rm M}[d]$ with the simple estimator $\widehat{\sigma}_{\rm simple}^{\rm M}$, and show that the learning estimator converges faster. In addition, the learning estimator is shown to be robust against the choice of the sampling interval of trajectory data, which suggests the practical usefulness of the TUR-based estimators in Markov jump processes.\\ \indent
We consider a hopping dynamics between the states on a ring as illustrated in Fig.~\ref{fig: models}(c). There are $N_{\rm state}$ states on the ring labelled by $i\in\{1, 2, \cdots, N_{\rm state}\}$, and the transition rates between the states are given by:
\begin{eqnarray}
r(i, i+1) &=& \frac{D}{h^2} + \frac{A}{h}\left(-\cos\left[hi\right] + f\right)~~~~~\\
r(i+1, i) &=& \frac{D}{h^2},
\end{eqnarray}
where $h = 2\pi/N_{\rm state}$ is the distance between the neighboring states, and $r(i, j)$ is the transition rate from $i$ to $j$. In the limit of $h\rightarrow0$, the above dynamics converges to the following Langevin dynamics on the ring $x\in [0, 2\pi)$:
\begin{eqnarray}
\dot{x} = A(-\cos x + f) + \sqrt{2D}\xi_t,
\end{eqnarray}
where $\xi_t$ is the Gaussian white noise satisfying  $\left<\xi_t\xi_{t'}\right> = \delta(t-t')$. If we take the limit of $A\rightarrow0$, the stationary state is in equilibrium. Therefore, our hopping model is a good playground for testing the predicted behavior in both the limits.\\ \indent
In Fig.~\ref{fig: mj}, we show the results of numerical experiments of the one-dimensional hopping model with parameters $D = 1$ and $f = 3$. Figure.~\ref{fig: mj_eq} and \ref{fig: mj_Nstate} show the convergence of the optimal estimation $\tilde{\sigma}^{\rm M}[d^*]$ to the true entropy production rate in (i) the equilibrium limit ($A\rightarrow0$) and (ii) the Langevin limit ($h\rightarrow0$).
For the sake of comparison with our estimators, the calculations of the optimal value $\tilde{\sigma}^{\rm M}[d^*]$ and the true value $\sigma$ are conducted using the stationary distribution obtained by exact diagonalization of the transition matrix.
{\color{black}
Concretely, $\tilde{\sigma}^M[d^*]$ and $\sigma$ are calculated using the stationary distribution $p$ as follows:
\begin{eqnarray}
\tilde{\sigma}^M[d^*] &=& \sum_i \frac{2\left\{p(i)r(i, i+1) - p(i+1)r(i+1, i)\right\}^2}{p(i)r(i, i+1) + p(i+1)r(i+1, i)},~~~~~\\
\sigma &=& \sum_i \left\{p(i)r(i, i+1) - p(i+1)r(i+1, i)\right\}\\
&&~~~~~\times\ln\frac{p(i)r(i, i+1)}{p(i+1)r(i+1, i)}.
\end{eqnarray}
}
The results show that the short-time TUR-based estimator gives just a lower value of the true entropy production rate in Markov jump processes, while the true value can be obtained in both the two limits.\\ \indent
In Fig.~\ref{fig: mj_tau}, the performance of the learning estimator $\widehat{\sigma}^{\rm M}[d]$ is compared with the simple estimator $\widehat{\sigma}_{\rm simple}^{\rm M}$. Here, the estimation is conducted using trajectory data of length $\tau_{\rm obs}$ which are sampled every $\Delta t = 10^{-3}$, and the underlying dynamics is generated by the Gillespie algorithm \cite{Gillespie1977}. 
The convergence of the learning estimator is faster than the simple estimator, while both estimators converge fast compared to the other examples due to the simplicity of the present model.\\ \indent
In addition to the good convergence, we find another advantage of the learning estimator that it is robust against the value of the sampling interval of the trajectory data. In Fig.~\ref{fig: mj_dt}, we show the sampling interval dependence of the estimators with a fixed trajectory length $\tau_{\rm obs} = 10^3$. The simple estimator deviates from the true value as we increase the sampling interval $\Delta t$, while the learning estimator is not affected much. This is because, for the simple estimator, the sampling interval should be small enough so that it can detect all of the back and forth dynamics between states, which is necessary for the accurate estimation of the transition rates. On the other hand, the TUR-based estimator is not affected much by the coarse-graining of dynamics, because back and forth dynamics just cancel out in the calculation of the generalized current.

\section{Conclusions}
\label{sec: conclusions}
{\color{black}In this paper, we have developed a theoretical framework to apply machine learning to the estimation of the entropy production rate on the basis of the TUR. Our framework can treat both Langevin dynamics and Markov jump processes, and is relevant to biological systems that can be modeled by stochastic dynamics \cite{Ritort2006, Toyabe2010a}.}\\ \indent
First, we have analytically argued the short-time TUR. Specifically, we derived Eq.~(\ref{eq: ITUR}) and established its equality condition.
Equality is always achievable in Langevin dynamics even if the state is far from equilibrium, while this is not the case for Markov jump processes.
Our formulation includes the TUR with the partial entropy production rate of subsystems under autonomous interactions, which reveals the hierarchy of the estimation as represented in Eq.~(\ref{eq: hierarchy}) under limited availability of trajectory data.\\ \indent
On the basis of these analytical results, we have constructed the learning estimators [the binned learning estimator $\widehat{\sigma}^\lambda[{\bm d}_{\rm bin}]$ in Eq.~(\ref{eq: binned}) and the Gaussian learning estimator $\widehat{\sigma}[{\bm d}_{\rm Gauss}]$ in Eq.~(\ref{eq: Gaussian})] for Langevin dynamics, and have numerically shown that they can perform very well in several setups as presented in Fig.~\ref{fig: 2beads} to Fig.~\ref{fig: mj}.
Our learning estimators are useful under the practical condition that only finite-length trajectory data is available, because of the following properties: (i) good convergence, (ii) small computational cost and (iii) independence of the system parameters such as the diffusion constant.
For Markov jump processes, we have numerically demonstrated that the estimated values become exact in the equilibrium limit and the Langevin limit using the one-dimensional hopping model as shown in Fig.~\ref{fig: mj_eq} and (b).
We have also found another practical advantage of the TUR-based estimators in Markov jump processes: they are robust against the choice of the sampling interval of observation as shown in Fig.~\ref{fig: mj_dt}.\\ \indent
The foregoing results suggest that the maximization of Eq.~(\ref{eq: rhs}) is a good definition of the entropy production rate in Langevin dynamics from the learning perspective.
It is an interesting question to ask whether the maximized lower value of the short-time TUR has meaning as an indicator of dissipation in Markov jump processes as well, even
when it is not equal to the entropy production rate in general.\\ \indent
{\color{black}
We note that the exact estimation of the entropy production rate is also possible with the long-time TUR in Langevin dynamics, although it has not been explicitly claimed in the previous studies. This can be proved by following the fact that the rate function of the probability distribution and the empirical current $I(p, j)$ becomes quadratic in Langevin dynamics \cite{Gingrich2017}, and the proof in Ref.~\cite{Gingrich2016}. In addition, the optimized coefficient field should be proportional to the thermodynamic force field \cite{Busiello2019} as is the case for the short-time TUR. However, the short-time TUR seems to be better for the estimation of the entropy production rate, since it is not easy to prepare the ensemble of the long-time generalized current. For example, it may not be easy to determine the time length of the generalized current, since the exact estimation fails if it is not long enough \cite{Li2019}.
}\\ \indent
There remains room for improvement of the learning estimators, for example, in the choice of the analytical expression of the coefficient ${\bm d}({\bm x})$ in high dimensional setups. 
Lastly, the application of the learning method to more complex Markov jump processes with finite $\Delta t$ is a challenging but interesting problem, as the reconstruction of transitions becomes a non-trivial task.
We leave these questions for future consideration.\\ \indent
{\it Note added}. - After completion of our work, we became aware that Tan Van Vu and his collaborators had obtained similar results \cite{Tan2020}.

\begin{acknowledgments}
We thank Kazuya Kaneko, Hiroki Yamaguchi, Yuto Ashida, Shin-ichi Sasa, David H. Wolpert, Sreekanth K. Manikandan, Ralf Eichhorn, Supriya Krishnamurthy, Stefano Bo and Chun-Biu Li for fruitful discussions. {\color{black}We also thank Jordan M. Horowitz for the valuable comments on the manuscript.}
S. I. is supported by JSPS KAKENHI Grant No.~19H05796 and JST Presto Grant No.~JP18070368.
A. D. is supported by the World Premier International Research Center Initiative (WPI), MEXT, Japan.
T. S. is supported by JSPS KAKENHI Grant No.~16H02211 and 19H05796.
\end{acknowledgments}

\newpage
\appendix
\section{Details of the gradient ascent}
\label{sec: adam}
In this appendix, we explain the details of the gradient ascent.
We first introduce the algorithm called Adam \cite{Kingma2014}
and then explain the details of the data splitting.\\ \indent
\subsection{Adam}
We adopt Adam to improve the convergence of the gradient ascent in this study. Adam was recently proposed and has become popular in the field of deep learning because of its good convergence and simple algorithm. The update rule of Adam is as follows:
\begin{subequations}
\begin{eqnarray}
g_{t, i}  &\leftarrow& \partial_{a_i}f(a)\\
m_{t, i} &\leftarrow& \beta_1 m_{t-1, i} + (1-\beta_1)g_{t, i}\\
v_{t, i} &\leftarrow& \beta_2 v_{t-1, i} + (1-\beta_2)g_{t, i}^2\\
\widehat{m}_{t, i} &\leftarrow& m_{t, i} / (1-(\beta_1)^t)\\
\widehat{v}_{t, i} &\leftarrow& v_{t, i} / (1-(\beta_2)^t)\\
a_i &\leftarrow& a_i + \alpha\widehat{m}_{t, i} / (\sqrt{\widehat{v}_{t, i}} + \epsilon),
\end{eqnarray}
\end{subequations}
where $t$ is the number of current iterations, $i$ is the index of parameters $a$, and $m_{0, i}$, $v_{0, i}$ and $t$ are initialized with 0.
There are four hyperparameters $\alpha, \beta_1, \beta_2$ and $\epsilon$, which are suggested to be $\alpha = 10^{-3}, \beta_1 = 0.9, \beta_2 = 0.999$ and $\epsilon = 10^{-8}$ in the original paper. 
Among them, $\beta_1, \beta_2$ and $\epsilon$ are often kept unchanged from the suggested values, and thus we only tune $\alpha$ in this study.\\ \indent
Adam is considered to be efficient compared to the standard gradient ascent in two ways.
First, it determines the update vector depending not only on the current gradient but also on the past update vectors.
This gives inertia to update vectors, which is especially helpful to climb a function shaped like a mountain elongated in one direction, which gradually slopes to its maximum. Second, since Adam automatically tunes the step size for each parameter, it does not require a careful tuning of step size, which is not the case for the standard gradient ascent.\\ \indent

\subsection{Data splitting and hyperparameter tuning}
We next explain the details of the data splitting and the hyperparameter tuning here. 
Concretely, we divide the whole trajectory data ${\bm x}_0, {\bm x}_{\Delta t}, ... {\bm x}_{N\Delta t}$ into two parts, training ${\bm x}_0, ..., {\bm x}_{(N/2-1)\Delta t}$ and test data ${\bm x}_{N/2\Delta t}, ...,  {\bm x}_{(N-1)\Delta t}$. Here, we use the displacements $[{\bm x}_0, {\bm x}_1], ..., [{\bm x}_{(N/2-1)\Delta t}, {\bm x}_{N/2\Delta t}]$ to calculate $\widehat{\sigma}[d]$ for the case of ${\bm x}_0, ..., {\bm x}_{(N/2-1)\Delta t}$.
The number of data points in the training and the test data are aligned in this study for the sake of simplicity. Also, we do not consider the use of minibatches and the stochastic gradient ascent for the same reason.\\ \indent
The division by the middle point is important to minimize the leakage of information about the occurrence frequency in space.
If it is negligible, we can evaluate the performance of a learning estimator simply by checking the peak of the learning curve of $\widehat{\sigma}[d]|_{\rm test}$. An estimator with a higher peak of $\widehat{\sigma}[d]|_{\rm test}$ is assumed to be better because there is no way for $d(x)$ to be overfitted to the test data, and $\widehat{\sigma}[d]|_{\rm test}$ is expected not to exceed the true entropy production rate.\\ \indent
In reality, however, it might be possible that $\widehat{\sigma}[d]|_{\rm test}$ gives a larger value by chance. Indeed, the estimated values often become larger than the true entropy production rate when the data size is small, because (i) the correlation between the training and the test data are not negligible when the trajectory length is small, and (ii) the outliers of statistical fluctuations are picked up for the estimation when the fluctuation of the learning curve is large. Nonetheless, we find that following the above rules is an effective strategy to achieve fast convergence, because such effects soon vanish as the trajectory length increases.
\\ \indent
Therefore, we can conduct the hyperparameter tuning simply by finding the hyperparameters that maximize the peak of the learning curve of $\widehat{\sigma}[d]|_{\rm test}$. In this study, for the sake of simplicity, we tune the hyperparameters beforehand using other trajectories, and then calculate the mean and its standard deviation of the estimation results by using ten independent trajectories and adopting the tuned values for the hyperparameters.
In practice, it is also possible to conduct both the hyperparameter tuning and the estimation of the entropy production rate using the same trajectory data.\\ \indent

\section{Details of the estimators}
\label{sec: details}
\begin{figure}
\begin{tabular}{c}
	\subfigure[]{
		\includegraphics[width = 0.95\linewidth]{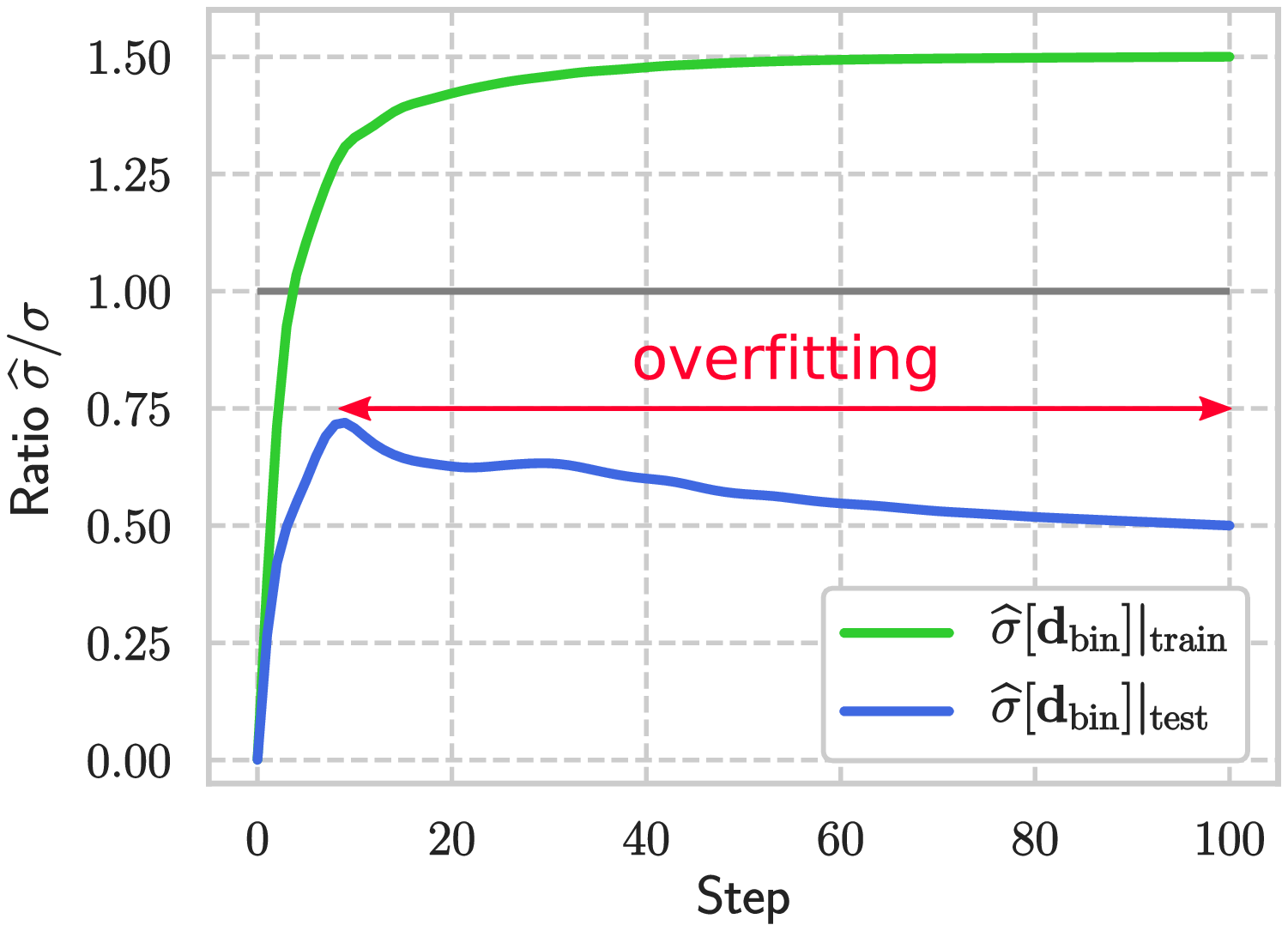}\label{fig: ex1}}\\
	\subfigure[]{
		\includegraphics[width = 0.95\linewidth]{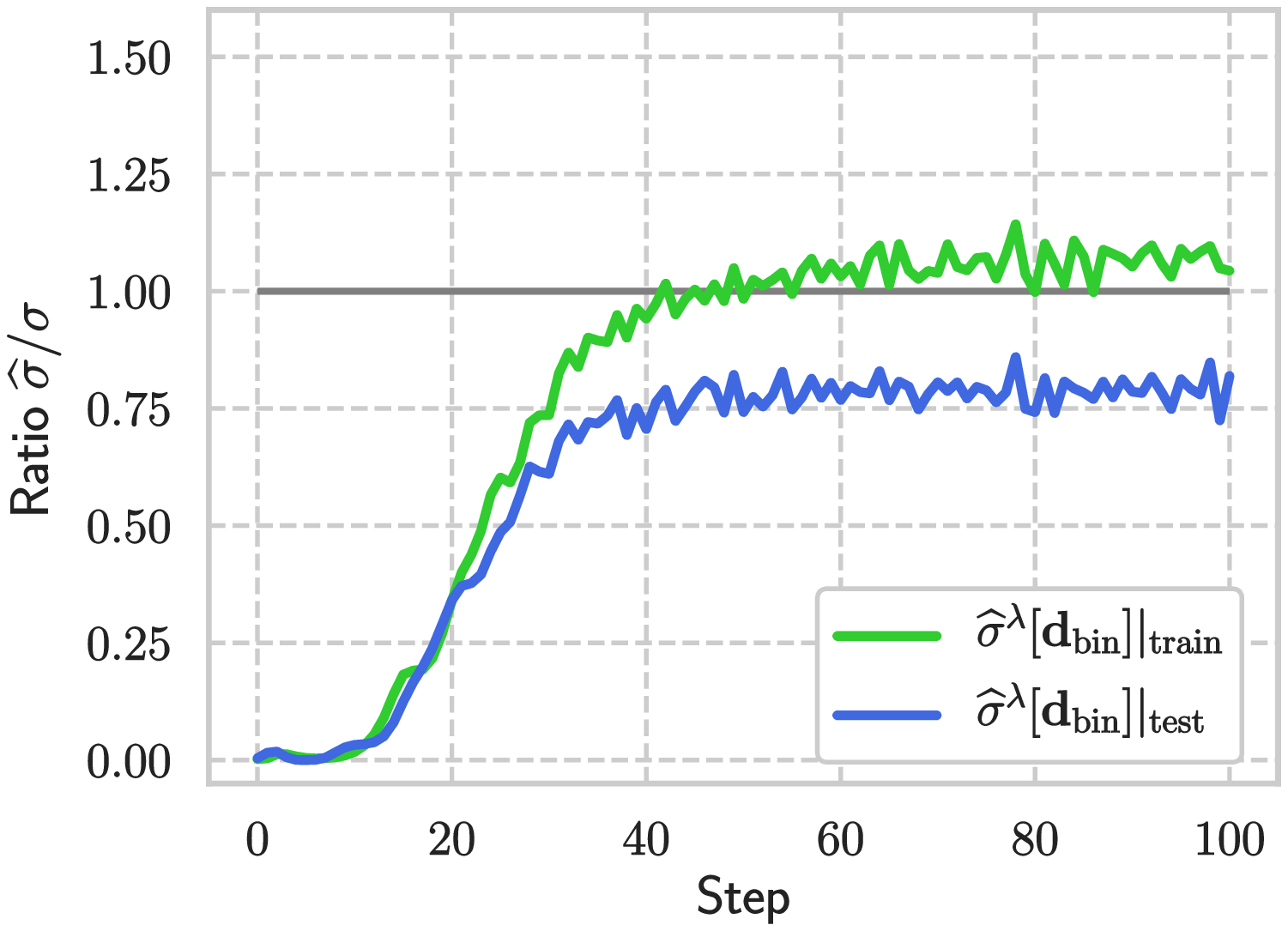}\label{fig: ex2}}\\
	\subfigure[]{
		\includegraphics[width = 0.95\linewidth]{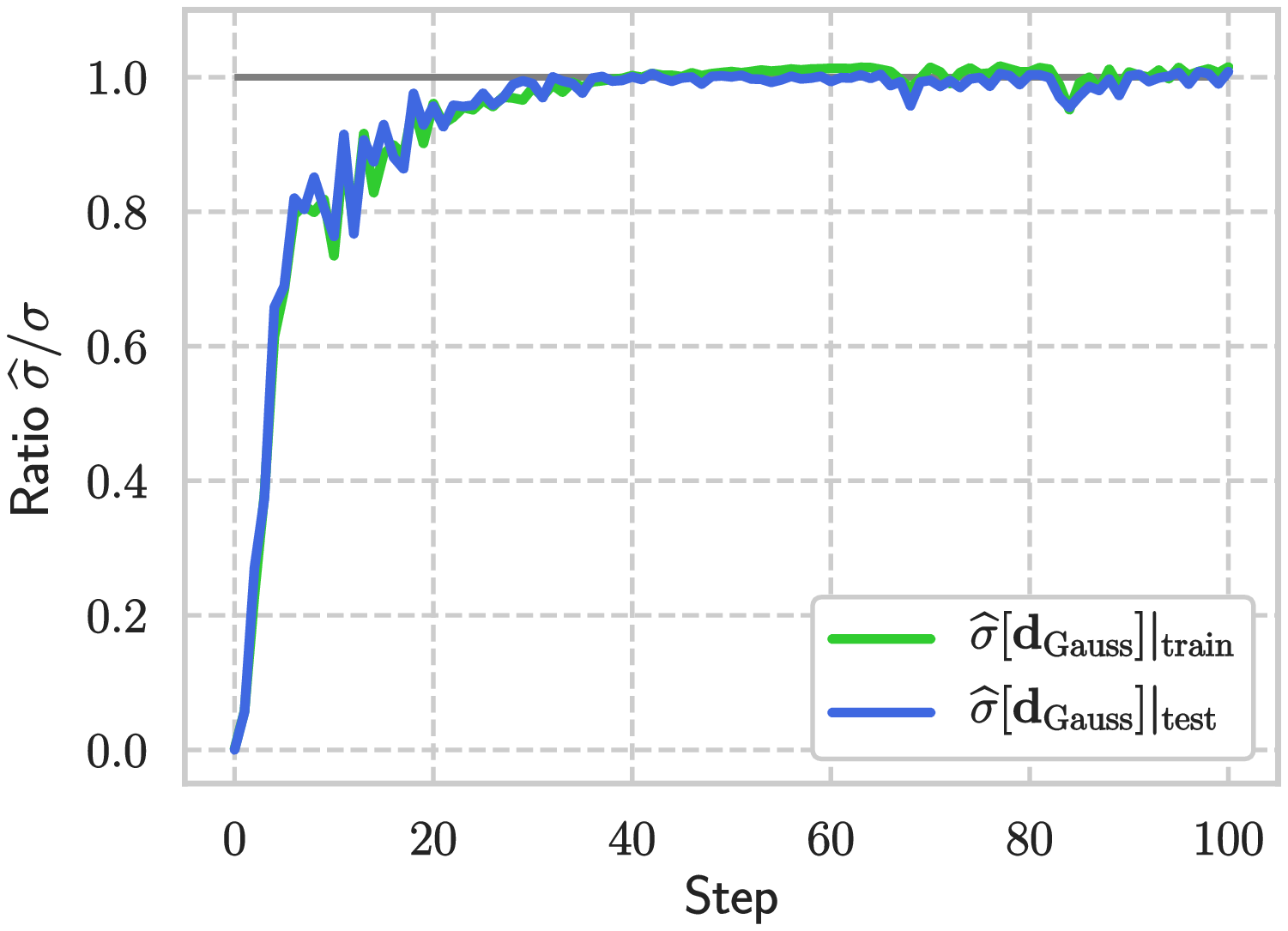}\label{fig: ex3}}\\		
\end{tabular}
\caption{Examples of the learning curves obtained using a trajectory generated by the two-beads model ($r = 0.1$, $\tau_{\rm obs} = 1000$): (a) Learning curves of $\widehat{\sigma}[d_{\rm bin}]$ (hyperparameters: $N_{\rm bin} = 20$, $\alpha = 1$, $\lambda = 0$). (b) Those of $\widehat{\sigma}^\lambda[{\bm d}_{\rm bin}]$ (hyperparamters: $N_{\rm bin} = 20$, $\alpha = 1$, $\lambda = 0.1$). (c) Those of $\widehat{\sigma}[{\bm d}_{\rm gauss}]$ (hyperparamters: $N_{\rm bin} = 6$, $\alpha = 10$). The same trajectory is used for all the experiments. The vertical axes are normalized by the true entropy production rate $\sigma$. The other system parameters are set in the same was as those in Fig.~\ref{fig: 2beads}. Since we adopt the maximum value of $\widehat{\sigma}[{\bm d}]|_{\rm test}$ for the estimation of the entropy production rate, the estimated values become (a) $0.72 \sigma$, (b) $0.86 \sigma$ and (c) $1.0 \sigma$.}
\label{fig: learning_curves}
\end{figure}
\begin{figure*}
\begin{center}
\begin{tabular}{cc}
	\subfigure[Convergence of the estimates]{
		\includegraphics[width = 0.45\linewidth]{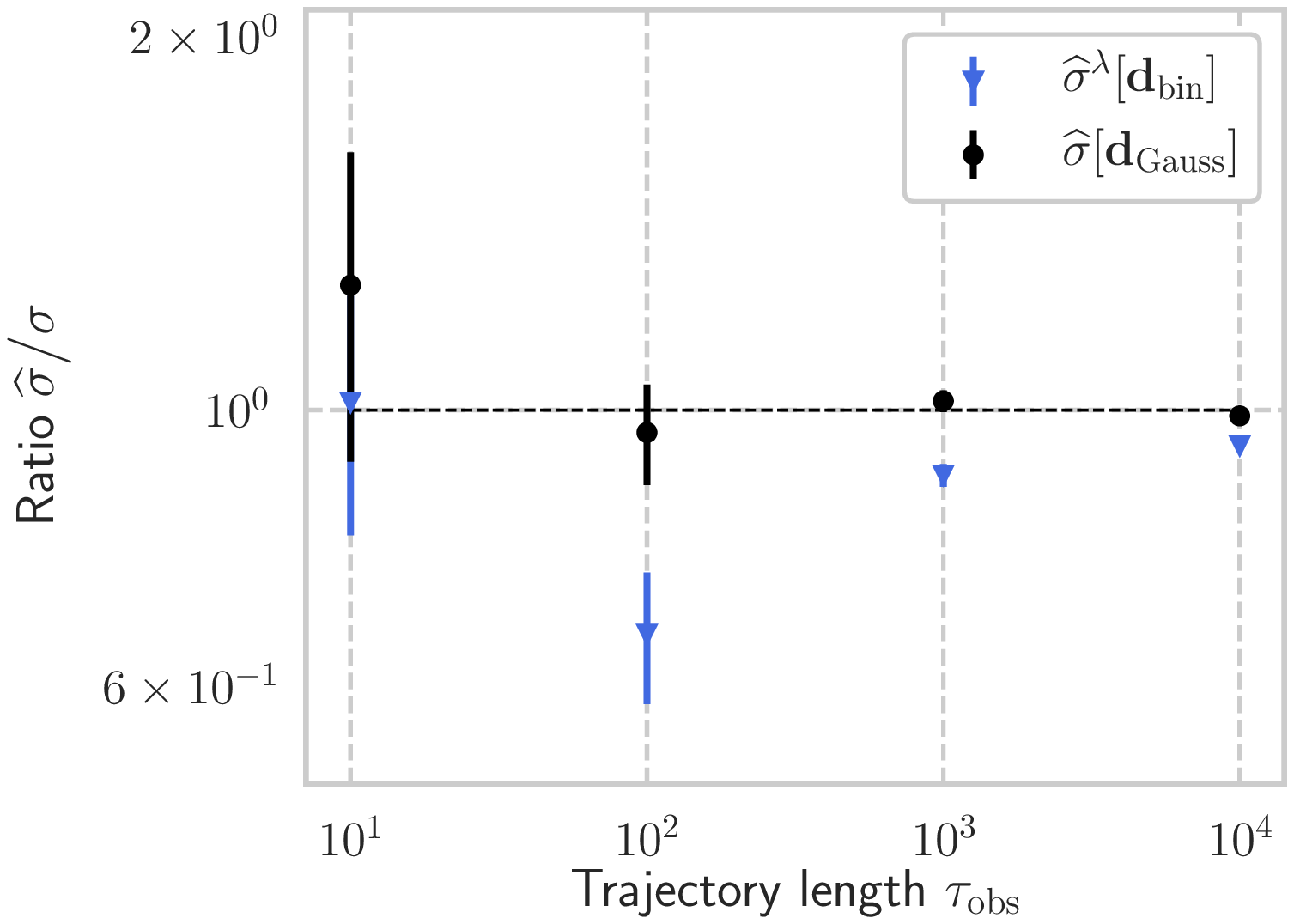}\label{fig: compare_convergence}}&
	\subfigure[Computation time]{
		\includegraphics[width = 0.45\linewidth]{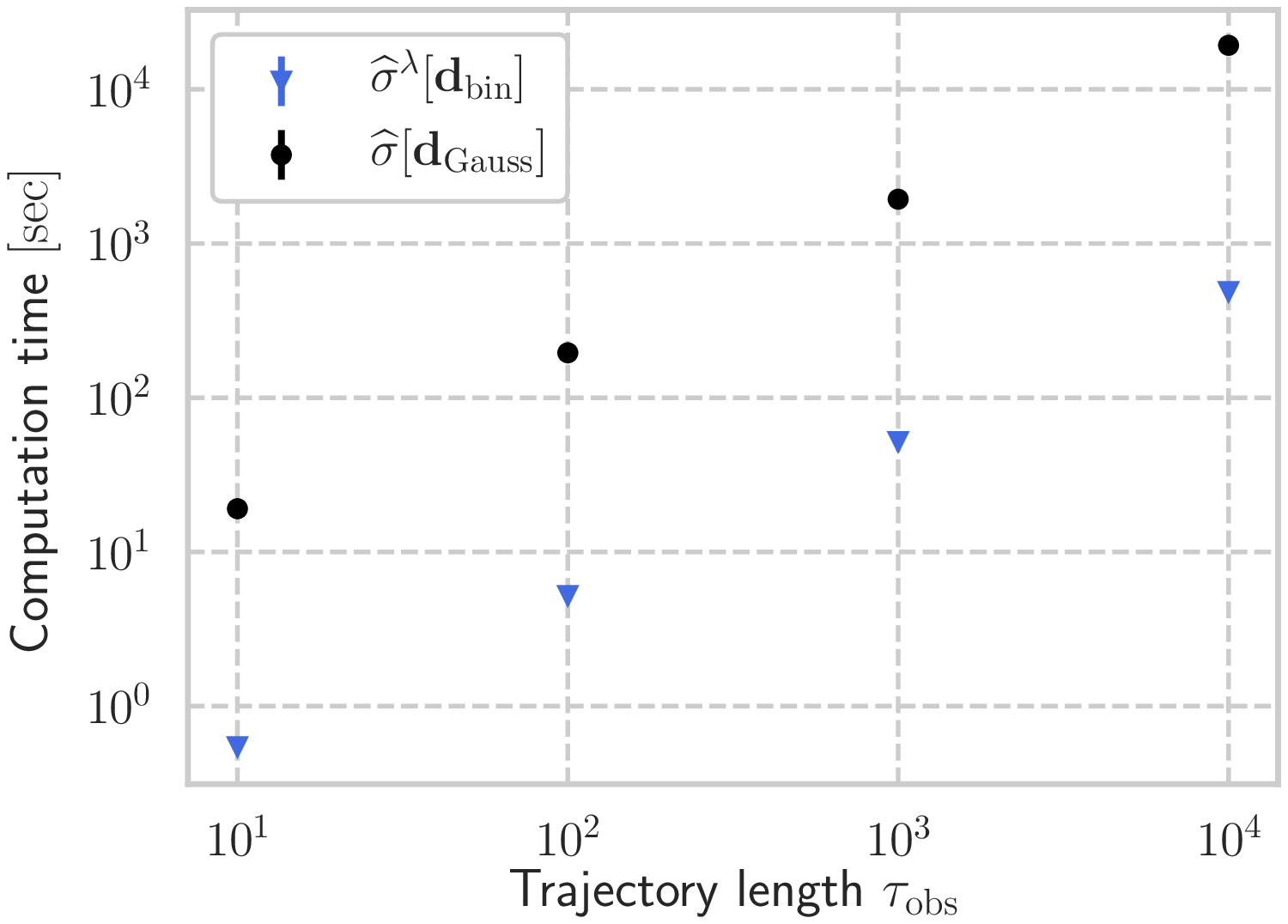}\label{fig: compare_time}}	
\end{tabular}
\caption{Comparison of the two learning estimators by using data generated by the two-beads model ($r = 0.1$) in terms of (a) the convergence speed and (b) the computation time. The mean and its standard deviation of ten independent trials are plotted. The computation time is measured as the time on a single core of a cluster computer. The other system parameters are set to the same as those in Fig.~\ref{fig: 2beads}.}
\label{fig: compare}
\end{center}
\end{figure*}

\begin{table*}
\begin{center}
\begin{tabular}{c c | c c c c c c} \hline
Model 				& $\tau_{\rm obs}$ 	& Algorithm 						&$N_{\rm bin}$	&$\alpha$	&$\lambda$	&$N_{\rm step}$	 \\ \hline
Two-beads ($r = 0.1$) 	& $10 - 10^4$		&$\widehat{\sigma}[{\bm d}_{\rm Gauss}]$		& 6			& 10		& 			& 100			 \\
					& $10^4$			&$\widehat{\sigma}^\lambda[{\bm d}_{\rm bin}]$	& 20			& 1		&$10^{-4}$	& 300			 \\
					& $10^3$			&$\widehat{\sigma}^\lambda[{\bm d}_{\rm bin}]$	& 12			& 1		&$10^{-2}$	& 300			 \\
					& $10^2$			&$\widehat{\sigma}^\lambda[{\bm d}_{\rm bin}]$	& 8			& 1		&$10^{-1}$	& 300			 \\
					& $10$			&$\widehat{\sigma}^\lambda[{\bm d}_{\rm bin}]$	& 8			& 1		&$10^{2}$		& 300			 \\ \hline
Two-beads ($r = 0.5$) 	& $10 - 10^4$		&$\widehat{\sigma}[{\bm d}_{\rm Gauss}]$		& 6			& 10		& 			& 100			 \\ \hline
Five-beads ($r = 0.1$)	& $10^4$			&$\widehat{\sigma}^\lambda[{\bm d}_{\rm bin}]$	& 2			& 1		&$10^{-5}$	& 300			 \\
					& $10^3$			&$\widehat{\sigma}^\lambda[{\bm d}_{\rm bin}]$	& 2			& 1		&$10^{-2}$	& 300			 \\
					& $10^2$			&$\widehat{\sigma}^\lambda[{\bm d}_{\rm bin}]$	& 2			& 1		&$10^{-1}$	& 300			 \\
					& $10$			&$\widehat{\sigma}^\lambda[{\bm d}_{\rm bin}]$	& 2			& 1		&$1$			& 300			 \\ \hline
Five-beads ($r = 0.5$)	& $10^4$			&$\widehat{\sigma}^\lambda[{\bm d}_{\rm bin}]$	& 2			& 1		&$10^{-3}$	& 300			 \\
					& $10^3$			&$\widehat{\sigma}^\lambda[{\bm d}_{\rm bin}]$	& 2			& 1		&$10^{-2}$	& 300			 \\
					& $10^2$			&$\widehat{\sigma}^\lambda[{\bm d}_{\rm bin}]$	& 2			& 1		&$10^{-1}$	& 300			 \\
					& $10$			&$\widehat{\sigma}^\lambda[{\bm d}_{\rm bin}]$	& 2			& 1		&$10$		& 300			 \\ \hline
Mexican-hat ($A=10^{-4}$)& $10 - 10^4$		&$\widehat{\sigma}[{\bm d}_{\rm Gauss}]$		& 6			& 10		& 			& 100			 \\ \hline
Mexican-hat ($A=1$) 	& $10 - 10^4$		&$\widehat{\sigma}[{\bm d}_{\rm Gauss}]$		& 6			& 1		& 			& 100			 \\ \hline				
Mexican-hat ($A=10^2$) 	& $10 - 10^4$		&$\widehat{\sigma}[{\bm d}_{\rm Gauss}]$		& 6			& 0.3		& 			& 100			 \\ \hline
One-dimensional hopping 	& 				&$\widehat{\sigma}^{\rm M}[d]$			& 			& 0.01	& 			& 300			 \\ \hline
\end{tabular}
\caption{Hyperparameters used for the learning estimators in this study. The details of hyperparameter tuning can be found in Supplymental Material. 
$N_{\rm step}$ of $\widehat{\sigma}^\lambda[{\bm d}_{\rm bin}]$ is set to be bigger than that of $\widehat{\sigma}[{\bm d}_{\rm Gauss}]$ because $\widehat{\sigma}^\lambda[{\bm d}_{\rm bin}]$ is computationally fast and the peak of the learning curves of $\widehat{\sigma}^\lambda[{\bm d}_{\rm bin}]$ sometimes comes at larger step number.}
\label{table: hyperparameters}
\end{center}
\end{table*}

{\color{black}
In this appendix, we give details of the estimators. We first define the learning estimators, and compare them in terms of convergence speed and computation time. Then, we give a detailed explanation on the KDE estimators \cite{Li2019}.
\subsection{Learning estimators}
In this subsection, we define the learning estimators by defining the model function of ${\bm d}({\bm x})$. For simplicity, we mainly focus on the case of two dimensional data ${\bm x} = (x, y)$, but the extension to the one or higher dimensional case is straightforward. Let $n_{\rm dim}$ be the dimension.\\ \indent
We first define the binned learning estimator $\widehat{\sigma}[{\bm d}_{\rm bin}]$. 
This estimator uses a coarse-grained function for ${\bm d} ({\bm x})$ which is binned into a square lattice.
Concretely, we define ${\bm d}_{\rm bin}({\bm x})$ as 
\begin{eqnarray}
&&{\bm d}_{\rm bin}({\bm x}) := {\bm d} (i(x), j(y)), \nonumber\\
&&~~{\rm ~with~}i(x) := \small\left\lceil\frac{x-x_{\rm min}}{b_{x}}\right\rceil, ~j(y) := \left\lceil\frac{y-y_{\rm min}}{b_{y}}\right\rceil\normalsize,~~~~\label{eq: binned}
\end{eqnarray}
where the indexes run over $i = 1,..., N_{\rm bin}, j = 1, ..., N_{\rm bin}$, $b_x$ and $b_y$ are the bin widths, $x_{\rm min}$ and $y_{\rm min}$ are the minimum of the binning and the brackets denote the ceiling function.
We determine these constants in the following manner. We first set $x_{\rm max}, x_{\rm min}, y_{\rm max}$ and $y_{\rm min}$ depending on the trajectory to include all the data points in the rectangle. Then, we determine $b_x$ and $b_y$ by dividing each direction by $N_{\rm bin}$, i.e.,
\begin{eqnarray}
b_x &=& \frac{x_{\rm max}-x_{\rm min}}{N_{\rm bin}},\\
b_y &=& \frac{y_{\rm max}-y_{\rm min}}{N_{\rm bin}}.
\end{eqnarray}
Thus, we consider $N_{\rm bin}$ as a hyperparameter to tune. The function ${\bm d}_{\rm bin}({\bm x})$ contains $n_{\rm dim}N_{\rm bin}^{n_{\rm dim}}$ parameters in total for the $n_{\rm dim}$ dimensional case.
The parameters are initialized by $\{{\bm d}(i, j)\}_k = {\rm uni}(-1, 1)$ before the gradient ascent, where ${\rm uni}(a, b)$ is a random variable that follows the uniform distribution in the range $a< x < b$.\\ \indent
Since ${\bm d}_{ij}$ are coupled with data points that lie in the same bin in the calculation of $\widehat{\sigma}[{\bm d}]$, ${\bm d}_{ij}$ are trained only with those data points.
In order to have ${\bm d}_{ij}$ trained in coordination with the surrounding parameters, we add a regularization term $\mathcal{R}({\bm d}_{\rm bin})$ in the objective function $\widehat{\sigma}[{\bm d}_{\rm bin}]$ of the gradient ascent as follows:
\begin{eqnarray}
f({\bm d}_{\rm bin}) &=& \widehat{\sigma}[{\bm d}_{\rm bin}]   -  \frac{\lambda}{4}\mathcal{R}({\bm d}_{\rm bin}), \\
\mathcal{R}({\bm d}_{\rm bin}) &:=& \sum_{i, j}  \!\sum_{\substack{i', j' \in {\rm nn}(i, j)}}  \!||{\bm d}(i, j) - {\bm d}(i', j')||^2,
\end{eqnarray}
where ${\rm nn}(i, j) := \{(i+1, j), (i-1, j), (i, j+1), (i, j-1)\}$ is the set of nearest neighbor indexes, and $||\cdot||$ is the $L^2$-norm whose definition is $||{\bm a}|| = \sqrt{\sum_i a_i^2}$, and $\lambda$ is another hyperparameter to tune in this estimator.
If we appropriately choose $\lambda$, the regularization term enhances the generalization capability of this estimator, because it requires the coefficient field to change smoothly over the space, and prevents the coefficient field from becoming overfitted to the training data. We denote this estimator with regularization as $\widehat{\sigma}^\lambda[{\bm d}_{\rm bin}]$.\\ \indent
Next, we define the Gaussian learning estimator $\widehat{\sigma}[{\bm d}_{\rm Gauss}]$. This estimator represents ${\bm d}({\bm x})$ as a linear combination of Gaussian functions whose centers are aligned to form a square lattice. Concretely, we define the $k$th element of ${\bm d}_{\rm Gauss}({\bm x})$ as
\begin{eqnarray}
&&\{ {\bm d}_{{\rm Gauss}}({\bm x}) \}_k:= \sum_{i=1}^{N_{\rm bin}}\sum_{j=1}^{N_{\rm bin}} \omega_k (i,j) K_k ({\bm x};i,j ),\nonumber\\
&&K_k ({\bm x};i,j ) := e^{ - \left({\bm x}-\bar{{\bm x}}(i,j)\right)^\mathsf{T} {{\bm M^{(k)}(i,j)}^{-1}}\left({\bm x}-\bar{{\bm x}}(i,j) \right) },~~~~~\label{eq: Gaussian}
\end{eqnarray}
where $\bar{{\bm x}}(i,j) = \left(x_{\rm min} + b_x \left(i - \frac{1}{2}\right), y_{\rm min} + b_y \left(j - \frac{1}{2}\right)\right)$ are the centers of the Gaussian functions ($i = 1,..., N_{\rm bin}, j = 1,..., N_{\rm bin}$), and $x_{\rm min}, y_{\rm min}, b_x$ and $b_y$ are determined in the same manner as before. 
Here, we assume that ${\bm M^{(k)}}(i,j)$ is a diagonal matrix whose $l$th element is ${\bm M^{(k)} (i,j)}_{ll}= ( {m^{(k)}_l (i,j)} )^2$ to make the matrix positive definite.
Therefore, ${\bm d}_{\rm Gauss}({\bm x})$ contains $n_{\rm dim}(n_{\rm dim} + 1)N_{\rm bin}^{n_{\rm dim}}$ parameters in total for the $n_{\rm dim}$ dimensional case: $n_{\rm dim}N_{\rm bin}^{n_{\rm dim}}$ from $\omega_{ij}^{(k)}$ and ${n_{\rm dim}}^2N_{\rm bin}^{n_{\rm dim}}$ from ${m^{(k)}_l (i,j)} $.
The parameters are initialized by $\omega_{ij}^{(k)}= {\rm uni}(-1, 1)$ and ${m^{(k)}_l (i,j)} = {\rm uni}(0, 1)$ before the gradient ascent.
\\ \indent
Unlike the binned learning estimator, the parameters of the Gaussian learning estimator are trained on the basis of all the data points. Therefore, ${\bm d}_{\rm Gauss}({\bm x})$ becomes automatically smooth over the space at the expense of additional computational cost. In addition, we emphasize that we do not assume that the state of the system itself is Gaussian, which guarantees its high performance for nonlinear dynamics with non-Gaussian distributions.\\ \indent
There are three hyperparameters $N_{\rm bin}, \lambda$ and $\alpha$ (step size of the gradient ascent) for the binned learning estimator, while there are two hyperparameters $N_{\rm bin}$ and $\alpha$ for the Gaussian learning estimator.
The details of the hyperparameter tuning are discussed in the Supplemental Material, and we summarize the results in TABLE~\ref{table: hyperparameters}.\\ \indent
In Fig.~\ref{fig: learning_curves}, we show examples of the learning curves, all of which are trained with the same trajectory generated by the two-beads model. In Fig.~\ref{fig: ex1}, there is a single peak in the curve of $\widehat{\sigma}[{\bm d}_{\rm bin}]|_{\rm test}$, which suggests that ${\bm d}_{\rm bin}(x)$ becomes overfitted to the training data from the peak. On the other hand, in Fig.~\ref{fig: ex2}, the overfitting is suppressed due to the regularization and the maximum of $\widehat{\sigma}^\lambda[{\bm d}_{\rm bin}]|_{\rm test}$ increases compared to that of Fig.~\ref{fig: ex1}. In Fig.~\ref{fig: ex3}, both of $\widehat{\sigma}[{\bm d}_{\rm Gauss}]|_{\rm train}$ and $\widehat{\sigma}[{\bm d}_{\rm Gauss}]|_{\rm test}$ converge to the true entropy production rate, which suggests that the Gaussian learning estimator is more data-efficient than the binned learning estimator.\\ \indent
Finally, we compare $\widehat{\sigma}^\lambda[{\bm d}_{\rm bin}]$ and $\widehat{\sigma}[{\bm d}_{\rm Gauss}]$ by using data generated by the two-beads model. We show the comparison results in Fig.~\ref{fig: compare}. We find that $\widehat{\sigma}[{\bm d}_{\rm Gauss}]$ is better in terms of the convergence speed, while $\widehat{\sigma}^\lambda[{\bm d}_{\rm bin}]$ is better for the computational cost. We confirmed that the relation between the learning estimators also holds in the other models and the parameter settings at least when data is two dimensional. On the basis of these observations, we adopt $\widehat{\sigma}[{\bm d}_{\rm Gauss}]$ for two dimensional data, and $\widehat{\sigma}^\lambda[{\bm d}_{\rm bin}]$ for higher dimensional data in the main text.
}

\subsection{Estimators with kernel density estimation}
In this subsection, we give a detailed description on the KDE estimators \footnotesize$\widehat{\dot{S}}_{\rm ss}^{\rm temp}$\normalsize and $\widehat{\sigma}[\widehat{{\bm F}}_{\rm sm}]$, both of which are introduced in the previous study \cite{Li2019}.\\ \indent
We first introduce the estimator \small$\widehat{\dot{S}}^{\rm temp}_{\rm ss}$\normalsize, which is based on the temporal average:
\small
\begin{eqnarray}
&&\widehat{\dot{S}}_{\rm ss}^{\rm temp}:= \frac{1}{\tau_{\rm obs}}\int_0^{\tau_{\rm obs}}\widehat{\bm F}_{\rm sm}({\bm x}(t))\circ d{\bm x}(t)\\
&&=\frac{1}{N\Delta t}\sum_{i=1}^{N}\widehat{\bm F}_{\rm sm}\left(\frac{{\bm x}_{i\Delta t} + {\bm x}_{(i-1)\Delta t}}{2}\right)\left[{\bm x}_{i\Delta t} - {\bm x}_{(i-1)\Delta t}\right],~~~~~~
\end{eqnarray}
where $\widehat{{\bm F}}_{\rm sm}({\bm x})$ is the thermodynamic force estimated by the kernel density estimation.
The thermodynamic force at ${\bm x}$ is calculated on the basis of displacements of data points which occurred around the position ${\bm x}$, by taking their distance from ${\bm x}$ into account. Concretely, $\widehat{{\bm F}}_{\rm sm}$ is obtained by
\begin{eqnarray}
&&\widehat{\bm F}_{\rm sm}({\bm x})=\frac{\widehat{\bm j}({\bm x})^\mathsf{T}{\bm B}^{-1}}{\widehat{p}({\bm x})}\\
&&:=\frac{1}{2\Delta t}\frac{\sum_{i=1}^{N-1}L({\bm x}_{i\Delta t}, {\bm x})\left[{\bm x}_{(i+1)\Delta t} - {\bm x}_{(i-1)\Delta t}\right]\cdot {\bm B}^{-1}}{\sum_{i=1}^{N-1}L({\bm x}_{i\Delta t}, {\bm x})},~~~~~~
\end{eqnarray}
\normalsize
where $L({\bm x}', {\bm x})$ is a kernel function which smoothly decreases as the distance between ${\bm x}$ and ${\bm x}'$ increases.
Here, we note that the KDE estimators rely on the knowledge of the diffusion matrix ${\bm B}$, while the other estimators introduced in this study are independent of such system parameters.\\ \indent
It was shown \cite{Li2019} that the Epanechnikov kernel realizes the fastest convergence:
\small
\begin{eqnarray}
L({\bm x}_{i\Delta t}, {\bm x}) \!\propto \! \begin{cases}
\prod_{j=1}^{d}\left(1 - \frac{(x_{i\Delta t;j} - x_j)^2}{b_j^2}\right), & \forall j~ |x_{i\Delta t, j} - x_j| < b_j, \nonumber\\
0, & {\rm otherwise}, \nonumber
\end{cases}~
\end{eqnarray}
where its bandwidth $b_j$ is determined by
\begin{eqnarray}
{\bm b} := \left(\frac{4}{N({\rm d}+2)}\right)^{\frac{1}{({\rm d}+4)}}\frac{\tilde{\bm \sigma}}{0.6745}.\label{eq: bandwidth}
\end{eqnarray}
\normalsize
Here, $\tilde{\bm \sigma}$ is a median absolute deviation:
\small
\begin{eqnarray}
\hspace{-0.3cm}
\tilde{\bm \sigma} := \sqrt{{\rm median}\left\{|v - {\rm median}(v)|\right\}{\rm median}\left\{|{\bm x} - {\rm median}({\bm x})|\right\}},~~ 
\end{eqnarray}
\normalsize
where $v$ is the magnitude of the velocities, i.e., $v_i = \sqrt{\sum_j (x_{i\Delta t, j} - x_{(i-1)\Delta t, j})^2}/\Delta t$.\\ \indent
Next, we introduce the estimator $\widehat{\sigma}[\widehat{\bm F}_{\rm sm}]$, which is based on the lower bound of the TUR.
We use the short-time TUR for this estimator, while the finite-time TUR is used in the original paper \cite{Li2019}.
Thus, we adopt the different notation from the original one \footnotesize$\widehat{\dot{S}}_{\rm TUR}^{(\widehat{F})}$\normalsize in this study.
$\widehat{\sigma}[\widehat{\bm F}_{\rm sm}]$ is simply defined by substituting $\widehat{\bm F}_{\rm sm}({\bm x})$ into $\widehat{\sigma}[{\bm d}]$. Since the thermodynamic force ${\bm F}({\bm x})$ becomes equivalent to the optimal coefficient ${\bm d}^*({\bm x})$ in the short-time TUR, $\widehat{\sigma}[\widehat{\bm F}_{\rm sm}]$ gives an exact estimate of the entropy production rate.\\ \indent
The expression of Eq.~(\ref{eq: bandwidth}) is usually derived assuming a Gaussian distribution for data points \cite{Bowman1997}, although its derivation seems not straightforward in this case because the kernel is used to estimate $\widehat{{\bm j}}/\widehat{p}$ which is not a density. In fact, Eq.~(\ref{eq: bandwidth}) was explained as a rule of thumb in \cite{Li2019}. Therefore, \footnotesize$\widehat{\dot{S}}_{\rm ss}^{\rm temp}$\normalsize and $\widehat{\sigma}[\widehat{\bm F}_{\rm sm}]$ would be optimized for data generated by linear Langevin equations. Indeed, in Sec.~\ref{sec: numerical experiment}, we show that their convergence become very slow for the nonlinear Langevin equation (\ref{eq: nonlinear}), while they achieve the good performance for linear Langevin equations (see Figs. \ref{fig: 2beads}, \ref{fig: 5beads} and \ref{fig: mx}).

{\color{black}
\section{Extension of the Gaussian learning estimator for higher dimensional setups}
\label{sec: move}
\begin{figure*}
\begin{center}
\begin{tabular}{cc}
	\subfigure[5-beads model with the deterministic method]{
		\includegraphics[width = 0.45\linewidth]{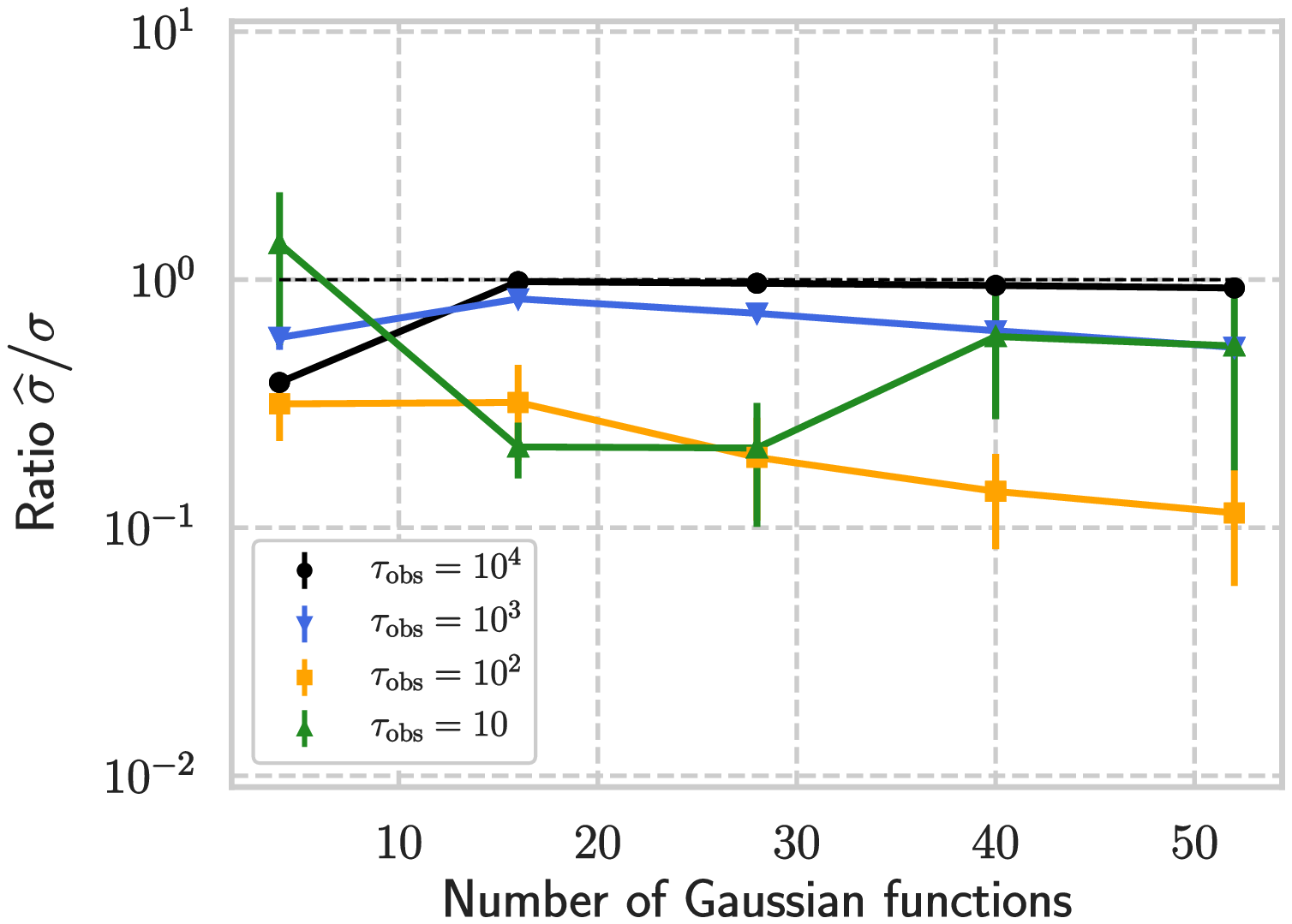}\label{fig: det_5}}&
	\subfigure[5-beads model with the gradient ascent]{
		\includegraphics[width = 0.45\linewidth]{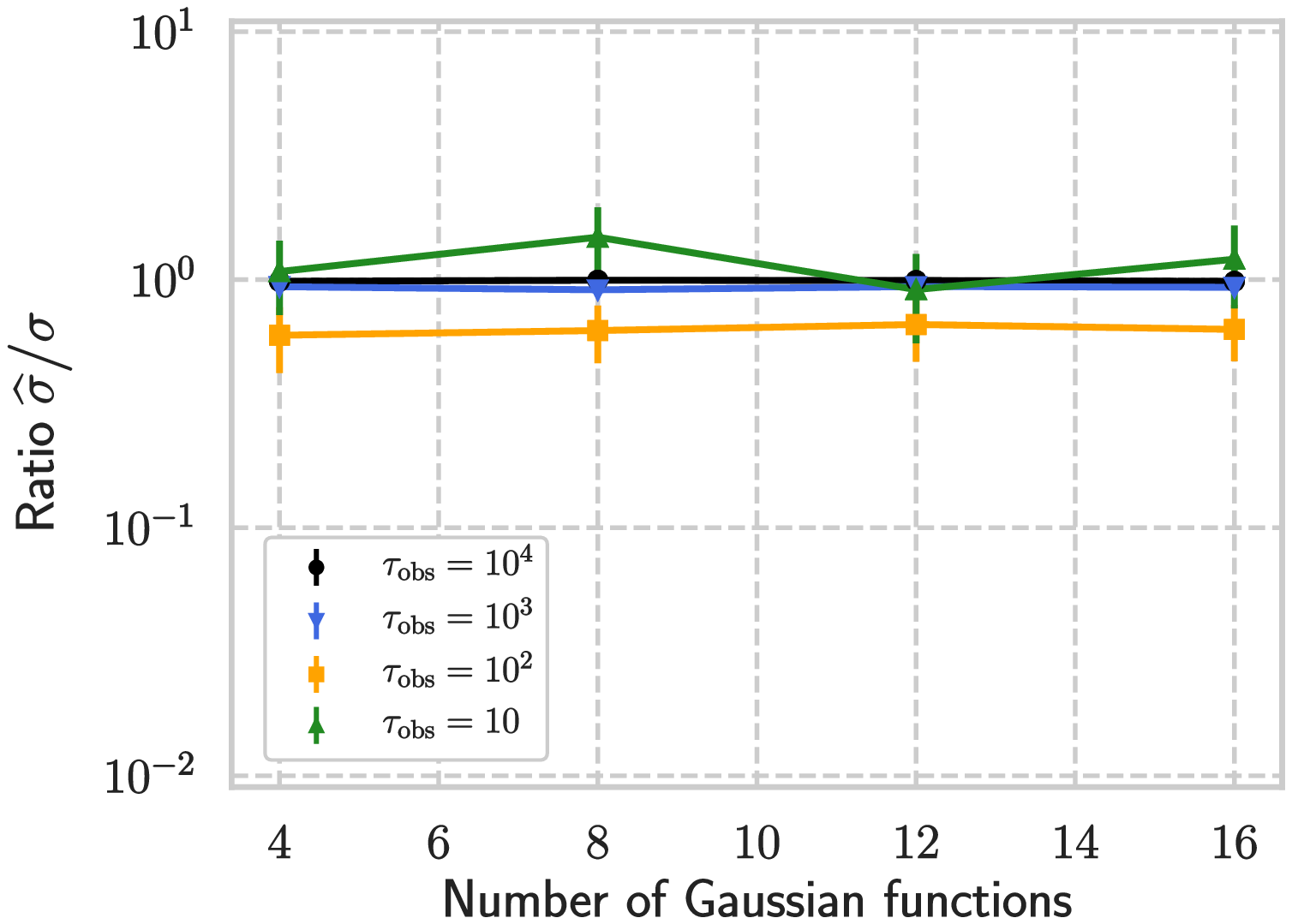}\label{fig: move_5}}\\
	\subfigure[10-beads model with the deterministic method]{
		\includegraphics[width = 0.45\linewidth]{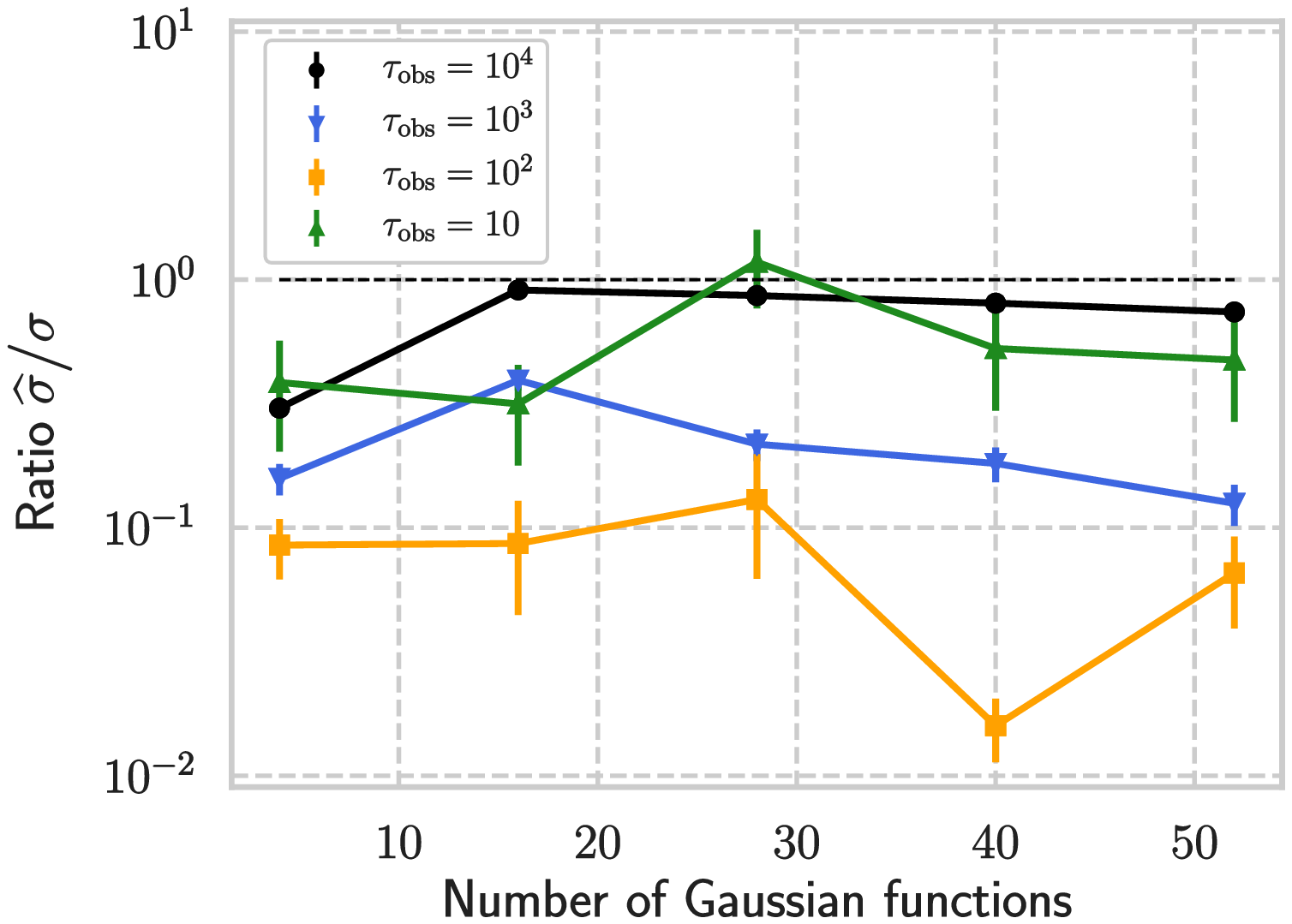}\label{fig: det_10}}&
	\subfigure[10-beads model with the gradient ascent]{
		\includegraphics[width = 0.45\linewidth]{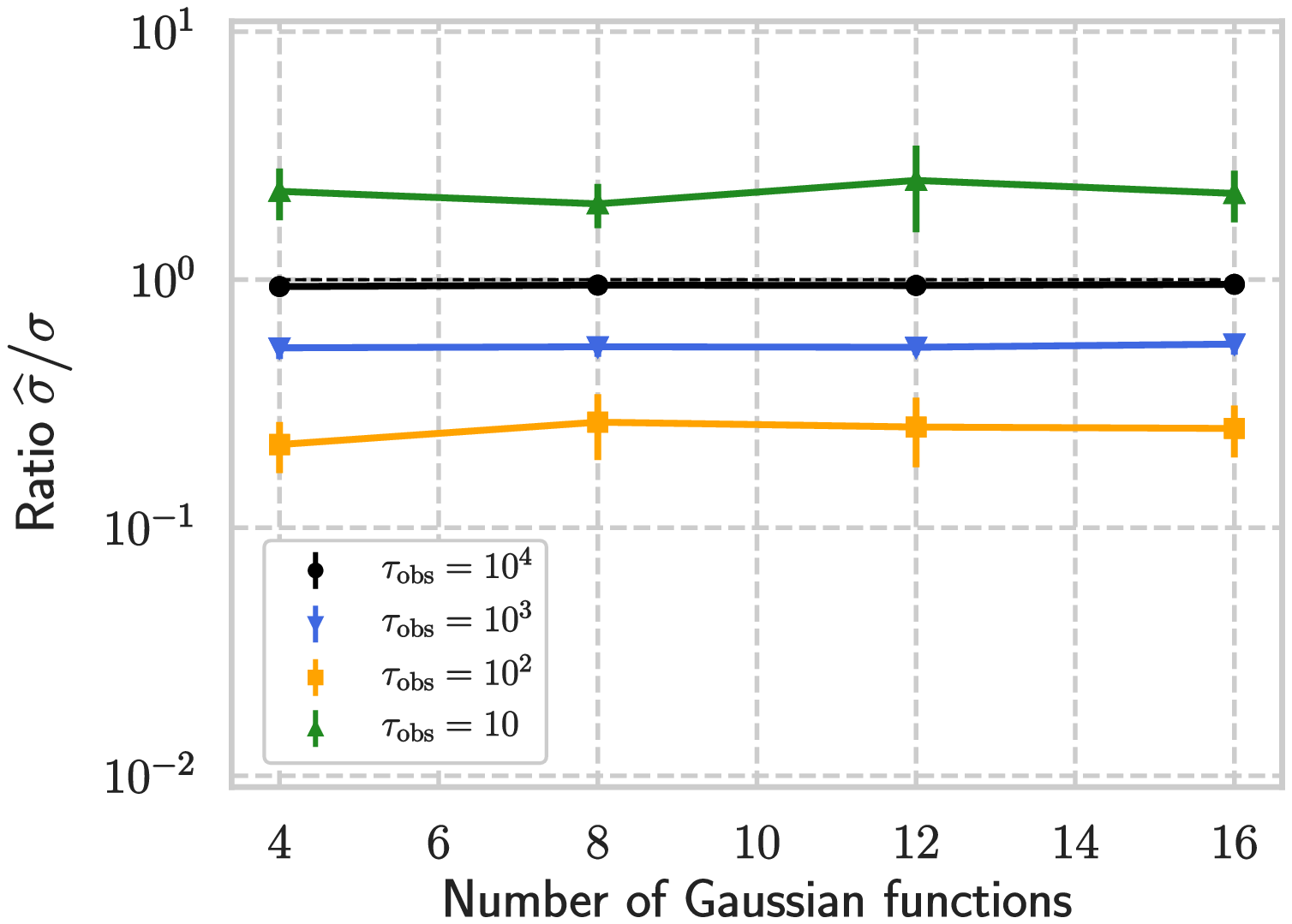}\label{fig: move_10}}\\
	\subfigure[15-beads model with the deterministic method]{
		\includegraphics[width = 0.45\linewidth]{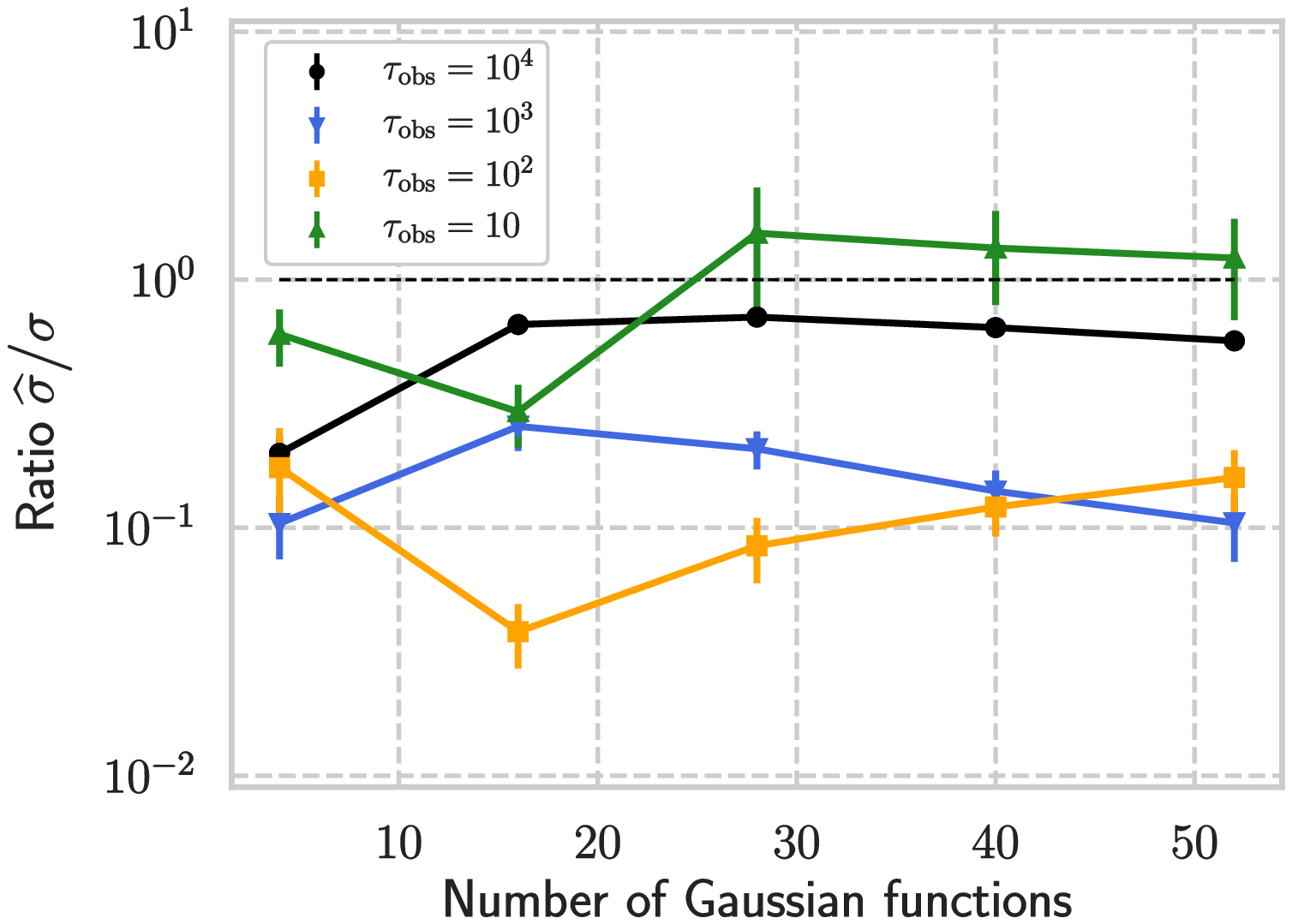}\label{fig: det_15}}&
	\subfigure[15-beads model with the gradient ascent]{
		\includegraphics[width = 0.45\linewidth]{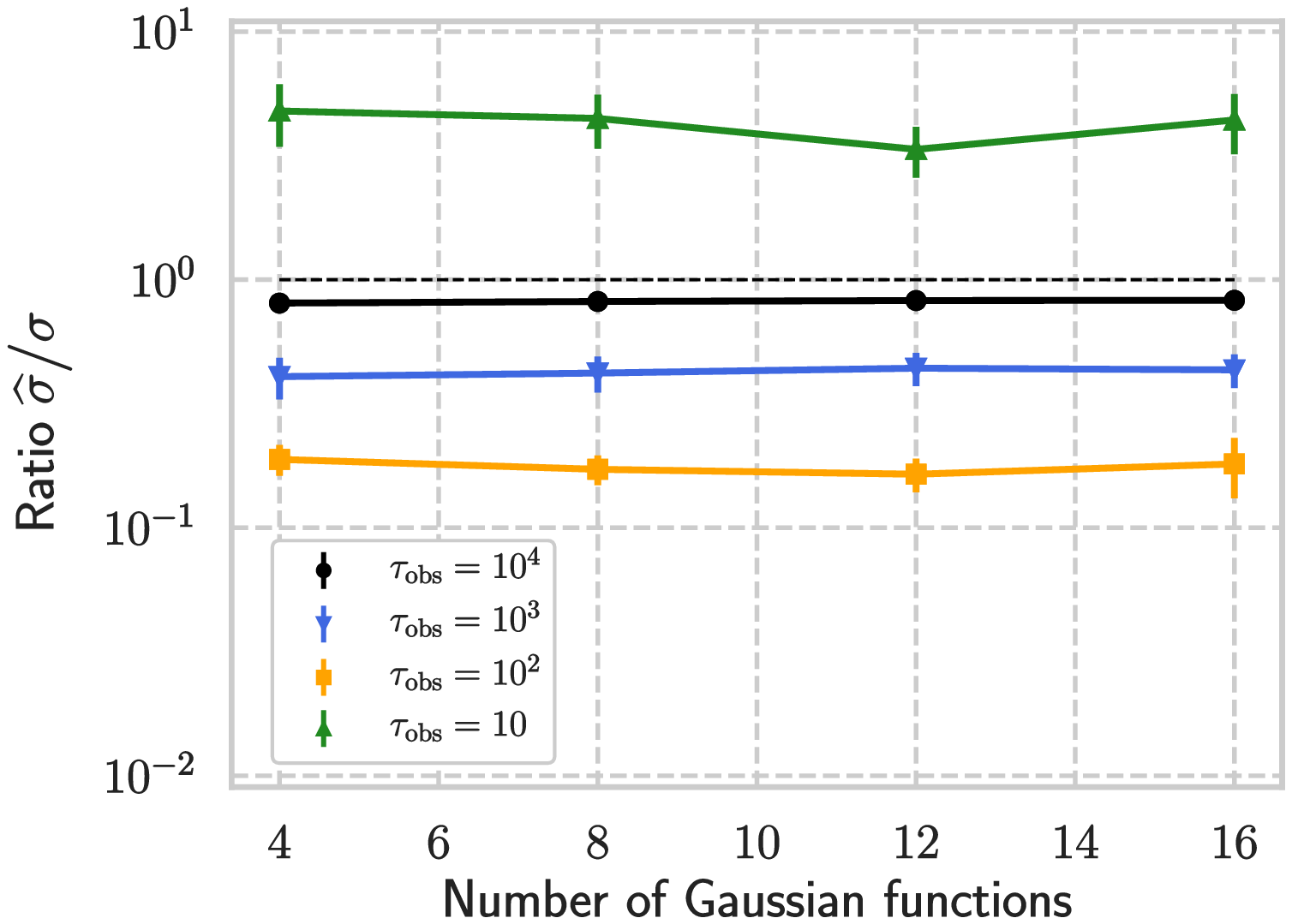}\label{fig: move_15}}
\end{tabular}
\caption{{\color{black}Numerical experiment with the $N$-beads model: (a)(c)(e) The dependence of $\widehat{\sigma}^{\rm det}[{\bm d}_{\rm Gauss, m}]$ on the number of Gaussian functions. (b)(d)(f) The dependence of the $\widehat{\sigma}[{\bm d}_{\rm Gauss, m}]$ on the number of Gaussian functions. The five-beads (a)(b), the 10-beads (c)(d) and the 15-beads (e)(f) models are used. Here, the cases with a larger number of Gaussian functions are investigated for the deterministic method, since the number of parameters to optimize is small compared to that of $\widehat{\sigma}[{\bm d}_{\rm Gauss, m}]$ for each Gaussian function.
The mean and its standard deviation of ten independent trials are plotted. The system parameters are set as $k=\gamma=1$ and $T_h = 250$. The sampling interval of the trajectories is set as $\Delta t = 10^{-3}$, and thus the number of data points is $10^3\tau_{\rm obs}$, half of which is used for the training, and the other half for the estimation. $\alpha = 1$ is used for the gradient ascent.}}
\label{fig: move}
\end{center}
\end{figure*}
In this appendix, we address two remaining questions: (i) the scalability of the learning estimators for higher dimensional data, and (ii) how the representation ability of the model function affects the performance. We first explain the setup for numerical experiments, where we consider an extension of the Gaussian learning estimator for high dimensional case. 
Then, we compare the following two methods using the $N$-beads model ($N\geq5$): (1) optimize only the coefficients of the linear combination of Gaussian functions by the deterministic optimization \cite{Tan2020}, and (2) optimize both the coefficients of the linear combination and the parameters of Gaussian functions by the gradient ascent. 
Here, we aim to answer the above-mentioned questions, and at the same time to show an example where our learning estimators and the method that uses similar techniques \cite{Tan2020} show a difference in performance.
We show that the second method (2) indeed shows better performance in terms of the convergence speed, while the first approach is faster in computation time.\\ \indent
We first consider an extension of the Gaussian learning estimator. The Gaussian learning estimator introduced in Appendix.~\ref{sec: details} is not applicable to high dimensional data as it is, since the computational complexity is $O(NN_{\rm Gauss})$, where $N_{\rm Gauss}$ is the number of Gaussian functions $N_{\rm Gauss} = N_{\rm bin}^{n_{\rm dim}}$ and it increases exponentially as the dimension $n_{\rm dim}$ increases (see Supplemental Material for the details). In order to suppress the number of Gaussian functions, we consider the positions of Gaussian functions as variables. Concretely, we define the model function ${\bm d}_{\rm Gauss, m}({\bm x})$ as
\begin{eqnarray}
&&\{ {\bm d}_{{\rm Gauss, m}}({\bm x}) \}_k:= \sum_{i=1}^{N_{\rm Gauss}}\omega_k (i) K_k ({\bm x};i),\nonumber\\
&&K_k ({\bm x};i) := e^{ - \left({\bm x}-\bar{{\bm x}}^{(k)}(i)\right)^\mathsf{T} {{\bm M^{(k)}(i)}^{-1}}\left({\bm x}-\bar{{\bm x}}^{(k)}(i) \right) },
\end{eqnarray}
where ${\bm M}^{(k)}(i)_{lm} = \delta_{lm}\left(m_l^{(k)}(i)\right)^2$.\\ \indent
Here, we introduce two estimators $\widehat{\sigma}[{\bm d}_{\rm Gauss, m}]$ and $\widehat{\sigma}^{\rm det}[{\bm d}_{\rm Gauss, m}]$ using the model function ${\bm d}_{\rm Gauss, m}({\bm x})$ . In $\widehat{\sigma}[{\bm d}_{\rm Gauss, m}]$, we optimize ${\bm w}(i), \bar{{\bm x}}^{(k)}(i)$ and $m_l^{(k)}(i)$ by the gradient ascent. Here, the variables are initialized by
\begin{eqnarray}
w_k(i) &=& {\rm uni}(-1, 1),\\
\bar{x}_l^{(k)}(i) &=& {\rm uni}(x_{{\rm min}, l}, x_{{\rm max}, l}),\\
m_l^{(k)}(i) &=& x_{{\rm max}, l} - x_{{\rm min}, l},
\end{eqnarray}
where ${\rm uni}(a, b)$ is a random variable that follows the uniform distribution in the range $a< x< b$, and $x_{{\rm min}, l}$ and $x_{{\rm max}, l}$ are the minimum and the maximum of the $l$th element of all the data points.
On the other hand, in $\widehat{\sigma}^{\rm det}[{\bm d}_{\rm Gauss, m}]$, we optimize only $w_k(i)$ by the deterministic optimization method proposed in Ref.~\cite{Tan2020} with the other variables fixed by the initial values.
The deterministic optimization method is expected to compute faster since it is not necessary to conduct the gradient ascent, while the model functions are restricted to those which can be described by a linear combination of fixed basis functions similarly to Ref.~\cite{Frishman2018}.
\\ \indent
In Fig.~\ref{fig: move}, we compare these two estimators using the $N$-beads model ($N = 5, 10, 15$) whose equations are defined in the same manner as the two-beads and the five-beads models. The system parameters are set as: $\Delta t = 10^{-3}, k=\gamma=1$ and $T_h = 250$. Since we find that the performance of $\widehat{\sigma}[{\bm d}_{\rm Gauss, m}]$ is almost independent of the step size $\alpha$ of the gradient ascent when $\alpha$ is sufficiently small, $\alpha$ is fixed to $1$ for all the setups.
We use the data splitting scheme both for $\widehat{\sigma}[{\bm d}_{\rm Gauss, m}]$ and $\widehat{\sigma}^{\rm det}[{\bm d}_{\rm Gauss, m}]$.\\ \indent
The results show that $\widehat{\sigma}[{\bm d}_{\rm Gauss, m}]$ is better in terms of the convergence, and it also performs equally well for various choice of $N_{\rm Gauss}$. Surprisingly, $N_{\rm Gauss} = 4$ is enough for $\widehat{\sigma}[{\bm d}_{\rm Gauss, m}]$ in all the examples, which reflects the high representation ability of the model function. Therefore, we answer to the questions at the beginning of this section in the affirmative: (i) the learning estimator is scalable to higher dimensional data if we choose the model function properly, and (ii) the representation ability of the model function indeed makes a difference in the performance.\\ \indent
Finally, we remark on the computation time of $\widehat{\sigma}[{\bm d}_{\rm Gauss, m}]$ and $\widehat{\sigma}^{\rm det}[{\bm d}_{\rm Gauss, m}]$. Although the computational complexities of these estimators are $O(NN_{\rm Gauss})$ and $O(\max(NN_{\rm Gauss},  N_{\rm Gauss}^3))$ respectively, and thus similar, $\widehat{\sigma}^{\rm det}[{\bm d}_{\rm Gauss, m}]$ usually computes faster in constant factor when $N_{\rm Gauss}$ is small, since it does not require the iteration of the gradient ascent (see Supplemental Material for the details).
}


\widetext
\newpage
\begin{center}
\textbf{\large Supplemental Material}
\end{center}

\setcounter{equation}{0}
\setcounter{figure}{0}
\setcounter{table}{0}
\setcounter{page}{1}
\makeatletter
\renewcommand{\theequation}{S\arabic{equation}}
\renewcommand{\thefigure}{S\arabic{figure}}
\renewcommand{\bibnumfmt}[1]{[S#1]}

{\color{black}
In this Supplemental Material, we show supplementary numerical results on the hyperparameter tuning and the computation time of the learning estimators. In the first part, we discuss the hyperparameter dependence of the learning estimators. Then, we show the results of hyperparameter tuning in each setup. In the second part, we discuss the computational complexities of the estimators used in this study, and compare their computation time.\\

\noindent
\textbf{1. Hyperparameter tuning}\\ \indent

First, we discuss the hyperparameter dependence of the learning estimators. In Fig.~\ref{fig: gauss_hyperparameter}, we show the hyperparameter ($N_{\rm bin}$ and $\alpha$) dependence of the Gaussian learning estimator.
Figure~S\ref{fig: gauss10man} and (b) show the $N_{\rm bin}$ and $\alpha$ dependence, and we find that the $\alpha$ dependence is more significant than $N_{\rm bin}$. In order to understand the reason, we plot the $\alpha$ dependence of the peak of the learning curve of $\widehat{\sigma}[{\bm d}_{\rm Gauss}]|_{\rm test}$ and $\widehat{\sigma}[{\bm d}_{\rm Gauss}]|_{\rm train}$ in Fig.~S\ref{fig: gauss_alpha_dependence} and (d). We conclude that ${\bm d}_{\rm Gauss}$ becomes overfitted to the training data at small $\alpha$ because the gradient ascent can find the maximum of $\widehat{\sigma}[{\bm d}_{\rm Gauss}]|_{\rm train}$ more accurately, while both of $\widehat{\sigma}[{\bm d}_{\rm Gauss}]|_{\rm test}$ and $\widehat{\sigma}[{\bm d}_{\rm Gauss}]|_{\rm train}$ become small at large $\alpha$ because the gradient ascent does not work well due to the large step size. On the basis of these results, we first tune $N_{\rm bin}$ with fixed $\alpha$ and sufficiently large $\tau_{\rm obs}$ (here, $\alpha$ should be roughly tuned beforehand), then tune $\alpha$ with the tuned $N_{\rm bin}$ for each $\tau_{\rm obs}$ in this study.\\ \indent
In Fig.~\ref{fig: binned_hyperparameter}, we show the hyperparameter ($N_{\rm bin}$, $\alpha$ and $\lambda$) dependence of the binned learning estimator. 
We show the $\alpha$ and $\lambda$ dependence in Fig.~S\ref{fig: bin10_10man} to (d), and write the top five values in the corresponding squares. On the contrary to the Gaussian learning estimator, the $\alpha$ dependence is subtle at $\lambda = 0$, while the peak values distribute along the line of constant $\alpha\lambda$. This can be explained by the fact that the regularization term appears in the gradient ascent with the coefficient $\alpha\lambda$. Therefore, we can fix $\alpha$ in this estimator, and tune the other hyperparameters $N_{\rm bin}$ and $\lambda$ for each $\tau_{\rm obs}$ in this study. Concretely, we first tune $N_{\rm bin}$ with $\lambda = 0$ as in Fig.~S\ref{fig: binN_dependence}, and then tune $\lambda$ with the tuned $N_{\rm bin}$ as in Fig.~S\ref{fig: lambda_dependence}.\\ \indent
We show the results of hyperparameter tuning for the following setups: (i) the two-beads model (Fig.~\ref{fig: hyperparam1}), (ii) the five-beads model (Fig.~\ref{fig: hyperparam2} and \ref{fig: hyperparam3}) and (iii) the Mexican-hat potential model (Fig.~\ref{fig: hyperparam4}, \ref{fig: hyperparam5} and \ref{fig: hyperparam6}).
On the basis of these results, we determine the values of the hyperparameters as summarized in TABLE.~\ref{table: hyperparameters} in the main text.
\\ \indent

\begin{figure}
\begin{center}
\begin{tabular}{cc}
	\subfigure[$\tau_{\rm obs} = 100$]{
		\includegraphics[width = 0.45\linewidth]{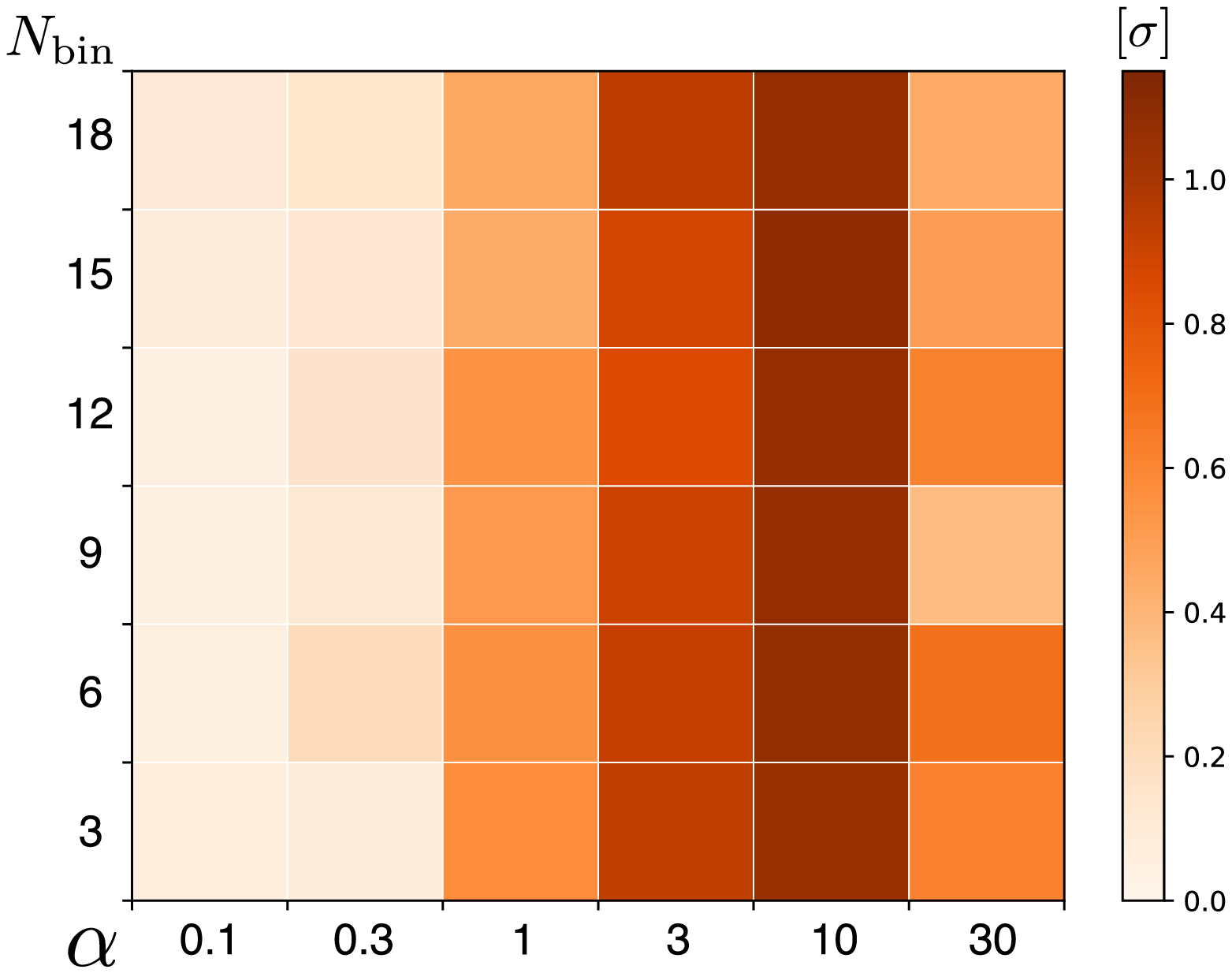}\label{fig: gauss10man}}&
	\subfigure[$\tau_{\rm obs} = 1000$]{
		\includegraphics[width = 0.45\linewidth]{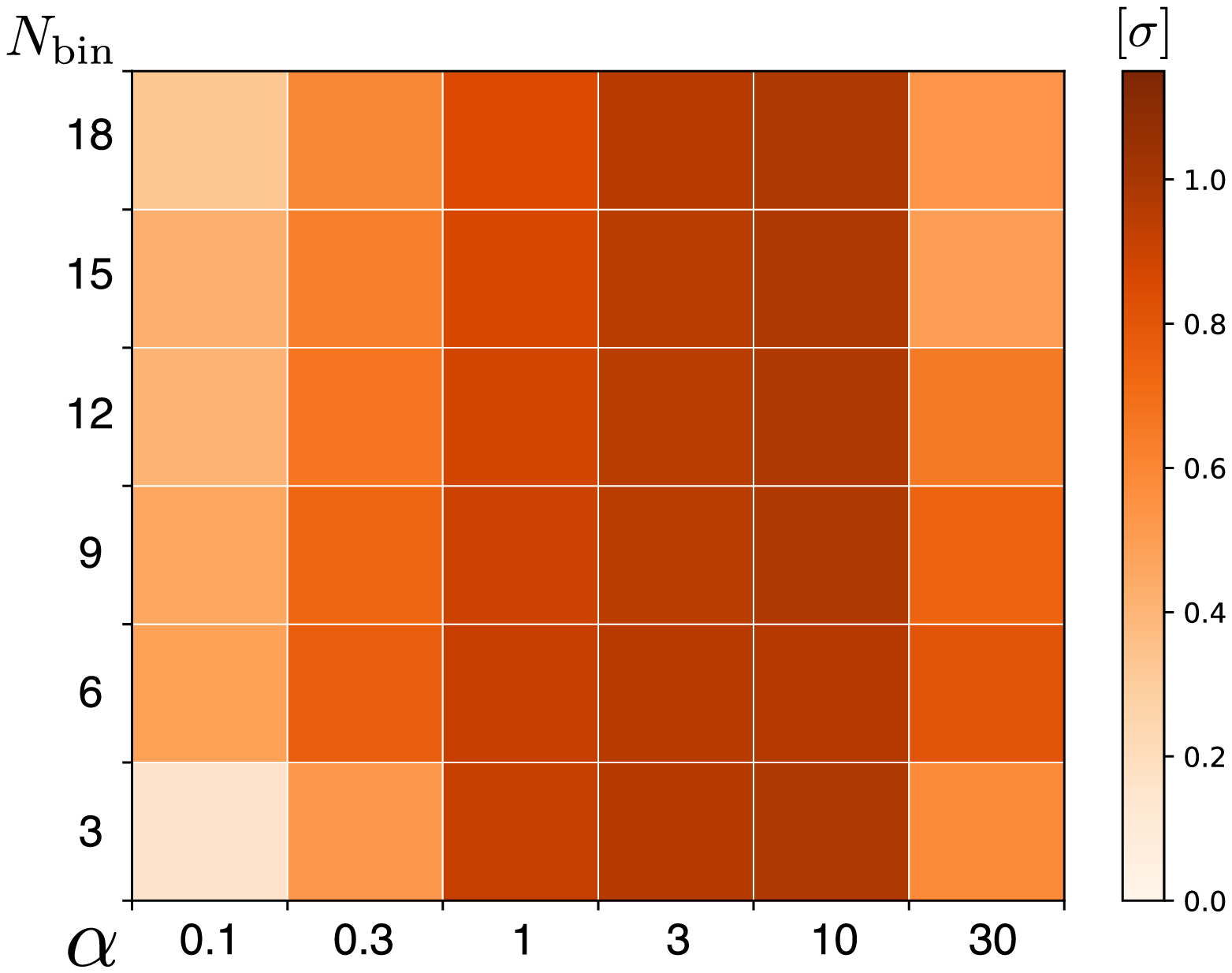}\label{fig: gauss100man}}\\
	\subfigure[$N_{\rm bin} = 6$]{
		\includegraphics[width = 0.45\linewidth]{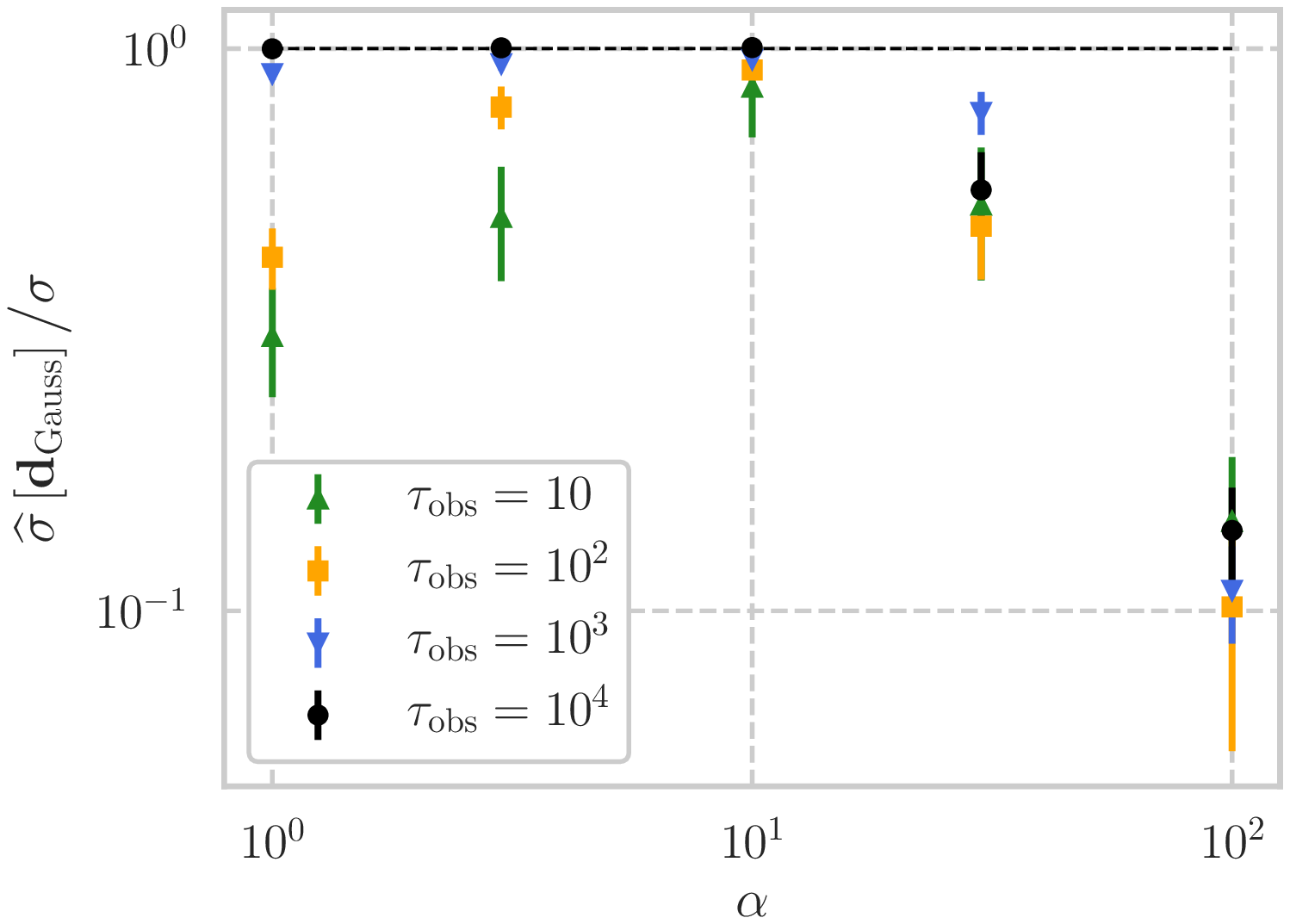}\label{fig: gauss_alpha_dependence}}&
	\subfigure[$N_{\rm bin} = 6$]{
		\includegraphics[width = 0.45\linewidth]{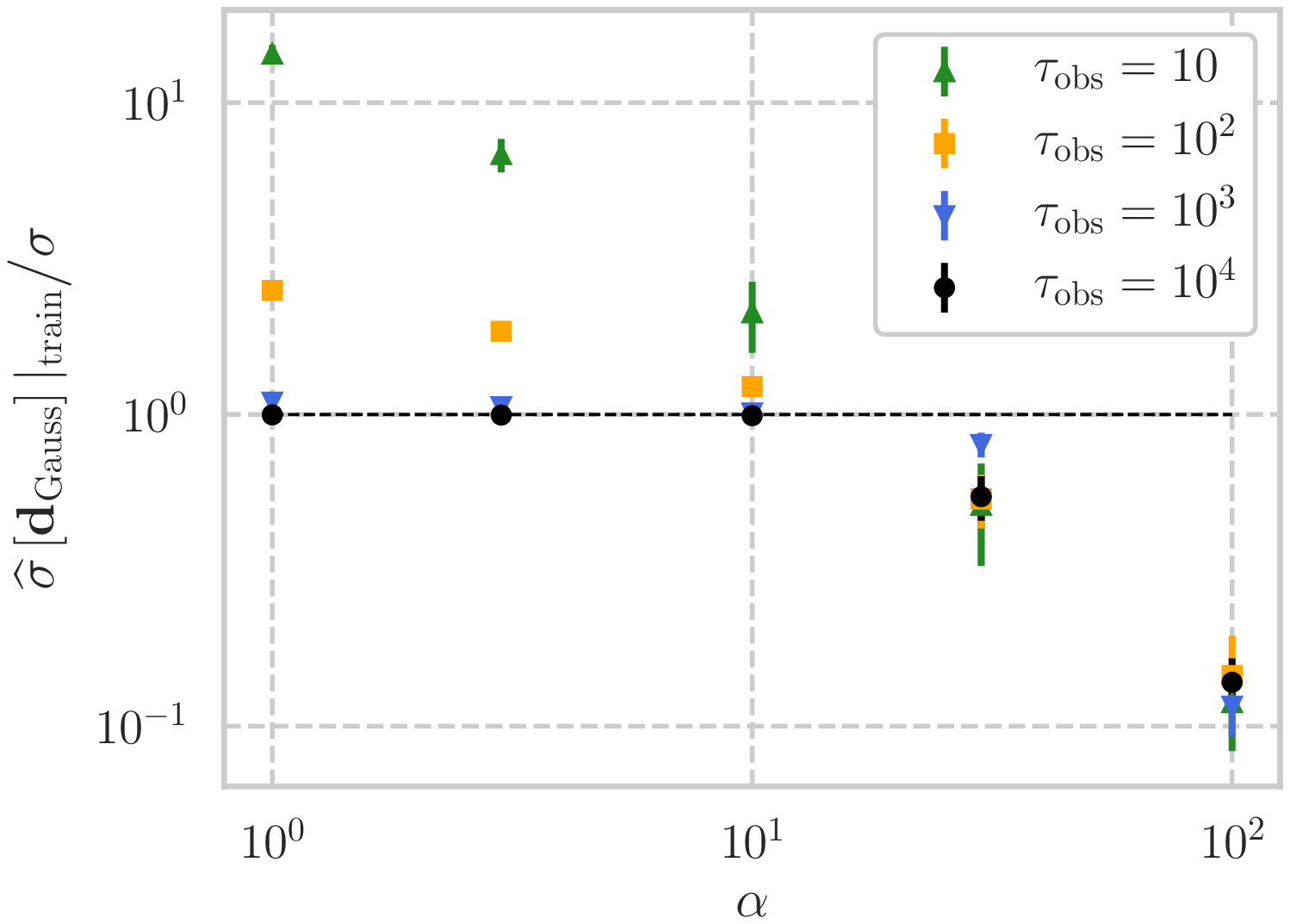}\label{fig: gauss_alpha_dependence_train}}
\end{tabular}
\caption{The hyperparameter dependence of the Gaussian learning estimator using data generated by the two-beads model ($r=0.1$, (a) $\tau_{\rm obs} = 10$, (b) $\tau_{\rm obs} = 100$). 
(a)(b) The $N_{\rm bin}$ and $\alpha$ dependence with the trajectory length $\tau_{\rm obs} = 10^2$ and $10^3$. (c) The $\alpha$ dependence for the trajectory length $\tau_{\rm obs}$ from $10$ to $10^4$. (d) The $\alpha$ dependence of the peak of the training curve for the trajectory length $\tau_{\rm obs}$ from 10 to $10^4$. 
In (c) and (d), the mean and its standard deviation of ten independent trials are plotted. The other system parameters are set to the same as those in Fig.~\ref{fig: 2beads}.}
\label{fig: gauss_hyperparameter}
\end{center}
\end{figure}

\begin{figure}
\begin{center}
\begin{tabular}{cc}
	\subfigure[$N_{\rm bin} = 10, \tau_{\rm obs} = 100$]{
		\includegraphics[width = 0.45\linewidth]{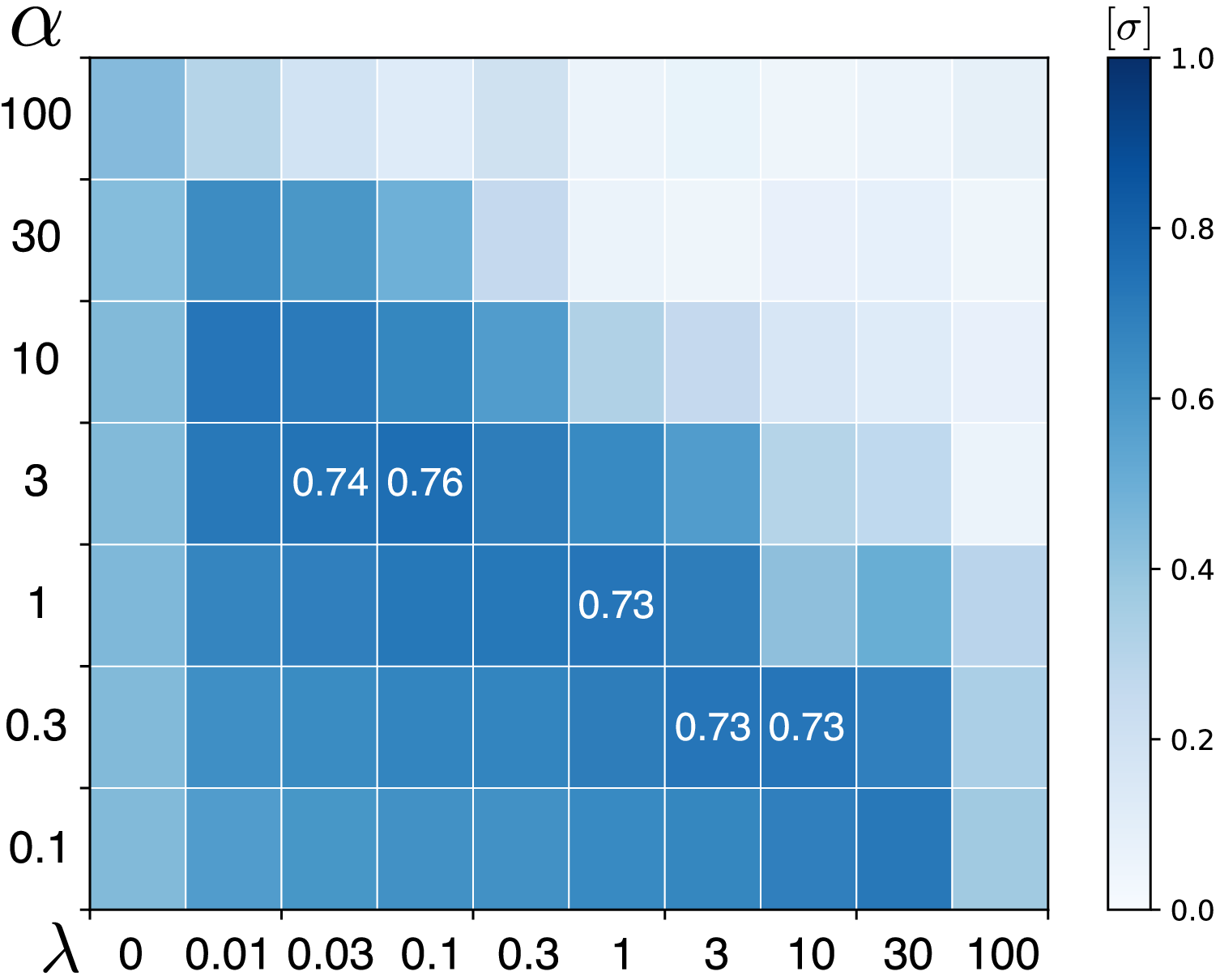}\label{fig: bin10_10man}}&
	\subfigure[$N_{\rm bin} = 20, \tau_{\rm obs} = 100$]{
		\includegraphics[width = 0.45\linewidth]{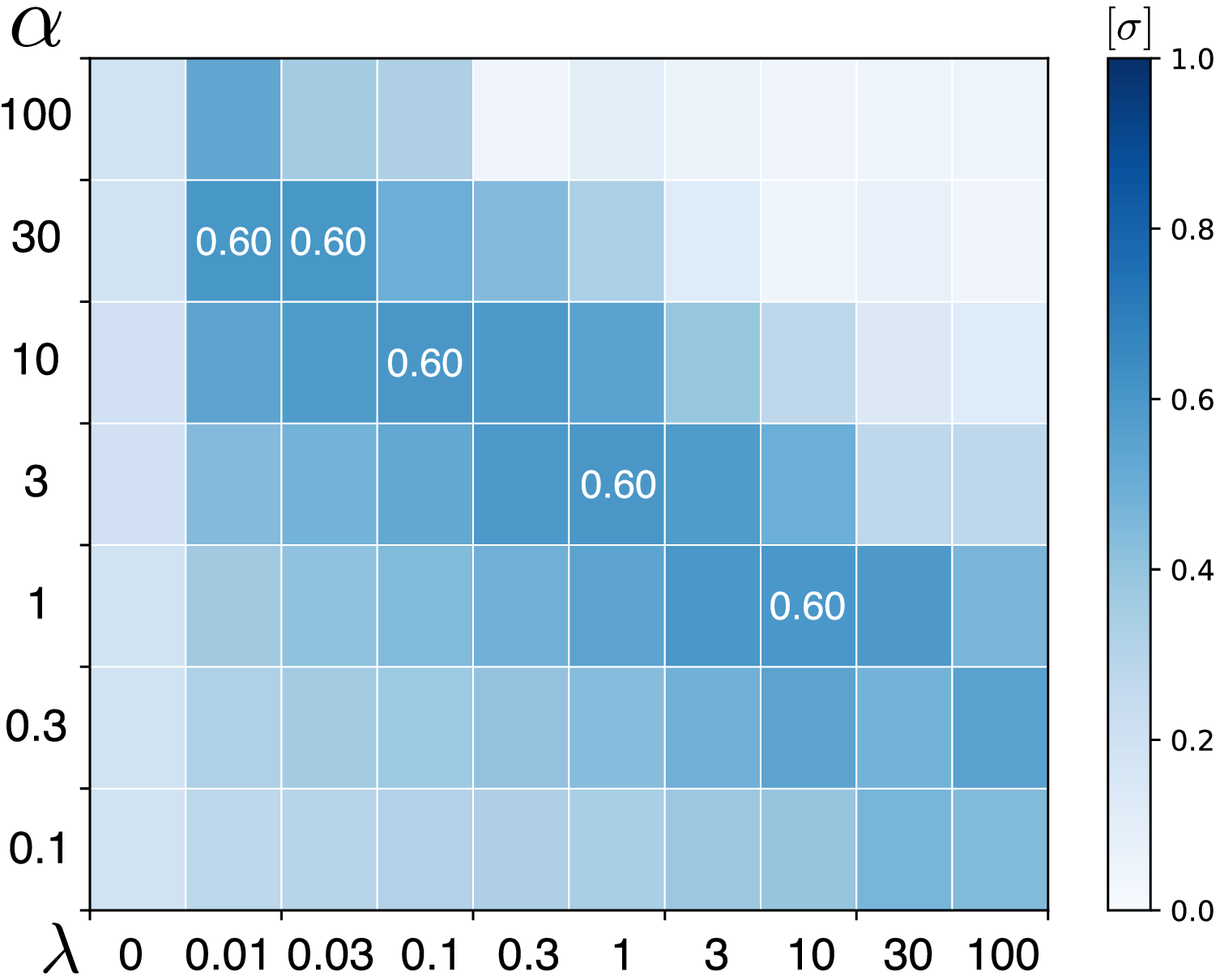}\label{fig: bin20_10man}}\\	
	\subfigure[$N_{\rm bin} = 10, \tau_{\rm obs} = 1000$]{
		\includegraphics[width = 0.45\linewidth]{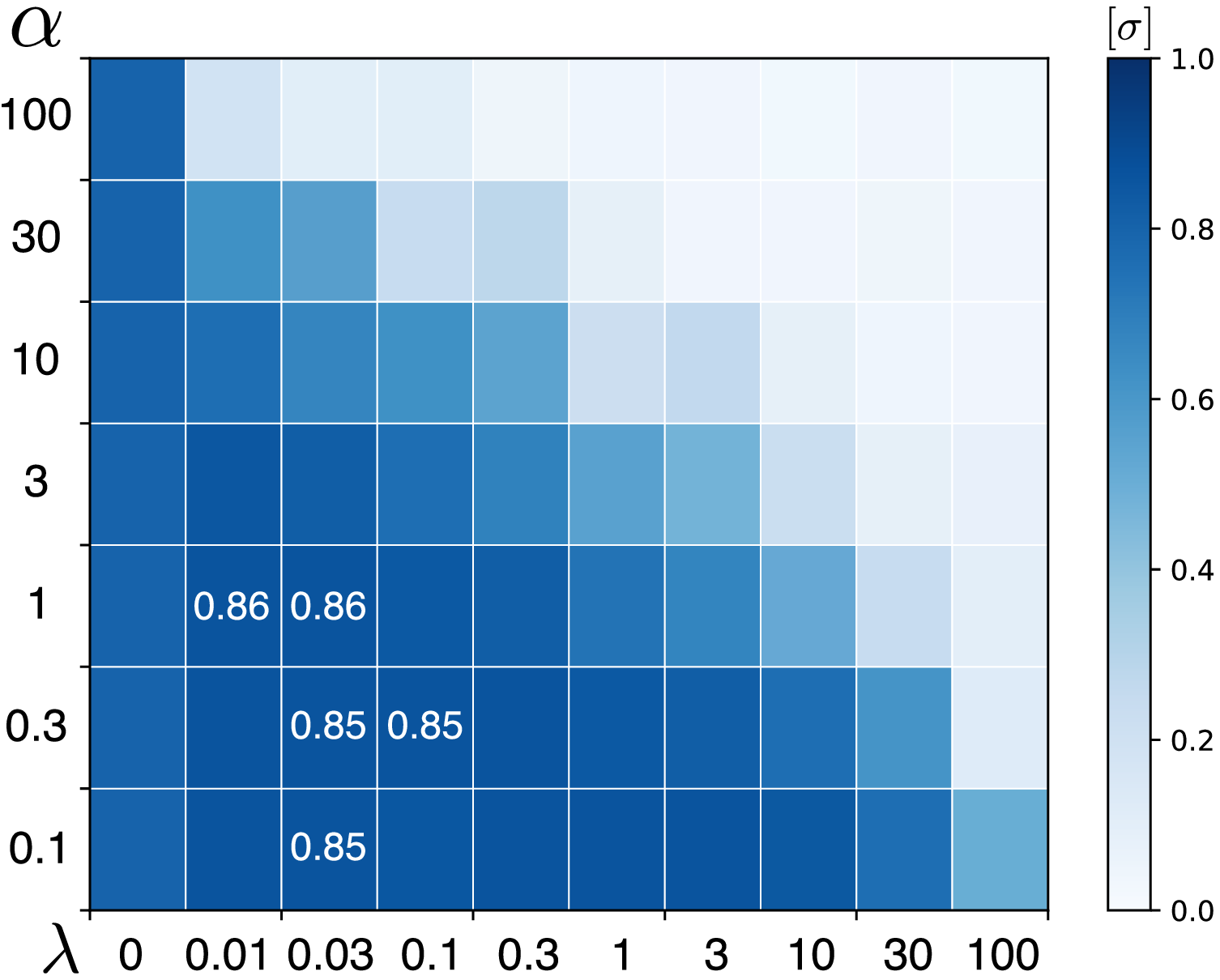}\label{fig: bin10_100man}}&
	\subfigure[$N_{\rm bin} = 20, \tau_{\rm obs} = 1000$]{
		\includegraphics[width = 0.45\linewidth]{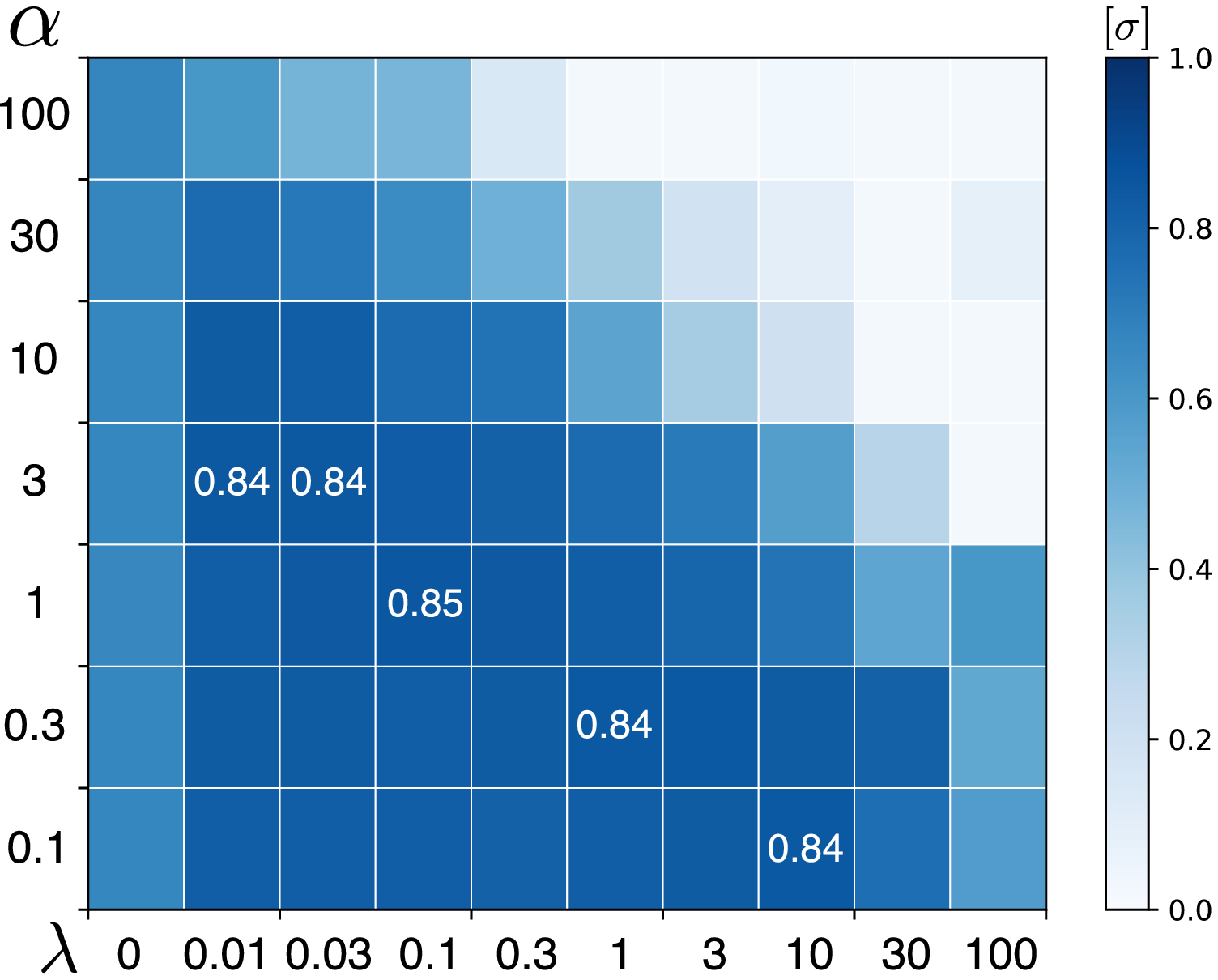}\label{fig: bin20_100man}}\\
	\hspace{-0.085\linewidth}
	\subfigure[$\alpha = 1, \lambda = 0$]{
		\includegraphics[width = 0.45\linewidth]{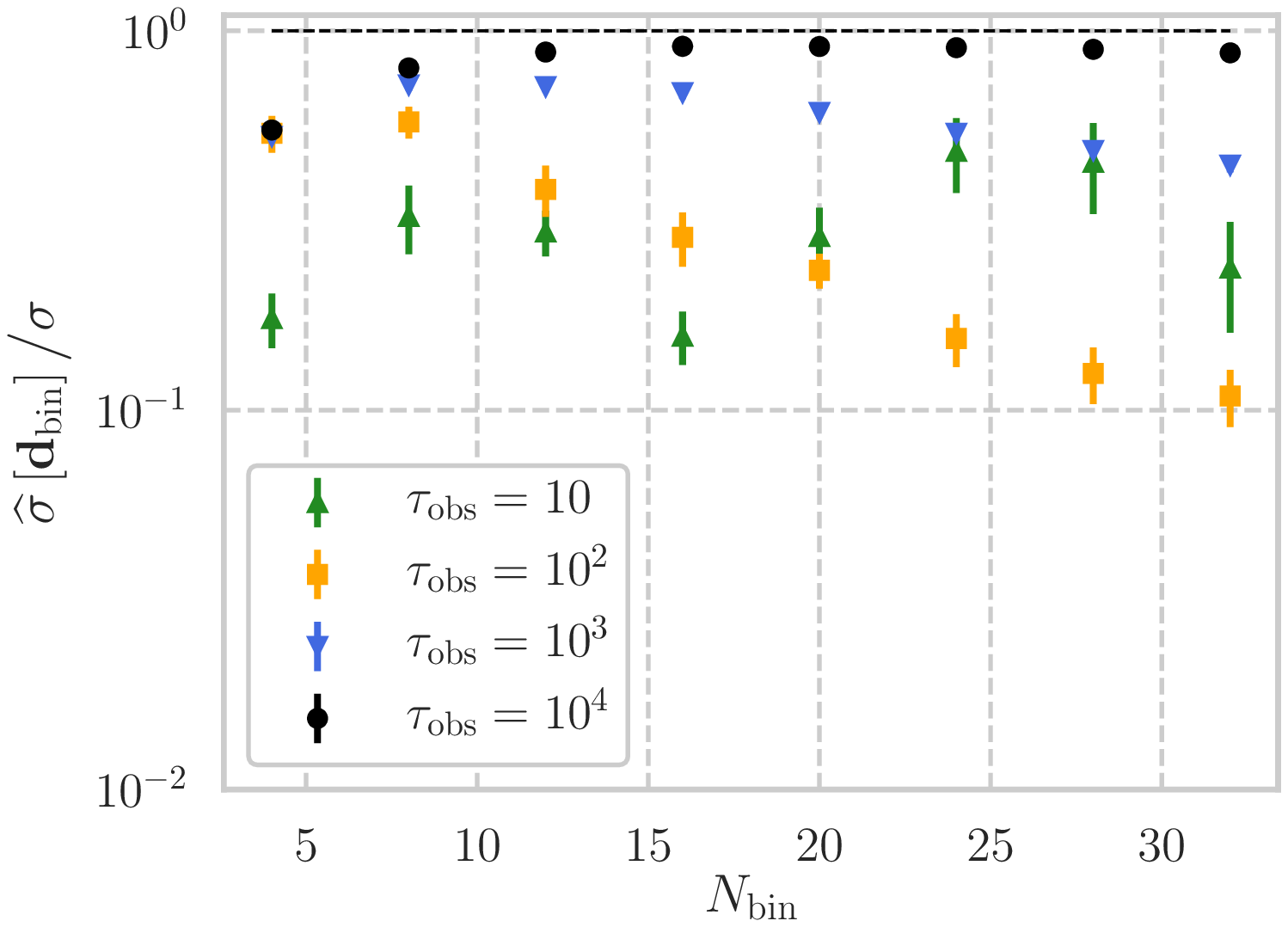}\label{fig: binN_dependence}}&
	\hspace{-0.05\linewidth}
	\subfigure[$\alpha = 1$, $N_{\rm bin} = 8, 8, 12, 20$ (for $\tau_{\rm obs} = 10$ to $10^4$)]{
		\includegraphics[width = 0.45\linewidth]{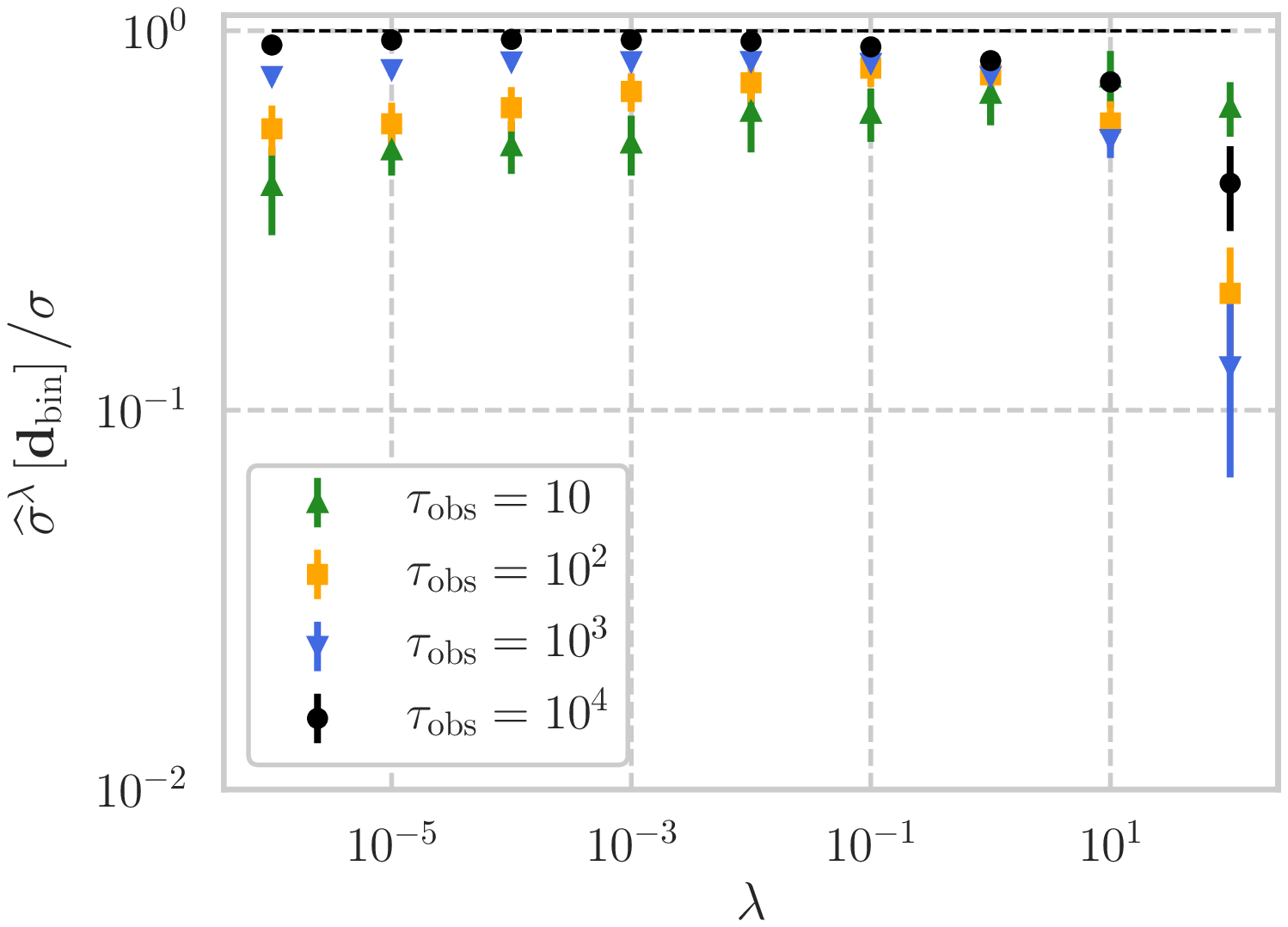}\label{fig: lambda_dependence}}
\end{tabular}
\caption{The hyperparameter dependence of the binned learning estimator using data generated by the two-beads model ($r=0.1$, (a)(b) $\tau_{\rm obs} = 100$, (c)(d) $\tau_{\rm obs} = 1000$). We show the $\alpha$, $\lambda$ dependence in (a)-(d), the $N_{\rm bin}$ dependence in (e) and the $\lambda$ dependence in (f) by fixing the other parameters as described in the subcaption. In (e) and (f), the mean and its standard deviation of ten independent trials are plotted. We show five values from the largest in (a)-(d). The other system parameters are set to the same as those in Fig.~\ref{fig: 2beads}.}
\label{fig: binned_hyperparameter}
\end{center}
\end{figure}

\begin{figure}
\begin{center}
\begin{tabular}{cc}
	\subfigure[$\alpha = 1$]{
		\includegraphics[width = 0.45\linewidth]{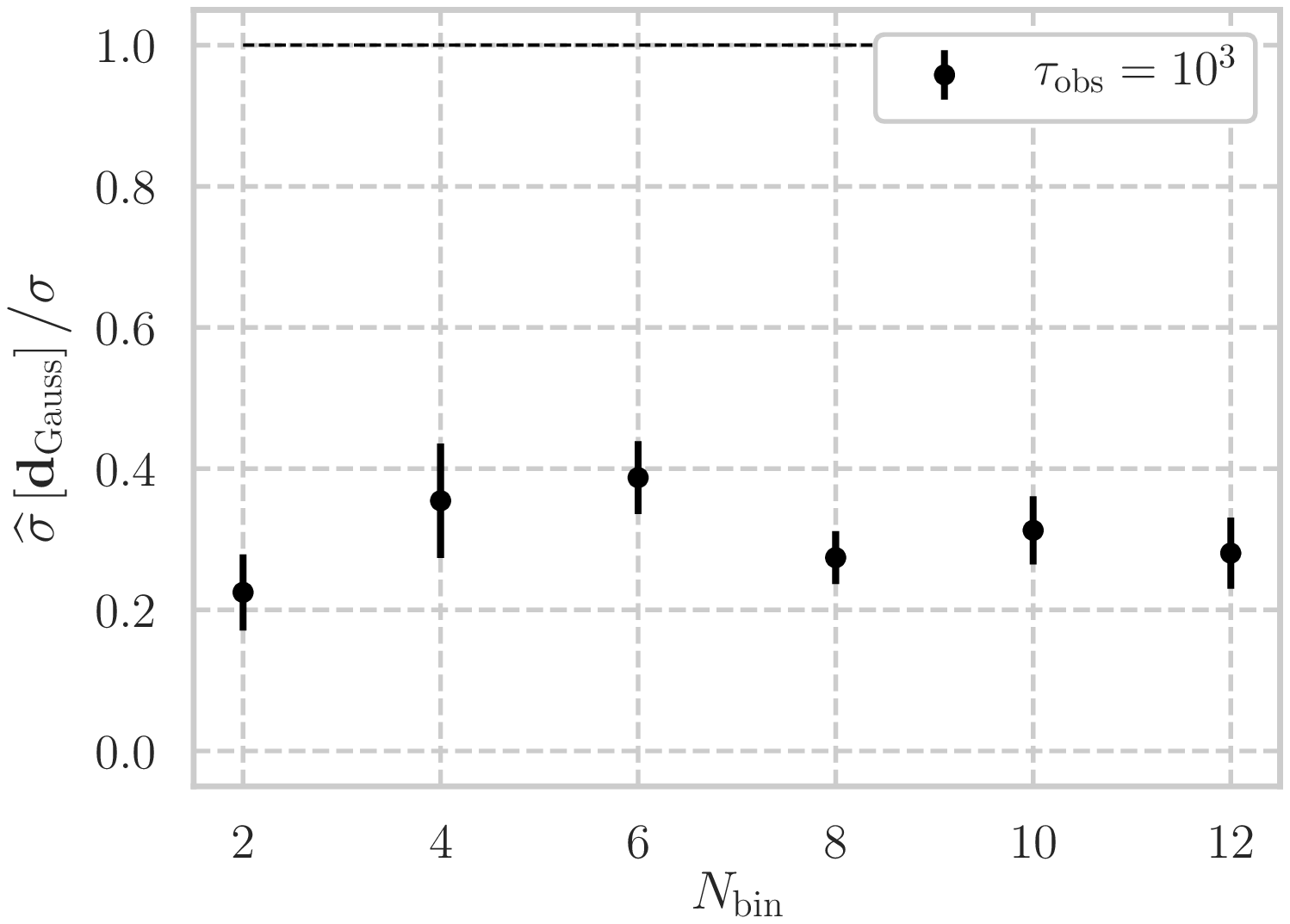}\label{fig: binN_dependence_2beads_r05}}&
	\subfigure[$N_{\rm bin} = 6$]{
		\includegraphics[width = 0.45\linewidth]{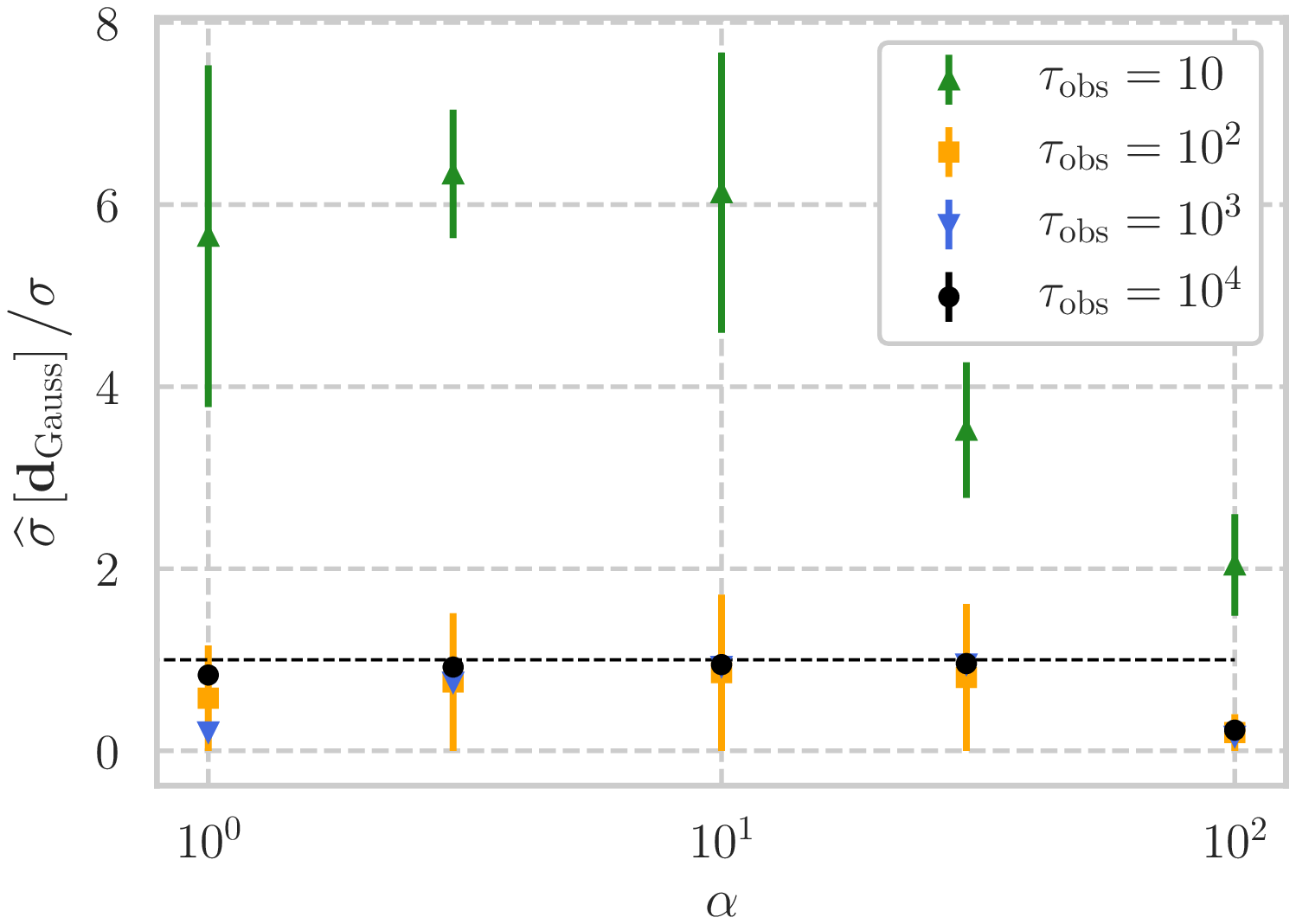}\label{fig: alpha_dependence_2beads_r05}}	
\end{tabular}
\caption{Hyperparameter tuning of the Gaussian learning estimator for the two-beads model ($r = 0.5$). (a) The $N_{\rm bin}$ dependence with the trajectory length $\tau_{\rm obs} = 10^3$. (b) The $\alpha$ dependence for the trajectory length $\tau_{\rm obs}$ from $10$ to $10^4$. }
\label{fig: hyperparam1}
\end{center}
\end{figure}

\begin{figure}
\begin{center}
\begin{tabular}{cc}
	\subfigure[$\alpha = 1, \lambda = 0$]{
		\includegraphics[width = 0.45\linewidth]{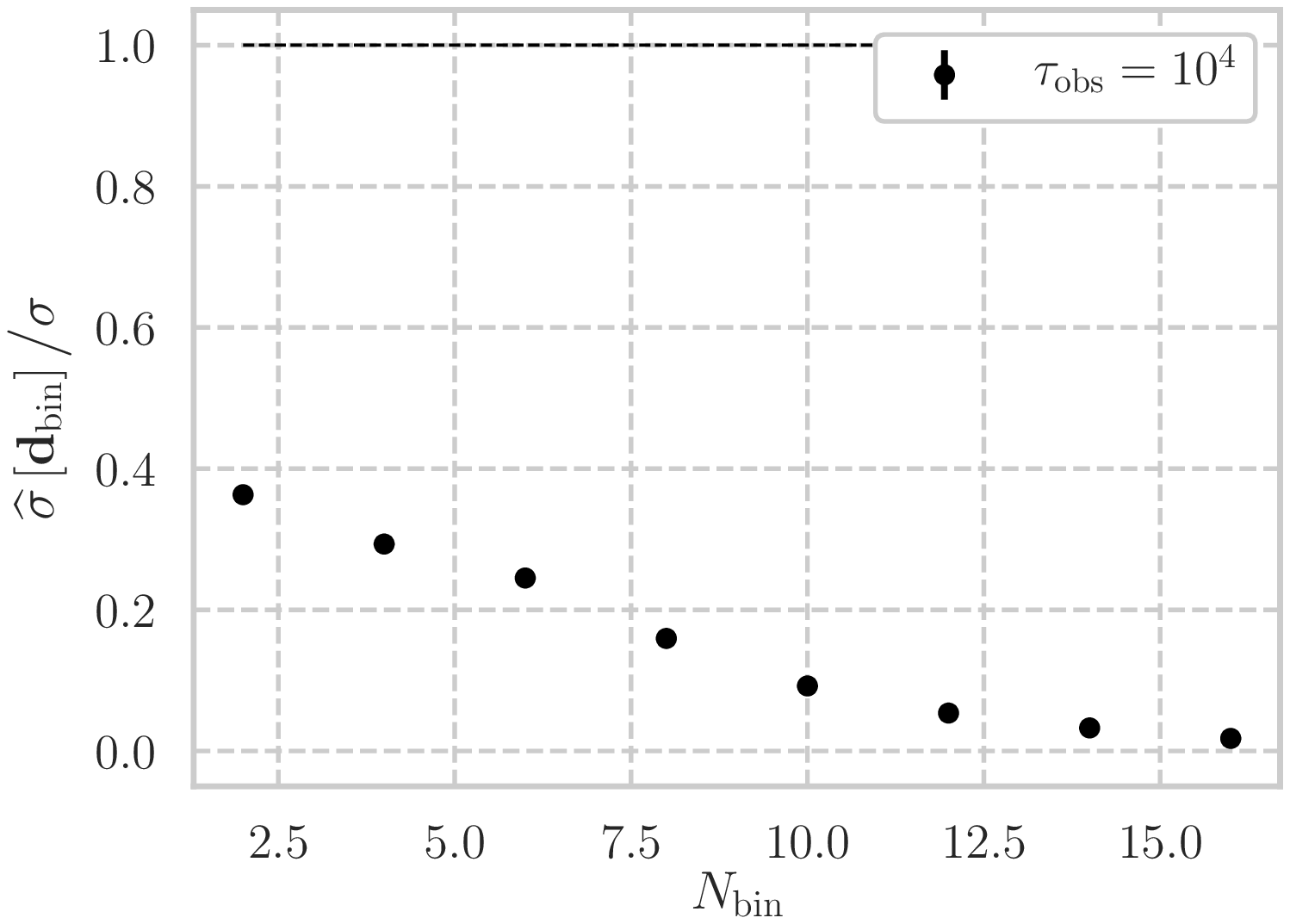}\label{fig: binN_dependence_5beads_r01}}&
	\subfigure[$N_{\rm bin} = 2, \alpha = 1$]{
		\includegraphics[width = 0.45\linewidth]{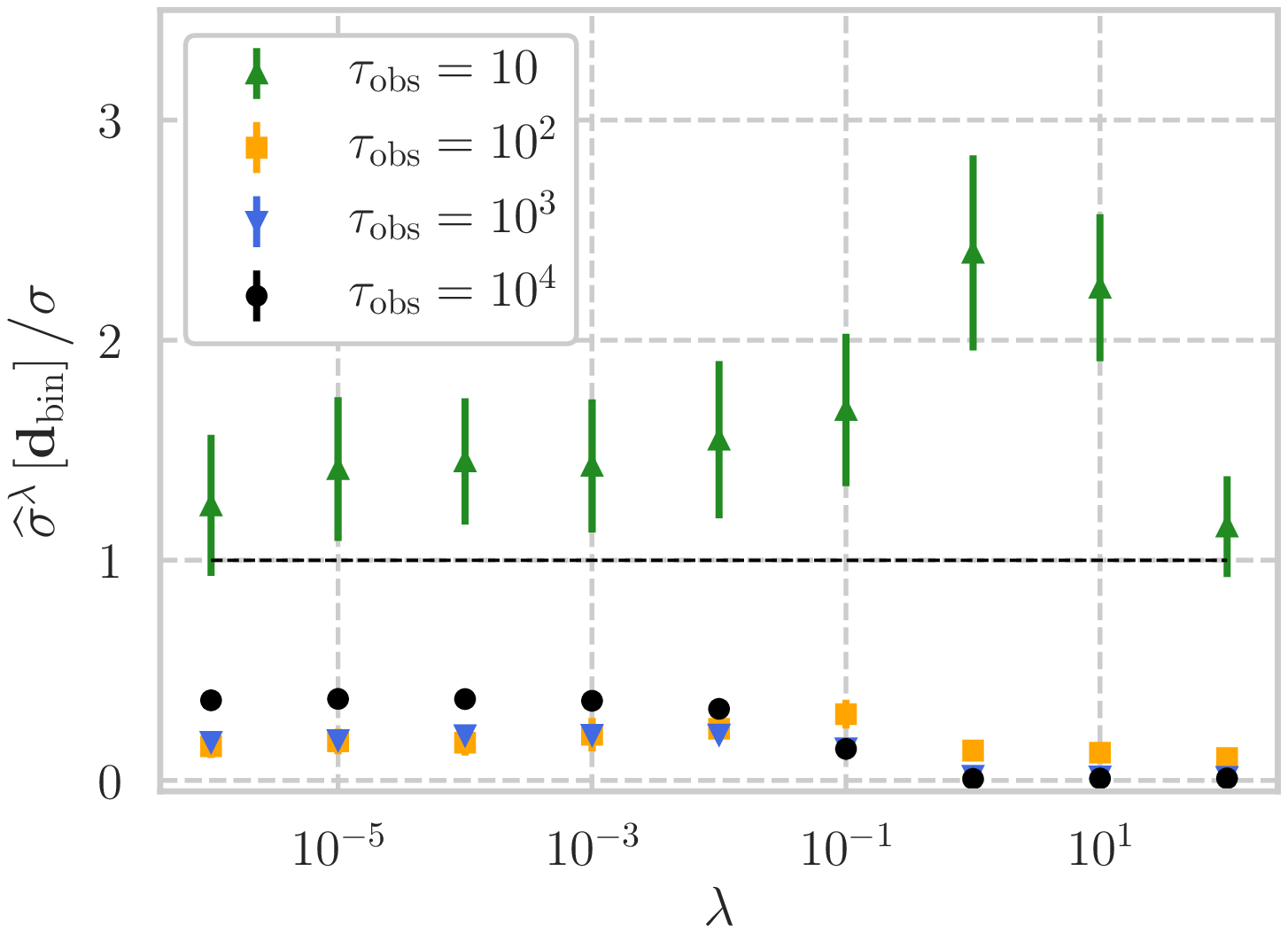}\label{fig: lambda_dependence_5beads_r01}}	
\end{tabular}
\caption{Hyperparameter tuning of the binned learning estimator for the five-beads model ($r = 0.1$). (a) The $N_{\rm bin}$ dependence with the trajectory length $\tau_{\rm obs} = 10^4$. (b) The $\lambda$ dependence for the trajectory length $\tau_{\rm obs}$ from $10$ to $10^4$. }
\label{fig: hyperparam2}
\end{center}
\end{figure}

\begin{figure}
\begin{center}
\begin{tabular}{cc}
	\subfigure[$\alpha = 1, \lambda = 0$]{
		\includegraphics[width = 0.45\linewidth]{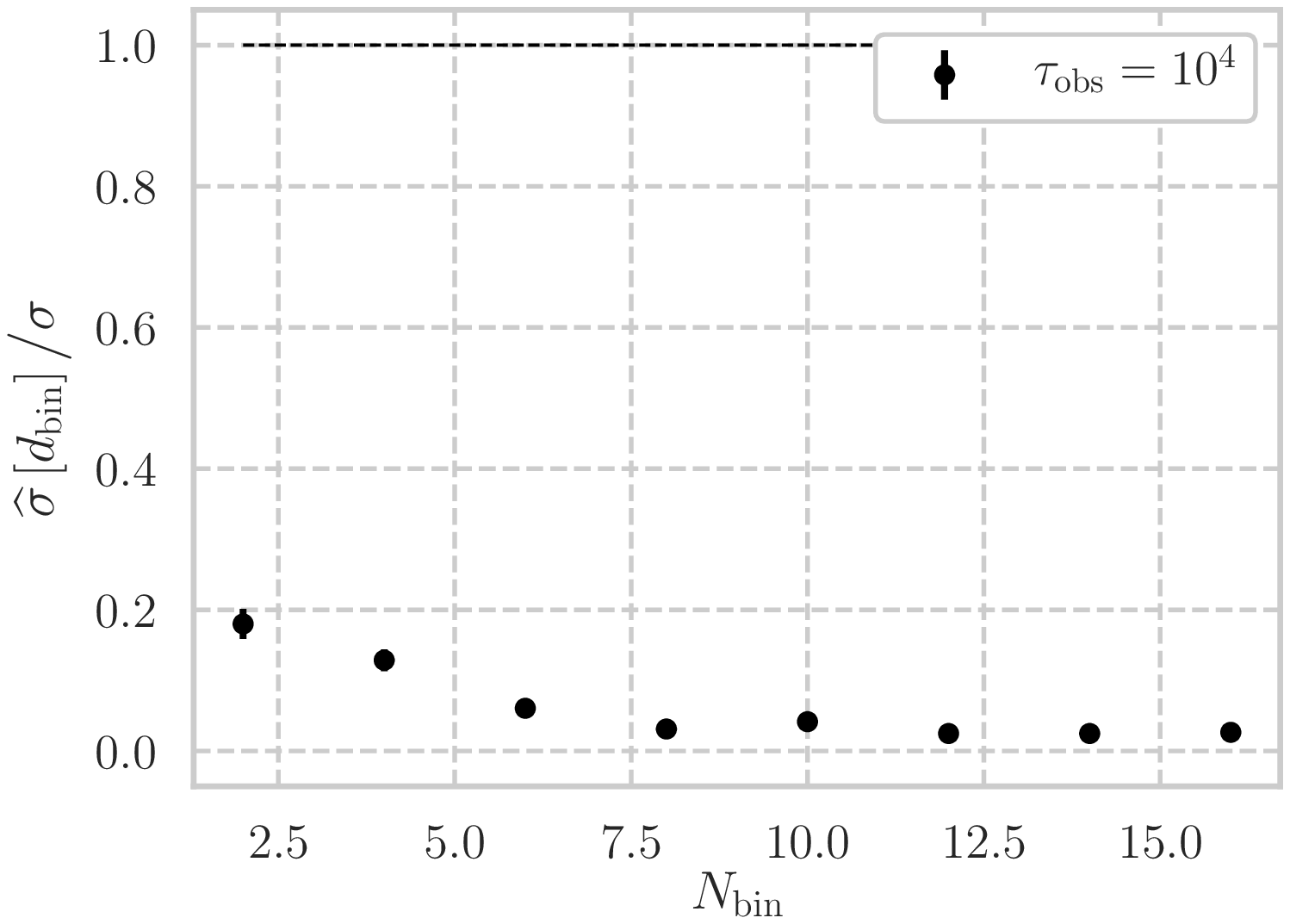}\label{fig: binN_dependence_5beads_r05}}&
	\subfigure[$N_{\rm bin} = 2, \alpha = 1$]{
		\includegraphics[width = 0.45\linewidth]{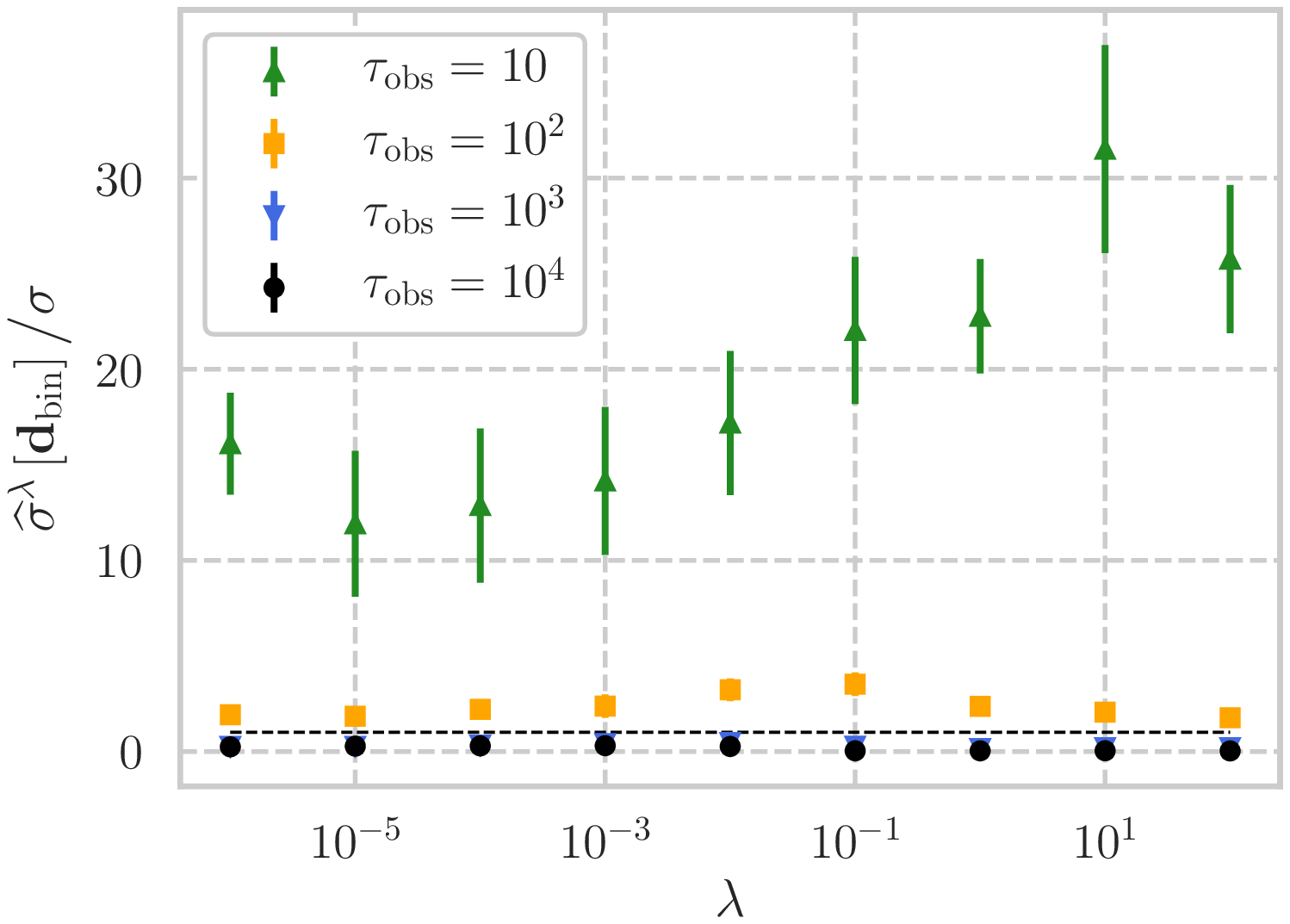}\label{fig: lambda_dependence_5beads_r05}}	
\end{tabular}
\caption{Hyperparameter tuning of the binned learning estimator for the five-beads model ($r = 0.5$). (a) The $N_{\rm bin}$ dependence with the trajectory length $\tau_{\rm obs} = 10^4$. (b) The $\lambda$ dependence for the trajectory length $\tau_{\rm obs}$ from $10$ to $10^4$. }
\label{fig: hyperparam3}
\end{center}
\end{figure}

\begin{figure}
\begin{center}
\begin{tabular}{cc}
	\subfigure[$\alpha = 10$]{
		\includegraphics[width = 0.45\linewidth]{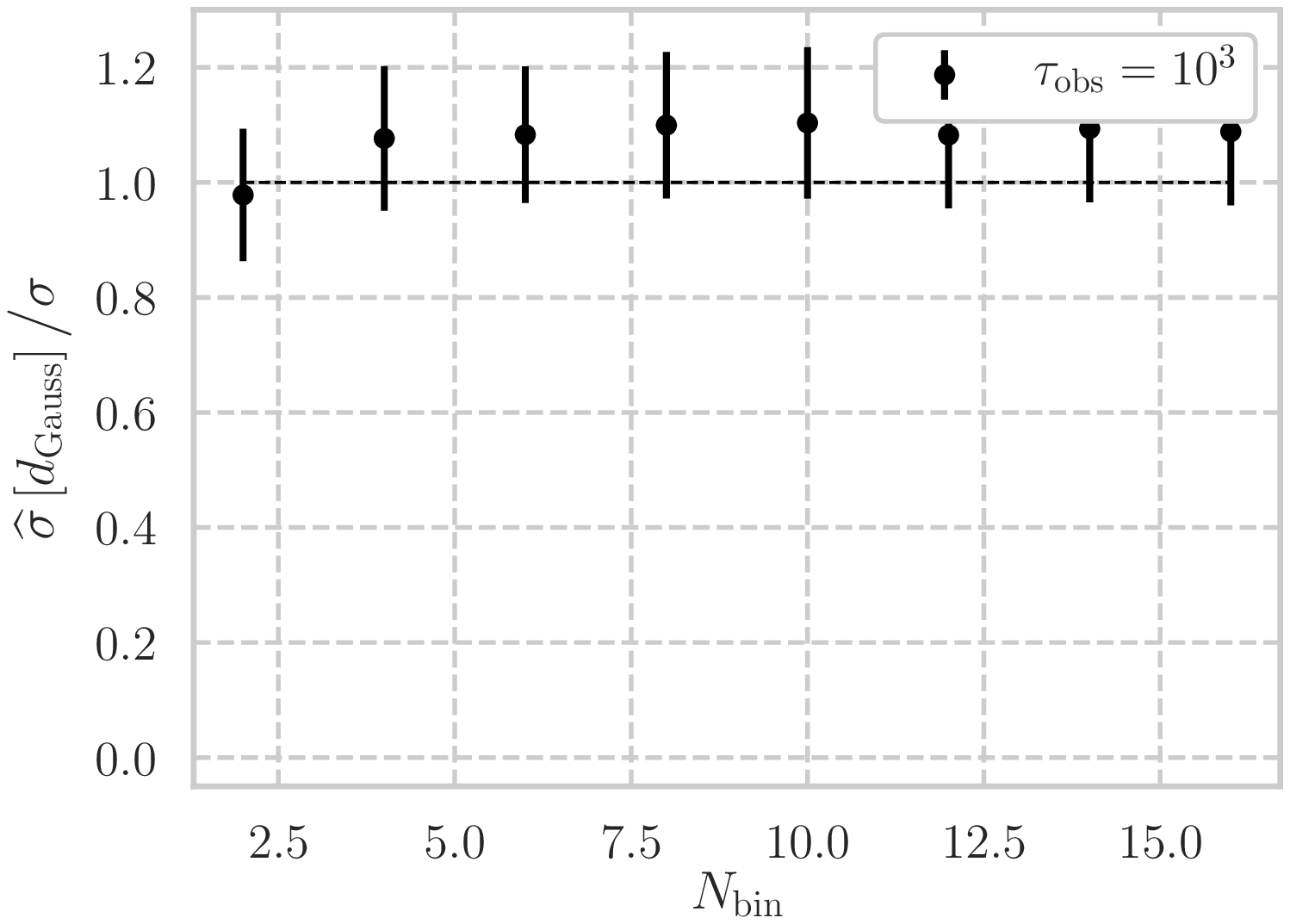}\label{fig: binN_dependence_mx_A0001}}&
	\subfigure[$N_{\rm bin} = 6$]{
		\includegraphics[width = 0.45\linewidth]{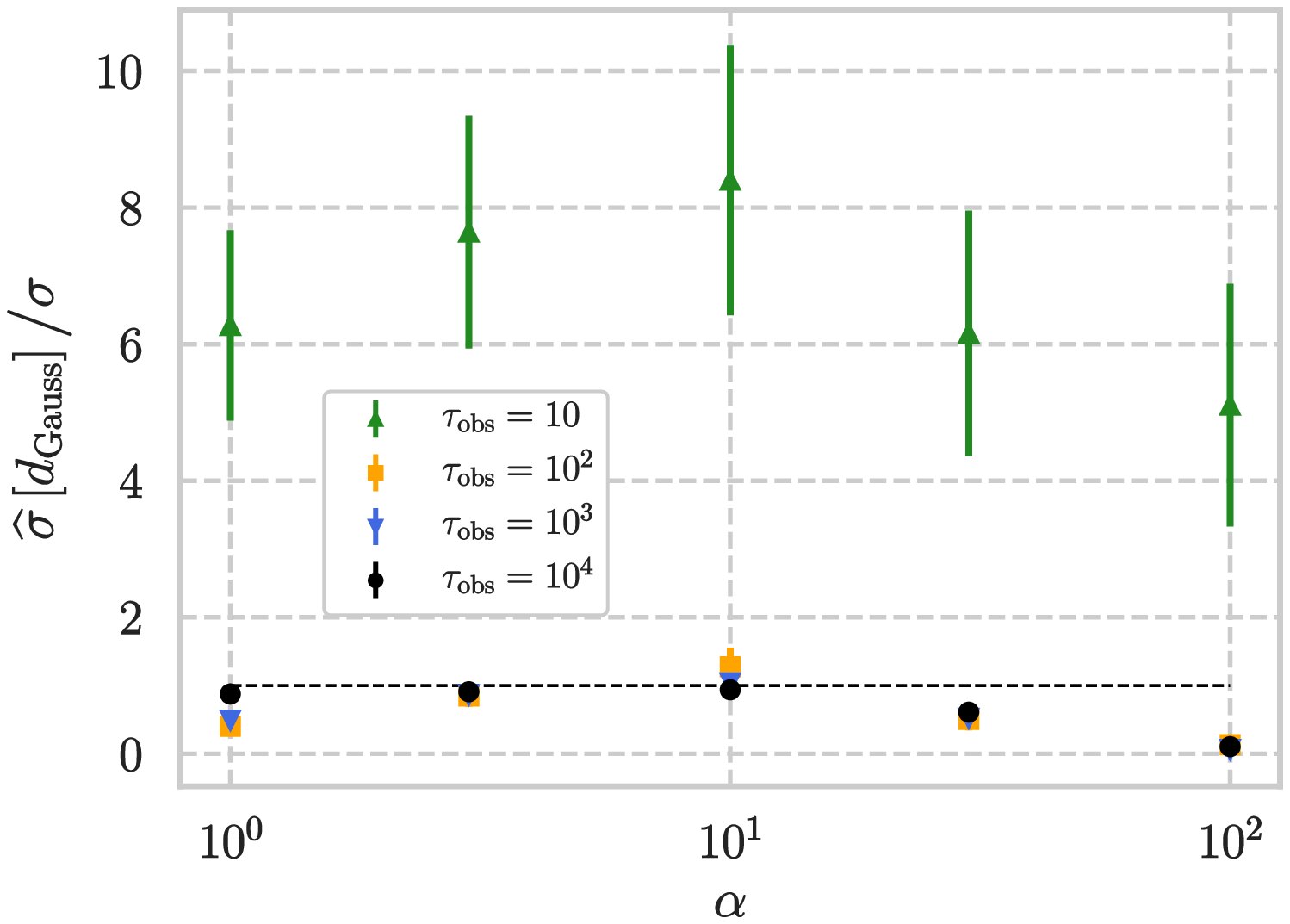}\label{fig: alpha_dependence_mx_A0001}}	
\end{tabular}
\caption{Hyperparameter tuning of the Gaussian learning estimator for the Mexican-hat potential model ($A= 10^{-4}$). (a) The $N_{\rm bin}$ dependence with the trajectory length $\tau_{\rm obs} = 10^3$. (b) The $\alpha$ dependence for the trajectory length $\tau_{\rm obs}$ from $10$ to $10^4$.}
\label{fig: hyperparam4}
\end{center}
\end{figure}

\begin{figure}
\begin{center}
\begin{tabular}{cc}
	\subfigure[$\alpha = 1$]{
		\includegraphics[width = 0.45\linewidth]{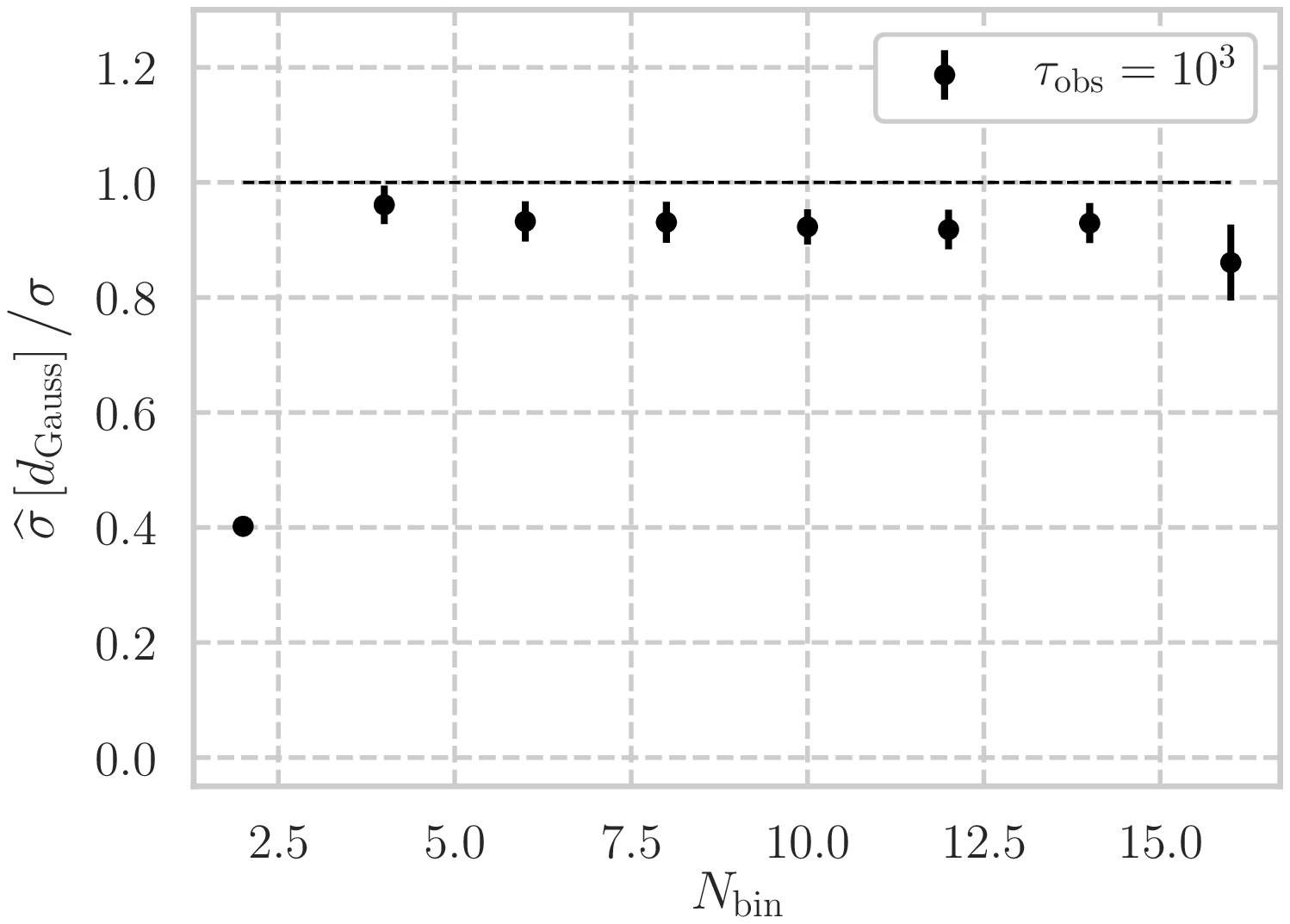}\label{fig: binN_dependence_mx_A1}}&
	\subfigure[$N_{\rm bin} = 6$]{
		\includegraphics[width = 0.45\linewidth]{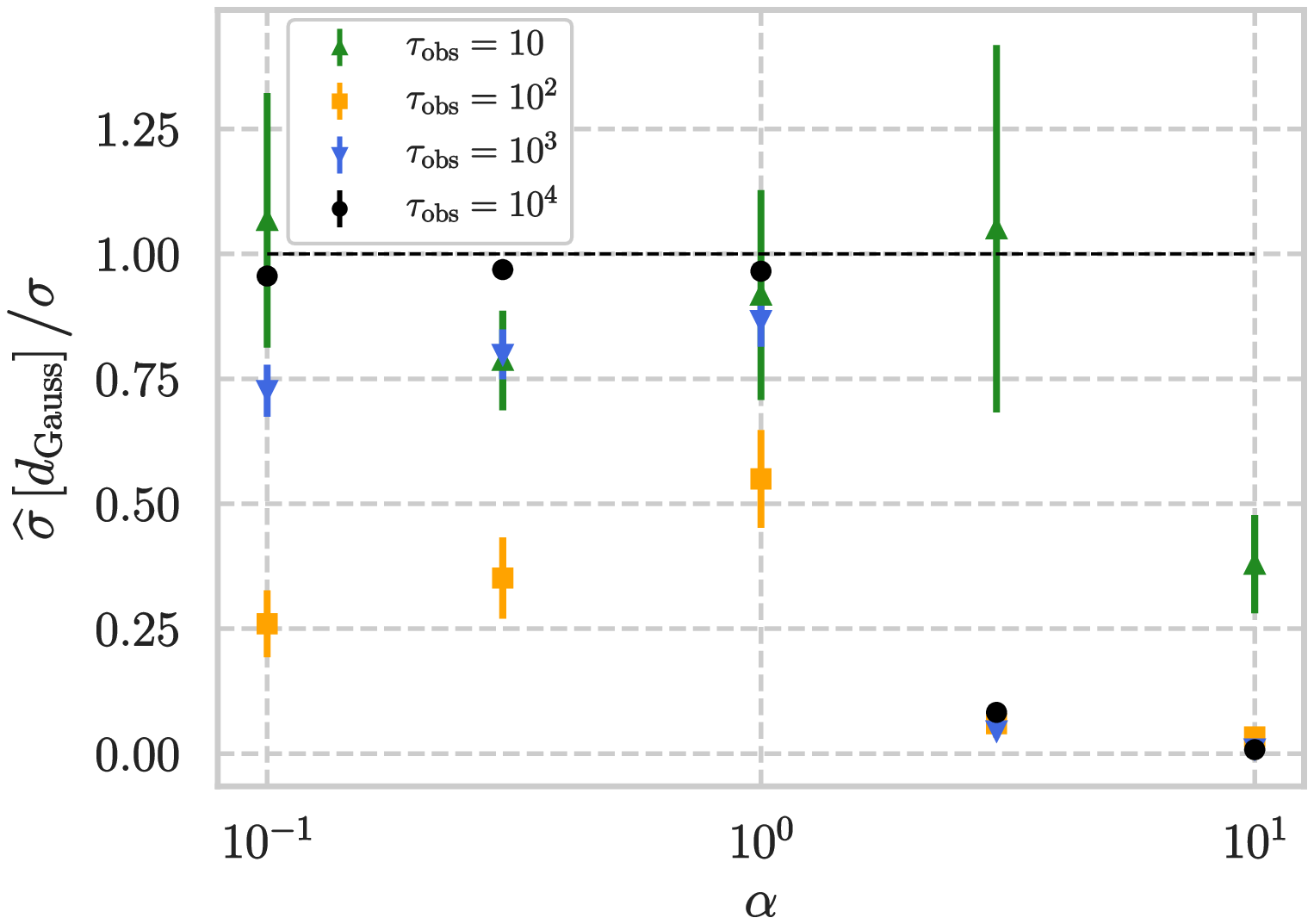}\label{fig: alpha_dependence_mx_A1}}	
\end{tabular}
\caption{Hyperparameter tuning of the Gaussian learning estimator for the Mexican-hat potential model ($A= 1$). (a) The $N_{\rm bin}$ dependence with the trajectory length $\tau_{\rm obs} = 10^3$. (b) The $\alpha$ dependence for the trajectory length $\tau_{\rm obs}$ from $10$ to $10^4$.}
\label{fig: hyperparam5}
\end{center}
\end{figure}

\begin{figure}
\begin{center}
\begin{tabular}{cc}
	\subfigure[$\alpha = 0.1$]{
		\includegraphics[width = 0.45\linewidth]{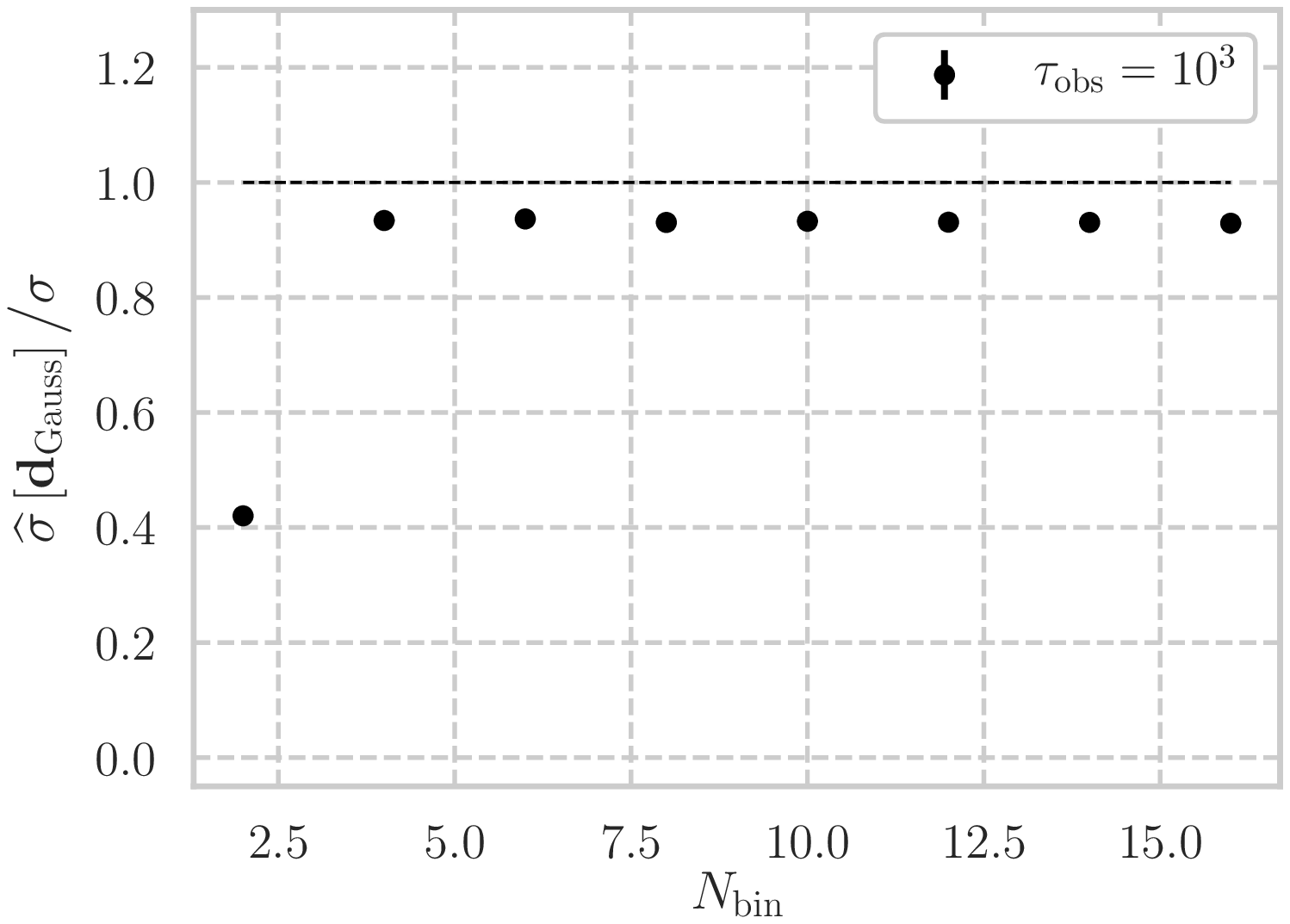}\label{fig: binN_dependence_mx_A100}}&
	\subfigure[$N_{\rm bin} = 6$]{
		\includegraphics[width = 0.45\linewidth]{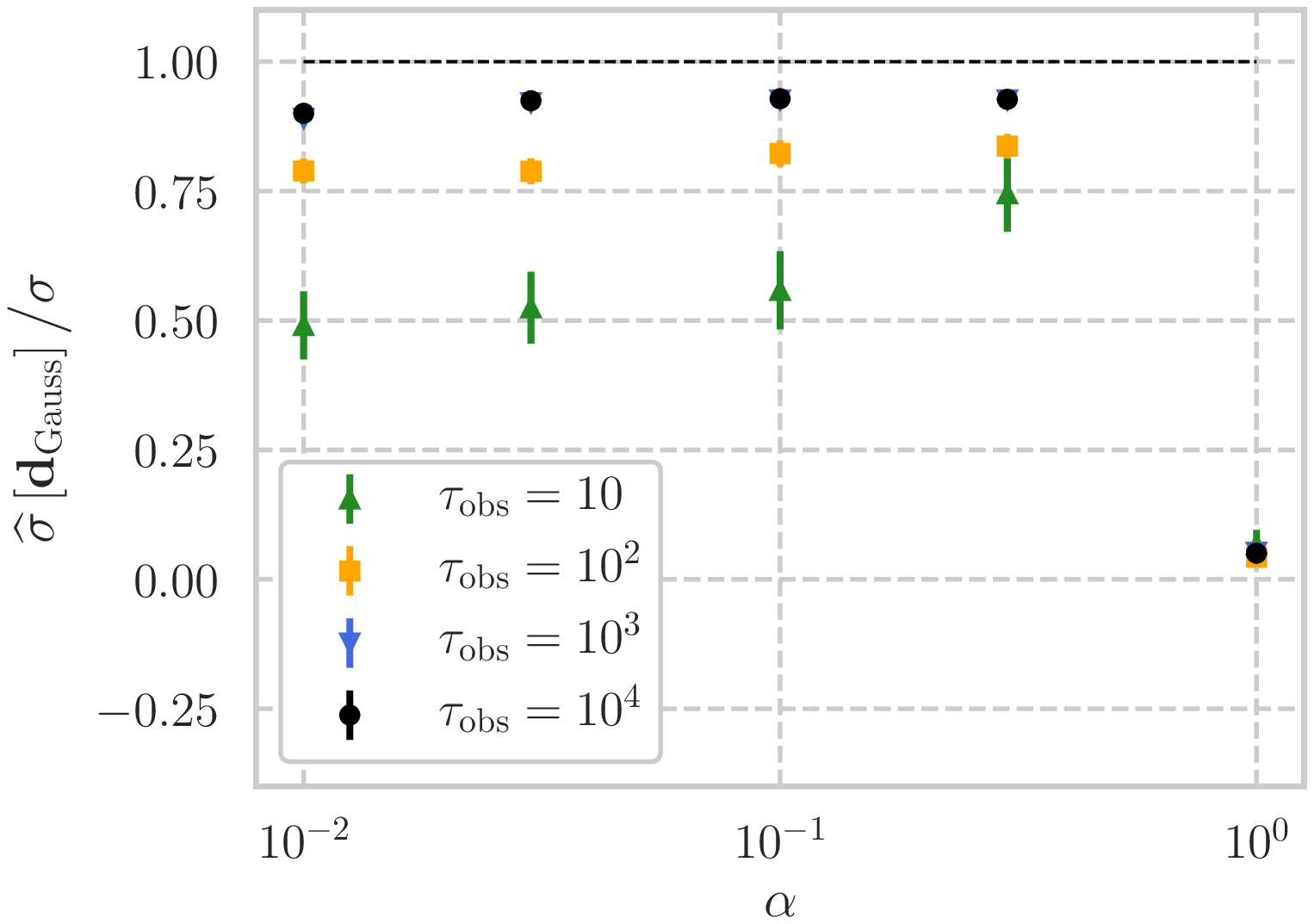}\label{fig: alpha_dependence_mx_A100}}	
\end{tabular}
\caption{Hyperparameter tuning of the Gaussian learning estimator for the Mexican-hat potential model ($A= 10^2$). (a) The $N_{\rm bin}$ dependence with the trajectory length $\tau_{\rm obs} = 10^3$. (b) The $\alpha$ dependence for the trajectory length $\tau_{\rm obs}$ from $10$ to $10^4$.}
\label{fig: hyperparam6}
\end{center}
\end{figure}

\noindent
\textbf{2. Computation time}\\ \indent

\begin{figure}
\begin{center}
\begin{tabular}{cc}
	\subfigure[Two-beads model]{
		\includegraphics[width = 0.45\linewidth]{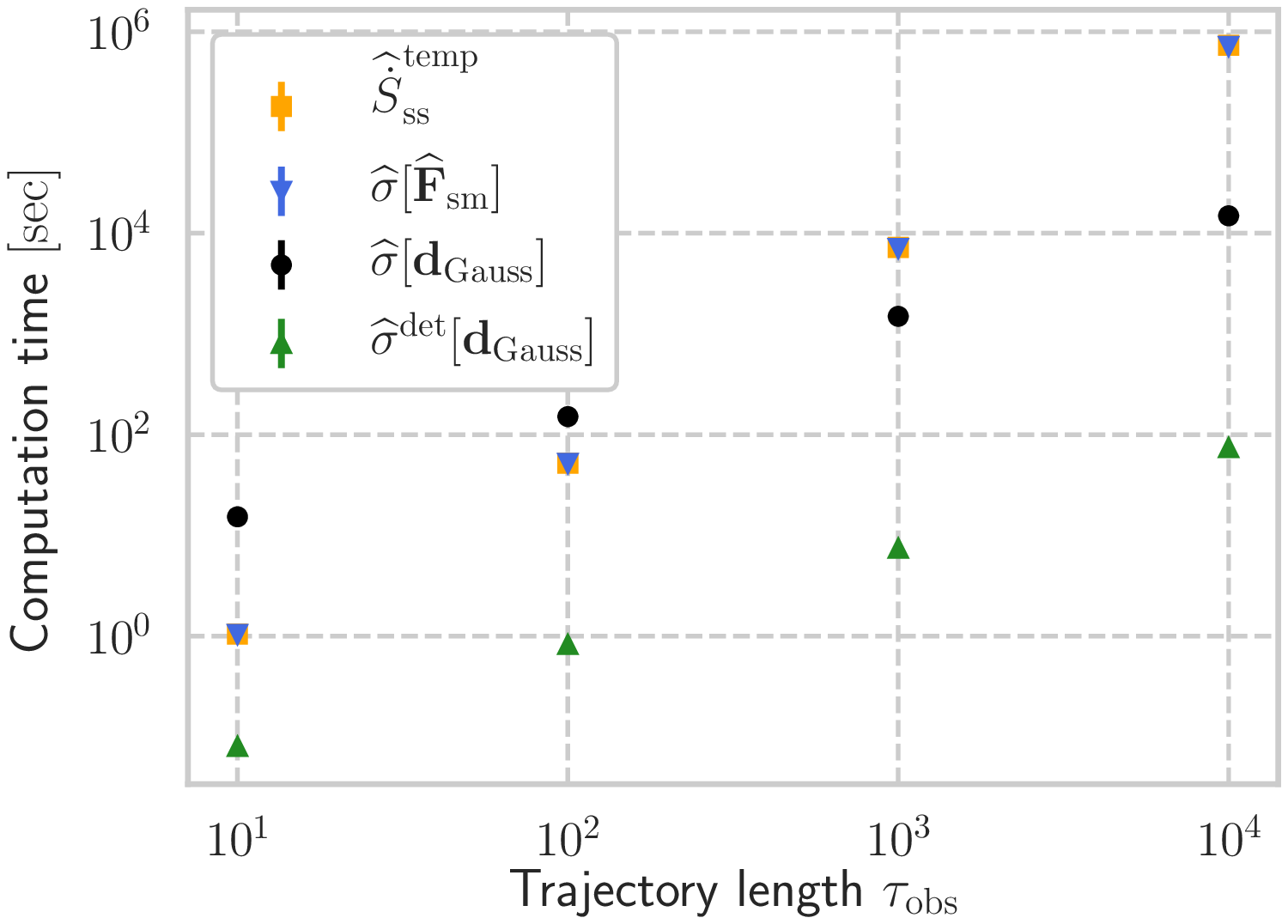}\label{fig: 2beads_time_r01}}&
	\subfigure[Five-beads model]{
		\includegraphics[width = 0.45\linewidth]{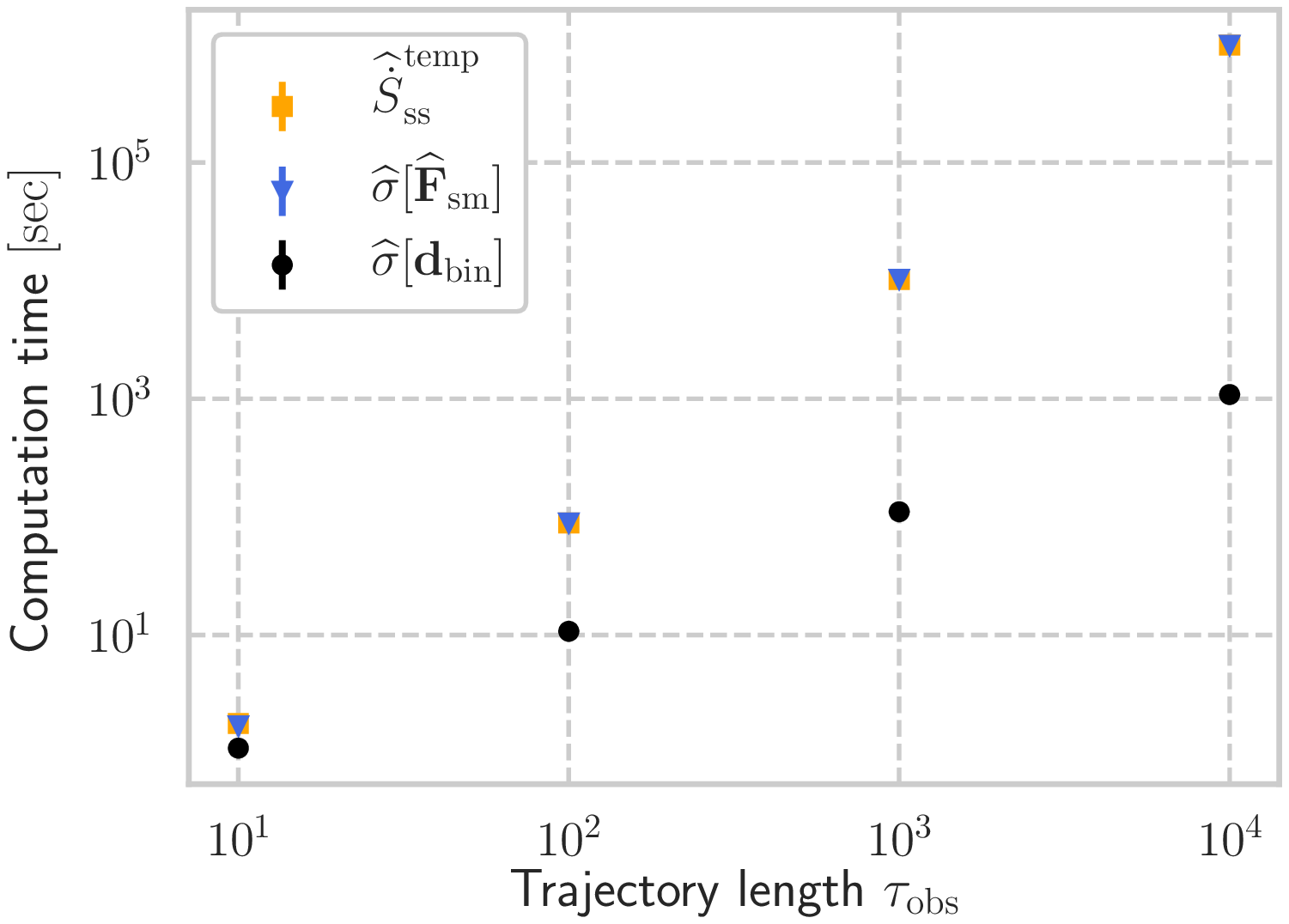}\label{fig: 5beads_time_r01}}
\end{tabular}
\caption{Comparison of the estimators in terms of the computation time: (a) The computational time of the estimators in the two-beads model ($T_c/T_h = 0.1$) with \footnotesize$\widehat{\dot{S}}_{\rm ss}^{\rm temp}$\normalsize (yellow squares), $\widehat{\sigma}[\widehat{\bm F}_{\rm sm}]$ (blue triangles), $\widehat{\sigma}[{\bm d}_{\rm Gauss}]$ (black circles) and $\widehat{\sigma}^{\rm det}[{\bm d}_{\rm Gauss}]$ (green triangles). (b) The computational time of the estimators in the five-beads model ($T_c/T_h = 0.1$) with \footnotesize$\widehat{\dot{S}}_{\rm ss}^{\rm temp}$\normalsize (yellow squares), $\widehat{\sigma}[\widehat{\bm F}_{\rm sm}]$ (blue triangles) and $\widehat{\sigma}^\lambda[{\bm d}_{\rm bin}]$ (black circles). The computation time is measured as the time on a single core of a cluster computer, while all the estimators could utilize parallel computation. The mean and its standard deviation of ten independent trials are plotted. The other system parameters are set to the same as those in Fig.~\ref{fig: 2beads} and Fig.~\ref{fig: 5beads}.}
\label{fig: computation}
\end{center}
\end{figure}
We compare the computation time of the four estimators used for Langevin dynamics in the main text.
First, we show the computational complexities of these estimators.
Then, we compare them in both the two-beads and the five-beads models.
We show that the learning estimators have smaller computational complexities, which means that they are suitable for long trajectory data, while they require the additional cost of the hyperparameter tuning.
We also discuss the computational complexity of $\widehat{\sigma}^{\rm det}[{\bm d}_{\rm Gauss}]$ which is studied in Appendix~\ref{sec: move}.
\\ \indent
We analyze the computational complexities of the learning estimators $\widehat{\sigma}^\lambda[{\bm d}_{\rm bin}]$ and $\widehat{\sigma}[{\bm d}_{\rm Gauss}]$ in terms of the data size $N = \tau_{\rm obs}/\Delta t$ and $N_{\rm bin}$, which includes the process of training and evaluation of $\widehat{\sigma}[{\bm d}]|_{\rm test}$. 
We fix the number of iterations of the gradient ascent as $N_{\rm step}$, which we found is not necessary to increase as $N$ or $N_{\rm bin}$ increases.
Therefore, the total computational complexity equals the computational complexity of the calculation of $\widehat{\sigma}[{\bm d}]$ and its gradient. In the case of the binned learning estimator, the calculation of $\widehat{\sigma}[{\bm d}_{\rm bin}]$ can be implemented with $O(N)$, while its gradient can be implemented with $O(\max(N, N_{\rm bin}^{n_{\rm dim}}))$, where $n_{\rm dim}$ is the dimension of data, and $N_{\rm bin}^{n_{\rm dim}}$ comes from the calculation of the regularization term. On the other hand, in the case of the Gaussian learning estimator, both the calculation of $\widehat{\sigma}[{\bm d}_{\rm Gauss}]$ and that of its gradient scale as $O(NN_{\rm Gauss})$, where $N_{\rm Gauss}$ is the number of Gaussian functions and satisfies $N_{\rm Gauss} = N_{\rm bin}^{n_{\rm dim}}$.\\ \indent
On the other hand, the computational complexities of \footnotesize$\widehat{\dot{S}}_{\rm ss}^{\rm temp}$\normalsize and $\widehat{\sigma}[\widehat{{\bm F}}_{\rm sm}]$ are $O(N^2)$, since the calculation of $\widehat{{\bm F}}_{\rm sm}(x)$ requires $O(N)$ computation for each position ${\bm x}$.\\ \indent
We compare their computation times in the two-beads and the five-beads models in Fig.~\ref{fig: computation}. The computation time is evaluated as the time on a single core of a cluster computer, while all the estimators can be implemented using parallel computation. The result is in accordance with the computational complexity analysis, and the learning estimators become better as the trajectory length increases. For example, the Gaussian (binned) learning estimator is around 50 (1000) times faster than \footnotesize$\widehat{\dot{S}}_{\rm ss}^{\rm temp}$\normalsize and $\widehat{\sigma}[\widehat{{\bm F}}_{\rm sm}]$ at $\tau_{\rm obs} = 10^4$.\\ \indent
We discuss the computational complexity of $\widehat{\sigma}^{\rm det}[{\bm d}_{\rm Gauss}]$ here.
For the comparison with $\widehat{\sigma}[{\bm d}_{\rm Gauss}]$, the same function is used for ${\bm d}_{\rm Gauss}({\bm x})$, while $\widehat{\sigma}^{\rm det}[{\bm d}_{\rm Gauss}]$ only optimizes the coefficients $w_k(i)$.
The computational complexity of $\widehat{\sigma}^{\rm det}[{\bm d}_{\rm Gauss}]$ is $O(\max(NN_{\rm Gauss},  N_{\rm Gauss}^3))$, where the latter term comes from the calculation of an inverse matrix \cite{Tan2020}.
Although the computational complexity is the same as that of the Gaussian learning estimator when $N_{\rm Gauss}$ is small, it can be expected that $\widehat{\sigma}^{\rm det}[{\bm d}_{\rm Gauss}]$ computes around $N_{\rm step}$ times faster because it does not require the iteration of the gradient scent. In Fig.~\ref{fig: computation}(a), we compare the computation time of $\widehat{\sigma}^{\rm det}[{\bm d}_{\rm Gauss}]$ with $\widehat{\sigma}[{\bm d}_{\rm Gauss}]$. The result is consistent with the discussion above, and $\widehat{\sigma}^{\rm det}[{\bm d}_{\rm Gauss}]$ is faster than $\widehat{\sigma}[{\bm d}_{\rm Gauss}]$ with a constant factor around $200$.\\ \indent
We note that the cost of the hyperparameter tuning is not taken into account in the computation time in Fig.~\ref{fig: computation}, while one may argue that the
hyperparameter tuning should be taken into account as an additional computational cost.
Such a cost might depend on the way that we implement the hyperparameter tuning and on the precision of the estimation required for our task, and can be small enough such that it does not compensate for the advantage of our machine learning method when the trajectory length is large. For example, it would be a good strategy to start with the hyperparameter tuning with shorter-length trajectories to reduce computation time, because we can expect that the optimal values would not drastically change as the trajectory length increases.
Indeed, we numerically confirmed that the optimal hyperparameters for the Gaussian learning estimator are almost independent of the trajectory length (see TABLE~\ref{table: hyperparameters}). It is an interesting future issue to give a theoretical foundation of this observation.

\end{document}